\documentclass{article}

\usepackage{PRIMEarxiv}

\usepackage[utf8]{inputenc} 
\usepackage[T1]{fontenc}    
\usepackage{hyperref}       
\usepackage{url}            
\usepackage{booktabs}       
\usepackage{amsfonts}       
\usepackage{nicefrac}       
\usepackage{microtype}      
\usepackage{lipsum}
\usepackage{fancyhdr}       
\usepackage{graphicx}       
\graphicspath{{media/}}     

\usepackage{amsmath,amsthm,amssymb}
\usepackage{stmaryrd}

\usepackage{physics}
\usepackage[version=4]{mhchem}
\usepackage{chemist}
\usepackage{siunitx}

\usepackage{array}
\usepackage{empheq}
\usepackage{enumitem}
\usepackage{scalerel}
\usepackage{subcaption}

\usepackage{upgreek}
\usepackage{stmaryrd}
\SetSymbolFont{stmry}{bold}{U}{stmry}{m}{n}
\usepackage{bm}

\usepackage{listings}

\usepackage{xcolor}
\definecolor{linkcol}{rgb}{0,0,0.4}
\definecolor{citecol}{rgb}{0.5,0,0}
\definecolor{blue_sketch}{rgb}{0,0,0.4}
\definecolor{red_sketch}{rgb}{0.5,0,0}
\definecolor{light-gray}{gray}{0.8}
\usepackage{hyperref}
\hypersetup{
    colorlinks=true,
    citecolor=black,
    linkcolor=linkcol,
    citecolor=citecol,
    urlcolor=red,
    linktoc=all
}

\usepackage{amsthm}

\newtheorem{theorem}{Theorem}[section]

\usepackage{tikz}
\usetikzlibrary{shapes.geometric, arrows}
\tikzstyle{program} = [rectangle, rounded corners, minimum width=3cm, minimum height=1cm, text centered, draw=black, fill=red!30!white]

\tikzstyle{subroutine} = [rectangle, minimum width=3cm, minimum height=1cm, text centered, draw=black, fill=blue!30]

\tikzstyle{classmethod} = [rectangle, minimum width=3cm, minimum height=1cm, 
text centered, draw=black, fill=orange!30]

\tikzstyle{classmember} = [rectangle, minimum width=3cm, minimum height=1cm, 
text centered, draw=black, fill=green!30]

\tikzstyle{arrow} = [thick, ->, >=stealth]

\newcolumntype{P}[1]{>{\centering\arraybackslash}p{#1}}

\newcommand{\x}{\vb{x}}

\newcommand{\U}{\vb{U}}
\newcommand{\uv}{\vb{u}}

\newcommand{\vF}{\vb{F}}

\newcommand{\vU}{\vb{U}}

\DeclareMathSymbol{\upD}{\mathalpha}{operators}{1}
\DeclareMathSymbol{\upP}{\mathalpha}{operators}{5}
\DeclareMathSymbol{\upS}{\mathalpha}{operators}{6}
\DeclareMathSymbol{\upO}{\mathalpha}{operators}{10}

\newcommand{\veps}{\varepsilon}

\newcommand{\mrm}{\mathrm}
\newcommand{\mc}{\mathcal}

\newcommand{\md}{\mathrm{d}}

\newcommand{\dvi}{\partial V_i}
\newcommand{\Sij}{\vb{S}_{i, j}^\tau}
\newcommand{\Si}{\vb{S}_i^\tau}

\newcommand{\bGamma}{\bm{\Upgamma}}
\newcommand{\btau}{\bm{\uptau}}

\newcommand\ppartial{\scaleobj{0.7}{\partial}}

\makeatletter
\newskip\@bigflushglue \@bigflushglue = -200pt plus 1fil

\def\bigcentering{\let\\\@centercr\rightskip\@bigflushglue%
	\leftskip\@bigflushglue
	\parindent\z@\parfillskip\z@skip}

\title{AVIP: a low temperature plasma code}

\author{
  Lionel Cheng, Nicolas Barleon, Olivier Vermorel, Benedicte Cuenot \\
  CERFACS \\
  42 Avenue Gaspard Coriolis\\
  Toulouse, France\\
  \texttt{\{cheng, barleon, vermorel, cuenot\}@cerfacs.fr} \\
   \And
  Anne Bourdon \\
  Laboratoire de Physique des Plasmas (LPP) \\
  Ecole Polytechnique, Institut Polytechnique de Paris, \\
  91120 Palaiseau, France \\
  \texttt{anne.bourdon@lpp.polytechnique.fr} \\
}

\begin{document}
\maketitle

\begin{abstract}
  A new unstructured, massively parallel code dedicated to low temperature plasmas, AVIP, is presented to simulate plasma discharges in interaction with combustion. The plasma species are modeled in a drift-diffusion formulation and the Poisson equation is solved consistently with the charged species. Plasma discharges introduce stiff source terms on the reactive Navier-Stokes equations and Riemann solvers, more robust than the schemes available in AVBP, have been implemented in AVIP and reported back in AVBP to solve the reactive Navier-Stokes equations. The validation of all the numerical schemes is carried out in this paper where numerous validation cases are presented for the plasma drift-diffusions equations and the reactive Navier-Stokes equations.
\end{abstract}

\keywords{Plasma \and Combustion \and Plasma discharges \and Riemann solvers \and Total Variation Diminishing}

\section{Introduction}

AVIP is a non-structured code written in Fortran for the numerical simulation of weakly-ionized plasmas. Two types of weakly-ionized plasmas are targetted: Hall-effect thruster plasmas and plasma discharges for plasma assisted combustion (PAC). AVIP has a similar data structure compared to AVBP to which it is coupled for plasma assisted combustion (PAC). Note that a subsequent version including Hall-effect thruster simulation capabilities is in preparation.

Governing equations for plasma assisted combustion are first presented: plasma drift-diffusion equations \cite{raimbault} are considered to simulate charged species. In plasma assisted combustion, charged species interact with neutral particles so that the mixture reactive Navier-Stokes equations need to be solved. In a second part, Sections \ref{sec:integrating_poisson} and \ref{sec:num_schemes} deal with numerical integration of the Poisson equation and transport equations. Finally Sections.~\ref{sec:plasma_validation} and \ref{sec:ns_validation} showcase the different validation cases for the schemes presented. Plasma and combustion are separated in all these cases: each part is validated separately and the interaction of the two will be presented in future work.


\section{Governing equations}

\label{sec:governing_equations}

\subsection{Electromagnetism}

In plasma simulations, the electric $\mathbf{E}$ and magnetic $\vb{B}$ fields appear in the transport equations and need to be computed. Magnetic field $\vb{B}$ is only used in HE thrusters simulation where it is constant and is negligible for NRP discharges simulations so that the electric field can be computed from the electromagnetic potential $\vb{E} = - \nabla \phi$. The electromagnetic potential is governed by the Poisson equation supplemented by boundary conditions \cite{jackson_electrodynamics}:

\begin{align}
    \nabla^2 \phi &= - \frac{\rho}{\varepsilon_0} \qq{in} \dot{\upO} \label{eq:poisson}\\
    \label{eq:poisson_dirichlet}
    \phi &= \phi_D \qq{on} \partial \upO_D \\
    \label{eq:poisson_neumann}
    \nabla \phi \cdot \vb{n} &= - E_n \qq{on} \partial \upO_N
\end{align}

\noindent where $\dot{\upO}$ denotes the interior of the domain, $\partial \upO_D$ the Dirichlet boundary, $\phi_D$ its imposed potential, $\partial \upO_N$ the Neumann boundary and $E_n$ its associated imposed normal electric field.

\subsection{Transport equations}

\subsubsection{Drift-diffusion equations}

Each plasma species $i$ is modeled using the drift-diffusion approximation \cite{raimbault,celes2008,tholin2012}:

\begin{equation}
    \pdv{n_i}{t} + \nabla \cdot \bGamma_i = S_{0i} \qquad \mrm{with} \, \bGamma_i = n_i \mu_i \vb{E} - D_i \nabla n_i
\end{equation}

\noindent where $n_i$ is the number density $i$, $\mu_i$ the mobility and $D_i$ the diffusion coefficient of species $i$. $\vb{E}$ is the electric field solved consistently with the plasma species using the Poisson equation recalled above.

\subsubsection{Gas mixture of neutral particules}

The reactive Navier Stokes equation are solved for a mixture of $N$ species:

\begin{align}
    \begin{split}
    &\pdv{\rho}{t} + \nabla \cdot [\rho \vb{u}] = 0 \\
    &\pdv{\rho \vb{u}}{t} + \nabla \cdot [\rho \vb{u} \vb{u} + p \vb{I} - \btau] = \rho \sum_{k=1}^N Y_k \vb{f}_k \\
    &\pdv{\rho E}{t} + \nabla \cdot [(\rho E + p) \vb{u} - \btau \cdot \vb{u} + \vb{q}] = \dot{\omega}_T + \dot{Q} + \rho \sum_{k=1}^N Y_k \vb{f}_k \cdot (\vb{u} + \vb{V}_k) \\
    &\pdv{\rho Y_k}{t} + \nabla \cdot [\rho Y_k(\vb{u} + \vb{V}_k)] = \dot{\omega_k}
    \end{split}
    \label{eq:ns_reactive}
\end{align}

\noindent where $p$ is the pressure, $\btau$ the viscous stress tensor, $\vb{f}_k$ the mass force acting on species $k$, $\vb{q}$ the heat flux, $\dot{\omega}_T$ the heat release due to the chemical reactions, $\dot{Q}$ an external energy source, $\dot{\omega}_k$ the mass production rate of species $k$ and $\vb{V}_k$ the diffusion velocity of species $k$ into the mixture. The energy equation is derived from the total energy equation and several other formulations are possible \cite[Chap. 1.1.5]{tnc}. The total non chemical energy equation is the one that is used throughout the simulations of AVIP.

\section{Numerical methods}

The different numerical methods used for each set of equations are detailed in this section. The nomenclature related to the unstructured vertex-centered finite volume methods is first presented. The Poisson equation discretization is then detailed in cartesian and cylindrical frames. Next two robust numerical schemes for drift-diffusion equations are presented: the improved Scharfetter-Gummel \cite{Kulikovsky1995} and limited Lax-Wendroff \cite{hirsch} schemes. Finally the implementation of the Riemann solver HLLC and its high-order MUSCL reconstruction are explained in the last part.

\subsection{Nomenclature and notations}

All the nomenclature and notations used in the paper concerning the computational grid or mesh are summarized below:

\begin{itemize}[label={--}]
    \item $\upO$ is the open set which corresponds to the domain of integration. Nomenclature from topology is adopted to indicate the interior of the domain $\mathring{\upO}$ and the boundary of the domain $\partial\upO = \overline{\upO} / \mathring{\upO}$ where $\overline{\upO}$ is the closure of $\upO$
    \item $\tau$ is a cell of the domain, $V_\tau$ is the cell volume, $n_v^\tau$ the number of vertex of the cell
    \item $i, j, k$ are nodes of the domain. To each node $i$ is associated a volume $V_i$, $\dvi$ is the boundary of the nodal volume, $\mathring{\dvi} = \mathring{\upO} \cap \dvi$ is the interior boundary of the nodal volume, $\dvi^b =  \partial \upO \cap \dvi$ is the boundary of the nodal volume which is also a boundary of the domain
    \item $E(i)$ is the set of elements (or cells) for which $i$ is a vertex
    \item $f$ is a face of a nodal surface or an element
    \item $\Sij$ is the surface normal of the edge $(i, j)$ in cell $\tau$ which is oriented from $i$ to $j$. $\Si$ is the surface associated to node $i$ in cell $\tau$, pointing inward towards the cell center and defined as a linear combination of the adjacent faces:

    \begin{equation}
        \label{eq:node_normal}
        \Si = - \frac{n_d}{n_v^f} \vb{S}^\tau_f
    \end{equation}

    \noindent where $n_d$ is the number of dimensions of the computational grid, $n_v^f$ the number of vertices on the face $f$. The cell dependence $\tau$ of the normals is sometimes omitted to alleviate notations.
    \end{itemize}

\noindent For a scalar variable $u(\x, t)$ the nodal average is defined as

\begin{equation}
u_i(\x, t) = \frac{1}{V_i} \int_{V_i} u \, dV
\end{equation}

\begin{figure}[htbp]
    \centering
    \begin{subfigure}[b]{0.45\textwidth}
        \begin{tikzpicture}

    \draw[thick] (0, 0) node[anchor=east] {$i$} -- (6, 0) node[anchor=west] {$j$};
    \draw[thick] (6, 0) -- (3, 5.196) node[anchor=south] {$k$};
    \draw[thick] (0, 0) -- (3, 5.196);

    \draw[dashed, draw=blue_sketch] (3, 0) -- (3, 1.7);

    \node[anchor=north east, text=blue_sketch] at (3, 1.7){$\dvi \cap \tau$};
    \node[anchor=north west, text=blue_sketch] at (3, 1.7){$\partial V_j \cap \tau$};
    \node[anchor=south, text=blue_sketch] at (3, 1.9){$\partial V_k \cap \tau$};

    \draw[dashed, draw=blue_sketch] (1.5, 2.6) -- (3, 1.7);
    \draw[dashed, draw=blue_sketch] (4.5, 2.6) -- (3, 1.7);

    \draw[very thick, draw=red_sketch, ->] (3, 0.85) -- (4, 0.85) node[anchor=west, text=red_sketch]{$\vb{S}_{ij}^\tau$};
    \draw[very thick, draw=red_sketch, ->] (3.75, 2.15) -- (3.1, 3.2) node[anchor=west, text=red_sketch]{$\vb{S}_{jk}^\tau$};
    \draw[very thick, draw=red_sketch, ->] (2.25, 2.15) -- (1.6, 1.05) node[anchor=north west, text=red_sketch]{$\vb{S}_{ki}^\tau$};

    \draw[->] (3.5, 0) -- (3.5, 0.85);
    \node[anchor=center] at (3.8, 0.5) {$r_{ij}^\tau$};

\end{tikzpicture}
        \caption{Triangular cell}
        \label{fig:tri_cell}
    \end{subfigure}
    \begin{subfigure}[b]{0.45\textwidth}
        \begin{tikzpicture}

    \draw[thick] (0, 0) node[anchor=east] {$i$} -- (6, 0) node[anchor=west] {$j$};
    \draw[thick] (0, 4) node[anchor=east] {$l$} -- (6, 4) node[anchor=west] {$k$};
    \foreach \x in {0, 6}
        \draw[thick] (\x, 0) -- (\x, 4);

    \draw[dashed, draw=blue_sketch] (3, 0) -- (3, 4);
    \draw[dashed, draw=blue_sketch] (0, 2) -- (6, 2);

    \node[anchor=north east, text=blue_sketch] at (3, 2) {$\dvi \cap \tau$};
    \node[anchor=north west, text=blue_sketch] at (3, 2) {$\partial V_j \cap \tau$};
    \node[anchor=south east, text=blue_sketch] at (3, 2) {$\partial V_l \cap \tau$};
    \node[anchor=south west, text=blue_sketch] at (3, 2) {$\partial V_k \cap \tau$};

    \draw[very thick, draw=red_sketch, ->] (3, 1) -- (4, 1) node[anchor=west, text=red_sketch]{$\vb{S}_{ij}^\tau$};
    \draw[very thick, draw=red_sketch, ->] (4.5, 2) -- (4.5, 3) node[anchor=west, text=red_sketch]{$\vb{S}_{jk}^\tau$};
    \draw[very thick, draw=red_sketch, ->] (3, 3) -- (2, 3) node[anchor=east, text=red_sketch]{$\vb{S}_{kl}^\tau$};
    \draw[very thick, draw=red_sketch, ->] (1.5, 2) -- (1.5, 1) node[anchor=west, text=red_sketch]{$\vb{S}_{li}^\tau$};

    \draw[->] (3.5, 0) -- (3.5, 1);
    \node[anchor=center] at (3.8, 0.6) {$r_{ij}^\tau$};

    \end{tikzpicture}
        \caption{Quadrangular cell}
        \label{fig:quad_cell}
    \end{subfigure}
    \caption{Metrics definitions in 2D cells.}
    \label{fig:2D_cells}
\end{figure}
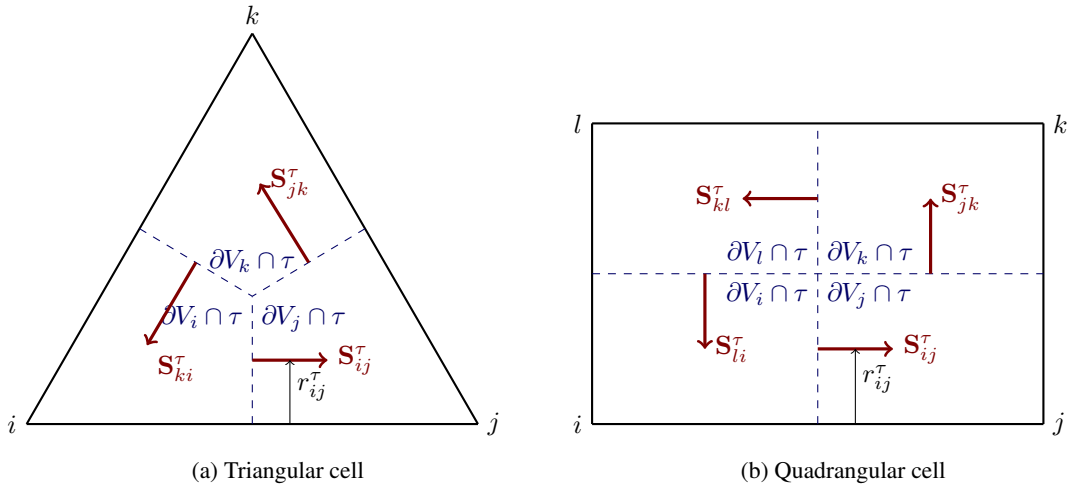

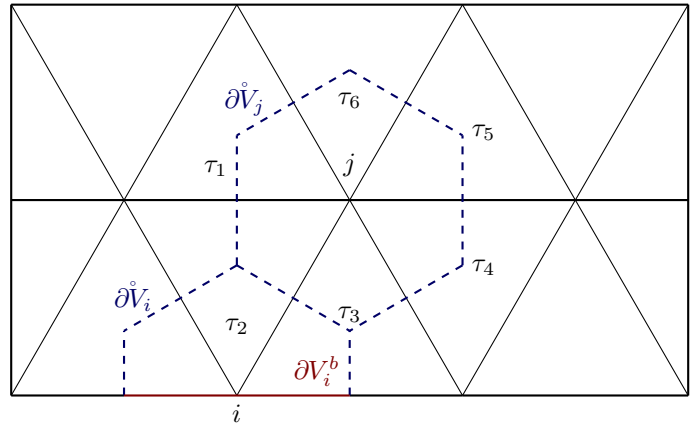
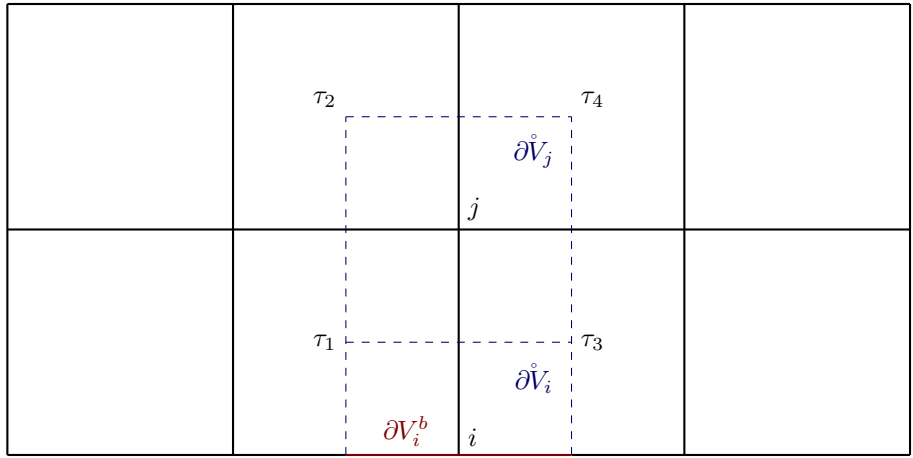
\begin{figure}[htbp]
    \centering
    \begin{subfigure}[b]{\textwidth}
        \centering
        \begin{tikzpicture}

    \draw[thick] (-3, 0) -- (-1.5, 0);
    \draw[thick] (1.5, 0) -- (3, 0);
    \draw[thick] (3, 0) -- (6.0, 0);
    \draw[thick] (-3, 2.6) -- (6, 2.6);
    \draw[thick] (-3, 5.2) -- (6, 5.2);
    \draw[thick] (-3, 0) -- (-3, 5.2);
    \draw[thick] (6, 0) -- (6, 5.2);

    \foreach \x in {-3, 0, 3}
        \draw (\x, 0) -- (\x + 3, 5.2);
    \foreach \x in {-3, 0, 3}
        \draw (\x + 3, 0) -- (\x, 5.2);

    \draw[thick, dashed, draw=blue_sketch] (1.5, 0) -- (1.5, 0.86);
    \draw[thick, dashed, draw=blue_sketch] (-1.5, 0) -- (-1.5, 0.86);
    \draw[thick, dashed, draw=blue_sketch] (0, 1.732) -- (-1.5, 0.86);
    \draw[thick, draw=red_sketch] (-1.5, 0) -- (1.5, 0) node[anchor=south east, text=red_sketch]{$\partial V_i^b$};
    \node[anchor=north] at (0, 0) {$i$};
    \node[anchor=south east, text=blue_sketch] at (-1, 1) {$\mathring{\dvi}$};

    \node[anchor=south] at (1.5, 2.8) {$j$};
    \draw[thick, dashed, draw=blue_sketch] (0, 1.732) -- (1.5, 0.86);
    \draw[thick, dashed, draw=blue_sketch] (3, 1.732) -- (1.5, 0.86);
    \draw[thick, dashed, draw=blue_sketch] (3, 1.732) -- (3, 3.461);
    \draw[thick, dashed, draw=blue_sketch] (0, 1.732) -- (0, 3.461);
    \draw[thick, dashed, draw=blue_sketch] (1.5, 4.33) -- (0, 3.461);
    \draw[thick, dashed, draw=blue_sketch] (1.5, 4.33) -- (3, 3.461);
    \node[anchor=south east, text=blue_sketch] at (0.5, 3.6) {$\mathring{\partial V_j}$};

    \node[anchor=south east] at (0, 2.8) {$\tau_1$};
    \node[anchor=south] at (0, 0.7) {$\tau_2$};
    \node[anchor=south] at (1.5, 0.85) {$\tau_3$};
    \node[anchor=west] at (3, 1.7) {$\tau_4$};
    \node[anchor=west] at (3, 3.5) {$\tau_5$};
    \node[anchor=south] at (1.5, 3.7) {$\tau_6$};

\end{tikzpicture}
        \caption{Triangular elements}
        \label{fig:tri_domain}
    \end{subfigure}
    \centering
    \begin{subfigure}[b]{\textwidth}
        \centering
        \begin{tikzpicture}

\draw[thick] (-6, 0) -- (-1.5, 0);
\draw[thick] (6, 0) -- (1.5, 0);
\foreach \y in {3, 6}
    \draw[thick] (-6, \y) -- (6, \y);
\foreach \x in {-6, -3, 0, 3, 6}
    \draw[thick] (\x, 0) -- (\x, 6);

\draw[dashed, draw=blue_sketch] (-1.5, 0) -- (-1.5, 4.5);
\draw[dashed, draw=blue_sketch] (1.5, 0) -- (1.5, 4.5);
\foreach \y in {1.5, 4.5}
    \draw[dashed, draw=blue_sketch] (-1.5, \y) -- (1.5, \y);
\node[anchor=south, text=blue_sketch] at (1, 0.7) {$\mathring{\dvi}$};
\node[anchor=south, text=blue_sketch] at (1, 3.7) {$\mathring{\partial V_j}$};

\draw[thick, draw=red_sketch] (-1.5, 0) -- (1.5, 0);
\node[anchor=south, text=red_sketch] at (-0.7, 0) {$\dvi^b$};

\node[anchor=south west] at (0, 0) {$i$};
\node[anchor=south west] at (0, 3) {$j$};

\node[anchor=east] at (-1.5, 1.5) {$\tau_1$};
\node[anchor=south east] at (-1.5, 4.5) {$\tau_2$};
\node[anchor=west] at (1.5, 1.5) {$\tau_3$};
\node[anchor=south west] at (1.5, 4.5) {$\tau_4$};

\end{tikzpicture}
        \caption{Quadrangular elements}
        \label{fig:quad_domain}
    \end{subfigure}
    \caption{2D meshes with nodes $i$ and $j$ belonging respectively to the interior and the boundary of the domain for triangular and quadrangular elements. Neighboring cells belonging to $E(j)$ are shown for both cases.}
\end{figure}

Individual triangular and quadrangular cells with the different normals defined above are shown in Figures \ref{fig:tri_cell} and \ref{fig:quad_cell}. Triangular and quadrangular domains are depicted in Figures \ref{fig:tri_domain} and \ref{fig:quad_domain} where nodal volumes are drawn.

\subsection{Cylindrical frames}

\begin{figure}[htbp]
    \centering
    \begin{tikzpicture}
    \draw[->] (0, 0) -- (8.5, 0) node[anchor=north] {$x$};
    \draw[->] (0, -2) -- (0, 3) node[anchor=east] {$r$};
    \draw[domain=-90:90, smooth, dashed, variable=\t] plot ({0.8 * cos(\t)}, {2 * sin(\t)});
    \draw[domain=-90:90, smooth, variable=\t] plot ({7 + 0.8 * cos(\t)}, {2 * sin(\t)});
    \draw[domain=90:270, smooth, variable=\t] plot ({7 + 0.8 * cos(\t)}, {2 * sin(\t)});
    \draw[domain=90:270, smooth, variable=\t] plot ({0.8 * cos(\t)}, {2 * sin(\t)});
    \draw (7, -2) -- (7, 2);
    \draw (0, 2) -- (7, 2);
    \draw (0, -2) -- (7, -2);
    \draw[fill=red_sketch, opacity=0.2] (0, 0) rectangle (7, 2);
    \draw (3, 0.5) rectangle (5, 1.5);
    \draw (3, -0.5) rectangle (5, -1.5);
    \draw[domain=-90:90, smooth, dashed, variable=\t] plot ({5 + 0.6 * cos(\t)}, {1.5 * sin(\t)});
    \draw[domain=-90:90, smooth, dashed, variable=\t] plot ({5 + 0.2 * cos(\t)}, {0.5 * sin(\t)});
    \draw[domain=-90:90, smooth, dashed, variable=\t] plot ({3 + 0.6 * cos(\t)}, {1.5 * sin(\t)});
    \draw[domain=-90:90, smooth, dashed, variable=\t] plot ({3 + 0.2 * cos(\t)}, {0.5 * sin(\t)});
    \node[anchor=center, red_sketch] at (3.2, 2.3) {Computational domain};
    \node[anchor=center, red_sketch] at (4.2, 1.) {Cell $\tau$};
\end{tikzpicture}
    \caption{Axisymmetric nodal volumes.}
    \label{fig:axi_sym_sketch}
\end{figure}
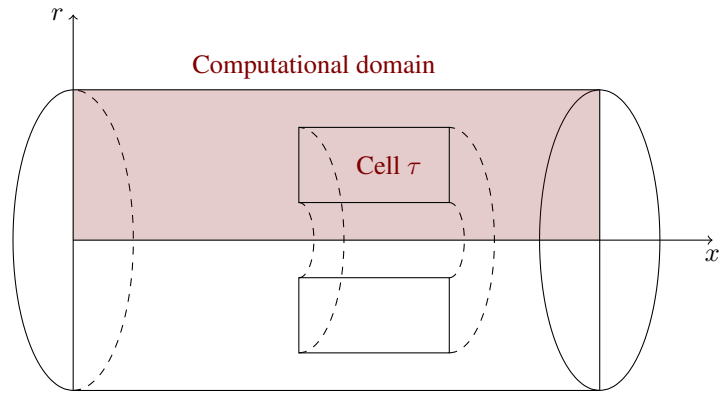

In this work, axisymmetric conditions, \textit{i.e.} $\pdv{\theta} = 0$, are often considered. This saves a lot computational time as a three-dimensional simulation can be carried out using a two-dimensional mesh. A typical cylindrical frame in the $(z, r)$ plane is shown in Fig.~\ref{fig:axi_sym_sketch}. In the following numerics chapters, the Poisson and plasma transport equations discretizations in cylindrical coordinates are thus presented along with their classical cartesian coordinates discretizations. Geometrical source terms appear for the Euler and Navier-Stokes equations which require special care for proper integration.

The elements and surfaces considered in this 2D mesh are actually tores in this setting as illustrated in Fig.~\ref{fig:axi_sym_sketch} and it should be always kept in mind that although we are solving the equations on a 2D mesh the geometry is in fact three dimensional.

\begin{figure}[htbp]
    \centering
    \begin{tikzpicture}
    \draw[domain=0:180, smooth, dashed, variable=\t] plot ({2 * cos(\t)}, {0.6 * sin(\t)});
    \draw[domain=-180:0, smooth, variable=\t] plot ({2 * cos(\t)}, {0.6 * sin(\t)});
    \draw[domain=0:360, smooth, variable=\t] plot ({2 * cos(\t)}, {4 + 0.6 * sin(\t)});
    \draw[domain=10:320, ->, smooth, red_sketch, variable=\t] plot ({1 * cos(\t)}, {2 + 0.3 * sin(\t)});
    \draw (2, 0) -- (2, 4);
    \draw (-2, 0) -- (-2, 4);
    \draw[->] (0, -1) -- (0, 5);
    \draw[<->] (0, 0) -- (-2, 0);
    \node[anchor=center] at (-1, -0.2) {$r$};
    \draw[<->] (0, 0) -- (1, 0);
    \node[anchor=center] at (0.5, -0.2) {$r / 2$};
    \draw[<->] (3, 0) -- (3, 4);
    \node[anchor=center] at (3.2, 2) {$h$};
    \draw[<->] (2.5, 0) -- (2.5, 2);
    \node[anchor=center] at (2.7, 1) {$\frac{h}{2}$};
    \draw[fill=red_sketch, opacity=0.3] (0, 0) rectangle (2, 4);
    \node[circle, fill, red_sketch, inner sep=2pt] at (1, 2) {};
    \node[anchor=center, red_sketch] at (-1, 1.5) {$d = 2 \pi \frac{1}{2}r$};
    \node[anchor=center, red_sketch] at (1, 4.2) {$A = h \times r$};
    \node[anchor=center, red_sketch] at (-3, 2) {$V = A \times d$};
\end{tikzpicture}
    \caption{Illustration of the first Pappus-Guldinus theorem with a cylinder volume.}
    \label{fig:pappus_guldinus_2}
\end{figure}
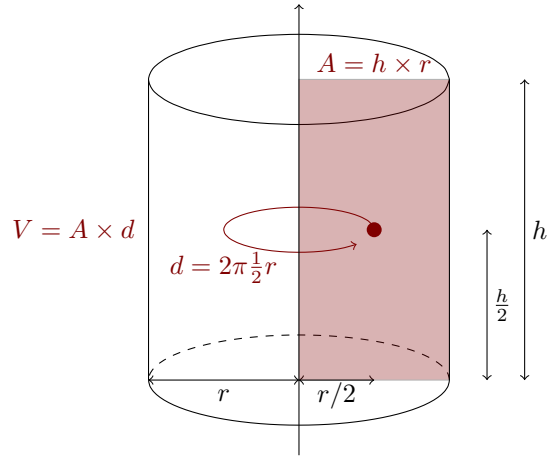

Two important results relevant to the finite volume method in cylindrical geometries are the following Pappus-Guldinus theorems illustrated in Figs.~\ref{fig:pappus_guldinus_1} and \ref{fig:pappus_guldinus_2} which deal with volumes and surfaces of revolution \cite{harris1998}. These theorems are used to compute fluxes and volumes in axisymmetric formulations of the Poisson equation and transports equations.

\begin{theorem}[First Pappus–Guldinus theorem]
    \label{theorem:pappus_first}
    The surface area A of a surface of revolution generated by rotating a plane curve C about an axis external to C and on the same plane is equal to the product of the arc length s of C and the distance d traveled by the geometric centroid of C:
\begin{equation}
    A = s\times d
\end{equation}
\end{theorem}

\begin{theorem}[Second Pappus–Guldinus theorem]
    \label{theorem:pappus_second}
    The volume V of a solid of revolution generated by rotating a plane figure F about an external axis is equal to the product of the area A of F and the distance d traveled by the geometric centroid of F. (Note that the centroid of F is usually different from the centroid of its boundary curve C.) That is:
\begin{equation}
    V = A\times d
\end{equation}
\end{theorem}

\begin{figure}[htbp]
    \centering
    \begin{tikzpicture}
    \draw[domain=0:180, smooth, dashed, variable=\t] plot ({2 * cos(\t)}, {0.6 * sin(\t)});
    \draw[domain=-180:0, smooth, variable=\t] plot ({2 * cos(\t)}, {0.6 * sin(\t)});
    \draw[domain=0:360, smooth, variable=\t] plot ({2 * cos(\t)}, {4 + 0.6 * sin(\t)});
    \draw[domain=10:320, ->, smooth, red_sketch, variable=\t] plot ({2 * cos(\t)}, {2 + 0.6 * sin(\t)});
    \draw[red_sketch] (2, 0) -- (2, 4);
    \draw (-2, 0) -- (-2, 4);
    \draw[->] (0, -1) -- (0, 5);
    \draw[<->] (0, -0.8) -- (2, -0.8);
    \draw[<->] (3, 0) -- (3, 4);
    \node[anchor=center] at (3.2, 2) {$h$};
    \draw[<->] (2.5, 0) -- (2.5, 2);
    \node[anchor=center] at (2.7, 1) {$\frac{h}{2}$};
    \node[anchor=center] at (1, -1) {$r$};
    \node[circle, fill, red_sketch, inner sep=2pt] at (2, 2) {};
    \node[anchor=center, red_sketch] at (-1, 1) {$d = 2 \pi r$};
    \node[anchor=center, red_sketch] at (1.5, 3) {$s = h$};
    \node[anchor=center, red_sketch] at (-3, 2) {$A = s \times d$};
\end{tikzpicture}
    \caption{Illustration of the first Pappus-Guldinus theorem with a cylinder surface.}
    \label{fig:pappus_guldinus_1}
\end{figure}
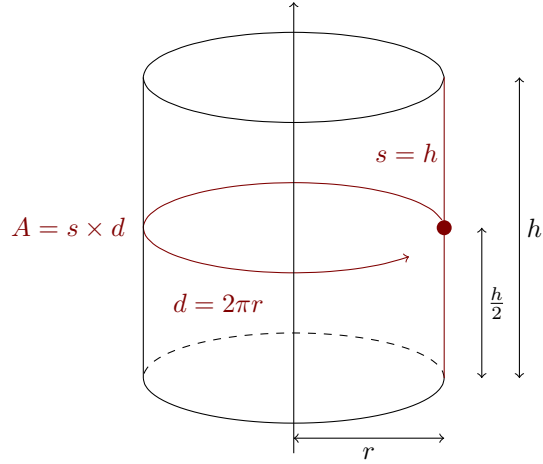

\subsection{Integrating the Poisson equation}

\label{sec:integrating_poisson}

The discretized equation for cartesian and cylindrical geometries are detailed in the vertex-centered finite volume formulation of AVIP. Validation cases are not shown in this paper and can be found in \cite{Cheng2022}.

\subsubsection{Cartesian geometry}

We integrate the Poisson equation Eq.~\eqref{eq:poisson} in the nodal volume $V_i$ for an interior node $i$ (\textit{i.e.} such that $\dvi^b = \emptyset$):

\begin{equation}
    \int_{V_i}\nabla^2 \phi \md V = \int_{\ppartial V_i} \nabla \phi \cdot \mathbf{n} \md S = \sum_{\tau \in E(i)} \int_{\ppartial V_i \cap \tau} \nabla \phi \cdot \mathbf{n} \md S
\end{equation}

The gradient is supposed to be constant inside each cell $\tau$ so that it can be extracted from the integral yielding

\begin{equation}
    \int_{V_i}\nabla^2 \phi dV = \sum_{\tau \in E(i)} \nabla \phi_{\tau} \cdot \int_{\ppartial V_i \cap \tau} \mathbf{n} dS = \sum_{\tau\in E(i)} \nabla\phi_{\tau} \cdot \sum_{j \in S(i)\cap\tau} \mathbf{S}_{i,j}^\tau
\end{equation}

The value of this constant gradient inside cell $\tau$ is computed from the Green-Ostrogradski theorem:

\begin{align}
    \nabla \phi_\tau & = \frac{1}{V_\tau} \int_{\ppartial V_\tau} \phi \, \mathbf{n} \, dS
    \\
    & = \frac{1}{V_\tau}\sum_{f\in F(\tau)} \phi_f \mathbf{S}_f
\end{align}

\noindent where $F(\tau)$ are the set of faces of an element $\tau$ and $\vb{n}$ denotes undimensionalized normals. Note that $\vb{S}_i$ has the dimension of a surface in 3D or a length in 2D. Following the definition of nodal normals Eq.\eqref{eq:node_normal} the gradient can be rewritten as \cite[Chap. 4]{lamarquethesis}:

\begin{equation}
    \nabla\phi_\tau = -\frac{1}{V_\tau n_d} \sum_{k\in\tau} \phi_k \mathbf{S}_k
\end{equation}
which in the end yields:
\begin{equation}
    \int_{V_i}\nabla^2 \phi dV = \sum_{\tau \in E(i)} -\frac{1}{V_\tau n_d} \sum_{k\in\tau} \phi_k \mathbf{S}_k \cdot \sum_{j \in S(i)\cap\tau} \mathbf{S}_{i,j}^\tau
\end{equation}

The following relation is exact in the case of triangular, tetrahedral and holds for regular quadrangular and hexahedral elements \cite{Auffray2007a}:
\begin{equation}
    \sum_{j \in S(i)\cap\tau} \mathbf{S}_{i,j}^\tau = \mathbf{S}^\tau_i / n_d
\end{equation}
which yields
\begin{equation}
    \int_{V_i}\nabla^2 \phi dV = \sum_{\tau \in E(i)} \sum_{k\in\tau} \left[-\frac{\mathbf{S}_k\cdot\mathbf{S}_i}{V_\tau n_d^2}\right] \phi_k.
\end{equation}

Finally with the integration of the charge density the following equation is obtained from the integration of both sides of the Poisson equation:

\begin{equation}
    \sum_{\tau \in E(i)} \sum_{k\in\tau} \left[-\frac{\mathbf{S}_k\cdot\mathbf{S}_i}{V_\tau n_d^2}\right] \phi_k = \frac{\rho_i V_i}{\veps_0}
    \label{eq:poisson_discretization_cartesian}
\end{equation}

Eq.~\eqref{eq:poisson_discretization_cartesian} defines a linear system, $i$ being the line and $k$ the column of the matrix for all $i$ and $k$ taking the values of all the inner nodes of the partition. One interesting feature of this matrix in cartesian coordinates is its symmetry. The Laplacian operator matrix $A$ is computed with node normal vectors and primal cell volumes where each coefficient $a_{ij}$ stands as:

\begin{equation}
\label{coeff_matriceA}
a_{ij} = \sum_{ij \in \tau} \left[-\frac{\mathbf{S}_i\cdot\mathbf{S}_j}{V_\tau n_d^2}\right]
\end{equation}

\noindent where the sum is performed over the cells for which $ij$ is an edge.

\textbf{Remark on the implementation: } The linear system solved in AVIP is in fact the one obtained from:
\begin{equation}
    -\nabla^2\phi = \frac{\rho}{\epsilon_0}
\end{equation}

\subsubsection{Axisymmetric configuration}

The solution of Poisson's equation in an $r-z$ axisymmetry geometry is needed in streamer simulations. First integration over the nodal volume is performed, here the volume is $dV = r \md r \md \theta \md z$:
\begin{equation}
    \int_{V_i}\nabla^2\phi \, \md V = \int_{V_i}\nabla^2 \phi \, r \md r \md\theta \md z.
\end{equation}
By assuming axisymmetric conditions \textit{i.e.} $\partial/\partial\theta = 0$, using Green's theorem, summing over
the neighboring cells and assuming constant gradient within a cell, it yields:
\begin{align}
    \int_{V_i}\nabla^2\phi dV &= 2\pi\int_{A_i}\nabla\cdot(r\nabla\phi)dA\\
     &= 2\pi\int_{\ppartial A_i} r\nabla\phi\cdot\vb{n}dl\\
     &= \sum_{\tau\in E(i)} 2\pi\int_{\ppartial A_i\cap\tau} r\nabla\phi\cdot d\vb{l}\\
     &= \sum_{\tau\in E(i)} 2\pi\nabla \phi_\tau\cdot\int_{\ppartial A_i\cap\tau} rd\vb{l}
\end{align}

In this section, all the normals denoted $\vb{n}$ are homogeneous to a length. The cell gradient is discretized as in the last section with nodal normals (note here that the gradient is really 2D here, all the axisymmetric information is contained with the radius):

\begin{equation}
    \nabla\phi_\tau = -\frac{1}{A_\tau n_d}\sum_{k\in\tau} \phi_k \vb{n}_k.
\end{equation}
Using the notations from Fig.~\ref{fig:tri_cell}, for a triangle the remaining integral yields:
\begin{equation}
    \int_{\ppartial A_i\cap\tau} rd\vb{l} = \frac{r_\tau+r_{ij}^\tau}{2}\vb{n}_{ij}^\tau + \frac{r_\tau+r_{ik}^\tau}{2}\vb{n}_{ik}^\tau
\end{equation}
From \cite{Auffray2007a}:
\begin{equation}
    \vb{n}_{ij}^\tau = \frac{1}{6}(\vb{n}_i-\vb{n}_j)
\end{equation}
In the end:
\begin{equation}
    \int_{\ppartial A_i\cap\tau} rd\vb{l} = \frac{1}{24}\left[(4r_\tau+2r_i+r_j+r_k)\vb{n}_i - (2r_\tau+r_i+r_j)\vb{n}_j-(2r_\tau+r_i+r_k)\vb{n}_k\right]
\end{equation}

For the quadrangular cell from Fig.~\ref{fig:quad_cell}, using similar relations the resulting integral yields:
\begin{equation}
    \int_{\ppartial A_i\cap\tau} rd\vb{l} = \frac{1}{16}\left[(4r_\tau+2r_i+r_j+r_k)\vb{n}_i - (2r_\tau+r_i+r_j)\vb{n}_j-(2r_\tau+r_i+r_k)\vb{n}_k\right]
\end{equation}

From these relations an equation similar to Eq.~\eqref{eq:poisson_discretization_cartesian} can be obtained yielding a different linear system: the radiuses introduced break the symmetry of the linear system matrix.

\subsection{Boundary conditions}

Two kind of boundary conditions are implemented in AVIP for the Poisson equation: Dirichlet and Neumann boundary conditions. In practice the laplacian discretization is computed for all nodes and then boundary conditions are applied on the Poisson matrix.

\subsubsection{Neumann boundary conditions}

For a Neumann boundary condition node $i$, \textit{e.g.} in Figs.~\ref{fig:tri_domain} and \ref{fig:quad_domain}, integration of the Poisson equation yields

\begin{equation}
    \int_{V_i}\nabla^2 \phi \md V = \int_{\mathring{\ppartial V_i}} \nabla \phi \cdot \mathbf{n} \md S + \int_{\ppartial V_i^b} \nabla \phi \cdot \mathbf{n} \md S
\end{equation}

\begin{figure}[htbp]
    \centering
    \begin{tikzpicture}

    \draw[thick] (0, -2) -- (0, 2);
    \draw[thick] (-2, -2) -- (-2, 2);
    \draw[thick] (-2, 2) -- (0, 2) node[anchor=west] {$j$};
    \draw[thick] (-2, 0) -- (0, 0) node[anchor=west] {$i$};
    \draw[thick] (-2, -2) -- (0, -2) node[anchor=north west] {$k$};
    \draw[dashed] (-1, -1) -- (-1, 1);
    \draw[dashed] (-1, 1) -- (0, 1);
    \draw[dashed] (-1, -1) -- (0, -1);

    \node[circle, draw=red_sketch, red_sketch, anchor=south] at (-1, 1) {$\tau_1$};
    \node[circle, draw=red_sketch, red_sketch, anchor=north] at (-1, -1) {$\tau_2$};

    \draw[thick, ->] (0, 1) -- (1, 1) node[anchor=west]{$\vb{S}_f^{\tau_1}$};
    \draw[thick, ->] (0, -1) -- (1, -1) node[anchor=west]{$\vb{S}_f^{\tau_2}$};
    \draw[thick, ->] (0, 0) -- (1, 0) node[anchor=west]{$\vb{S}_i = \frac{1}{2}(\vb{S}_f^{\tau_1} + \vb{S}_f^{\tau_2})$};

\end{tikzpicture}
    \caption{Boundary conditions normals.}
    \label{fig:bcs_normal}
\end{figure}
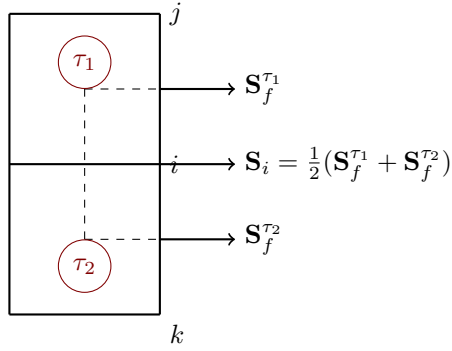

The first term in the right hand side is discretized as detailed in the sections above. Introducing the node normal boundary $\vb{S}_i^b$, which is the weighted sum of the face boundary normals shown in Fig.~\ref{fig:bcs_normal}, the second term reduces to

\begin{equation}
    \int_{\ppartial V_i^b} \nabla \phi \cdot \mathbf{n} \md S = \nabla \phi \cdot \vb{n} ||\vb{S}_i^b||
\end{equation}

The Neumann boundary condition is imposed by modifying the right hand side of the Poisson equation with the normal electric field $E_n$ from Eq.~\eqref{eq:poisson_neumann}. Using Eq.~\eqref{eq:poisson_discretization_cartesian}, the complete discretized line for a Neumann boundary condition node $i$ in the Poisson matrix is thus

\begin{equation}
    \sum_{\tau \in E(i)} \sum_{k\in\tau} \frac{\mathbf{S}_k\cdot\mathbf{S}_i}{V_\tau n_d^2} \phi_k = \frac{\rho_i V_i}{\veps_0} - E_n^i ||\vb{S}_i^b||
\end{equation}

Neumann boundary conditions are thus natural conditions for the Poisson equation in this vertex-centered formulation. Indeed when a boundary node is discretized using the laplacian discretization and nothing else is applied a zero Neumann boundary condition is implied.

\subsubsection{Dirichlet boundary conditions}

Dirichlet boundary conditions are enforced by imposing a one coefficient on the matrix and setting the imposed value in the right hand side of the linear system.

\subsection{Development of TVD schemes for plasma transport equations}

\label{sec:num_schemes}

Due to the large discrepancies in density involved in plasma simulations, robust numerical schemes are necessary. Thus the development of Total Variation Diminushing schemes has been done in AVIP for both drift-diffusion equations and euler equations.

\subsubsection{Integrating the drift-diffusion equations}

\label{subsec:scalar_implementation}

The model equation that needs to be solved in the case of drift-diffusion is simply the advection-diffusion equation

\begin{equation}
    \pdv{u}{t} + \nabla \cdot \vb{F} = 0
\end{equation}

\noindent where $\vb{F} = \vb{V} u - D \nabla u$. This equation is averaged over the dual volume $V_i$.

\begin{equation}
    \dv{u_i}{t} + \frac{1}{V_i} \int_{V_i} \nabla \cdot \vb{F} \, \md V = 0
\end{equation}

We integrate this equation in a vertex-centered formulation looping eventually on the edges of the dual volume. To preserve data structures close to AVBP \cite[Chap. 4]{lamarquethesis} the following discretization is used

\begin{equation}
    \label{eq:vertex_centered_flux}
    \int_{V_i} \nabla \cdot \vb{F} \, \md V = \sum_{\tau \in E(i)} \sum_{f \in \tau \cap \ppartial V_i} \int_f \vb{F} \cdot \vb{n} \, \md S
\end{equation}

The first sum is a loop on each neighboring cell of a given node and then we sum on the faces of the intersection of the neighboring cell $\tau$ and the nodal surface $\dvi$. In practice in AVIP the residual is computed \textit{at each edge} of the cells as shown in Fig.~\ref{fig:2D_cells} in 2D for triangular and quadrangular cells where three and four edges are present for each element type, respectively. The number of edges differs from the number of vertices in 3D: tetrahedra elements have six edges whereas hexahedral elements have twelve edges.

Axisymmetric conditions are also considered in this work. The flux formulation has to be slightly modified as the differential volume is now $\md V = r \md r \md \theta \md x$ where $(x, r, \theta)$ are the cylindrical coordinates. The integration on the azimuthal angle can be carried out since $\partial / \partial \theta = 0$ and therefore the averaged equation over the dual volume $V_i$ yields

\begin{equation}
    \dv{u_i}{t} + \frac{1}{r_iA_i} \int_{A_i} \nabla \cdot (r \vb{F}) \, \md A = 0
\end{equation}

\noindent where $V_i = 2\pi r_i A_i$ results from Theorem.~\ref{theorem:pappus_second}. Splitting this nodal surface flux along its edges yield the axisymmetric equivalent of Eq.~\eqref{eq:vertex_centered_flux} where the edges radiuses need to be included in the flux computation:

\begin{equation}
    \label{eq:vertex_centered_flux_axi}
    \int_{A_i} \nabla \cdot (r\vb{F}) \, \md A = \sum_{\tau \in E(i)} \sum_{f \in \tau \cap \ppartial A_i} \int_f \vb{F} \cdot r\vb{n} \, \md l
\end{equation}

The flux on the nodal surface associated to edge $ij$ in Fig.~\ref{fig:2D_cells} is finally simplified by invoking Theorem.~\ref{theorem:pappus_first}:

\begin{equation}
    \int_f \vb{F} \cdot r\vb{n} \, \md l = \vb{F}_{ij} r_{ij}^\tau \vb{S}_{ij}^\tau
\end{equation}

\noindent where the radius $r_{ij}^\tau$ is shown in Fig.~\ref{fig:2D_cells} for both elements and considering that the edge $ij$ is an axis edge.

Three TVD schemes will be presented to integrate the advection-diffusion equation in a finite volume formulation: the Scharfetter Gummel (SG) scheme \cite{scharfetter1969}, its improved version (ISG) \cite{Kulikovsky1995}, a simple upwind scheme \cite[Chap. 7]{hirsch} and the limited Lax-Wendroff (LLW) scheme \cite{lax_1960,sweby_1984}. The upwind and limited LW schemes are supplemented by a central differencing diffusive flux integration. These schemes has been originally developed for cartesian meshes and adaptations of these schemes on unstructured meshes with non-topologically dual elements has been carried out in AVIP and is presented.

\subsubsection{Scharfetter Gummel scheme}

Starting from \eqref{eq:vertex_centered_flux}, we can rewrite the flux term as

\begin{equation}
    \int_f \vb{F} \cdot \vb{n} \, \md S = \vb{F}_{ij} \cdot \vb{S}_{ij}
\end{equation}

\begin{figure}[htbp]
    \centering
    \begin{tikzpicture}

    \draw[thick] (0, 0) node[anchor=east] {$i$} -- (6, 0) node[anchor=west] {$j$};
    \draw[thick] (6, 0) -- (0, 6) node[anchor=south] {$k$};
    \draw[thick] (0, 0) -- (0, 6);

    \draw[dashed, draw=blue_sketch] (3, 0) -- (2, 2);
    \draw[dashed, draw=blue_sketch] (0, 3) -- (2, 2);
    \draw[dashed, draw=blue_sketch] (3, 3) -- (2, 2);

    \node[anchor=north east, text=blue_sketch] at (2, 1.7){$\dvi \cap \tau$};

    \draw[very thick, draw=red_sketch, ->] (2.5, 1.0) -- (3.8, 1.0) node[anchor=west, text=red_sketch]{$\vb{S}_{ij}^\parallel$};
    \draw[very thick, draw=red_sketch, ->] (2.5, 1.0) -- (3.8, 1.7) node[anchor=west, text=red_sketch]{$\vb{S}_{ij}^\tau$};
    \draw[very thick, draw=red_sketch, ->] (2.5, 1.0) -- (2.5, 1.7) node[anchor=west, text=red_sketch]{$\vb{S}_{ij}^\perp$};

\end{tikzpicture}
    \caption{Non topologically dual quadrangular cell}
    \label{fig:2D_cells_edges_tri_isorect}
\end{figure}
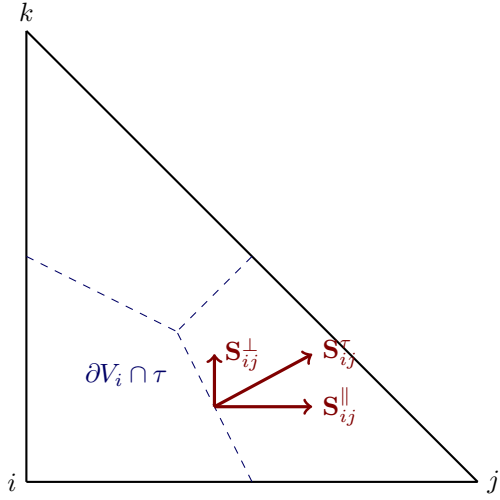

Splitting the nodal surface normal into a tangential and perpendicular part $\vb{S}_{ij} = \vb{S}_{ij}^\parallel + \vb{S}_{ij}^\perp$ with respect to the edge direction $\hat{\vb{ij}} = \vb{ij} / ij$

\begin{equation}
    \int_f \vb{F} \cdot \vb{n} \, \md S = \vb{F}_{ij} \cdot \vb{S}_{ij}^\parallel + \vb{F}_{ij} \cdot \vb{S}_{ij}^\perp
\end{equation}

\noindent where a definition of these normals for non-topologically dual meshes is shown in Fig.~\ref{fig:2D_cells_edges_tri_isorect}.

The Scharfetter Gummel scheme \cite{Kulikovsky1995} can be applied on the tangential part of the flux by projecting the flux along the tangential normal:

\begin{equation}
    \label{eq:dd_projected}
    \vb{F}_{ij} \cdot \vb{S}_{ij}^{\parallel} = u \vb{V}_{ij} \cdot \vb{S}_{ij}^{\parallel} - D \nabla u \cdot \vb{S}_{ij}^{\parallel}
\end{equation}

The SG scheme gives tangentiel flux projection but does not give any information about the \textit{normal} flux projection:

\begin{equation}
    \vb{F}_{ij} \cdot \vb{S}_{ij}^{\parallel} = \vb{V}_{ij} \cdot \vb{S}_{ij}^{\parallel} \frac{u_j - e^\alpha u_i}{1 - e^\alpha}
\end{equation}

\noindent where

\begin{equation}
    \alpha = \frac{h_{ij} \vb{V}_{ij} \cdot \hat{\vb{ij}}}{D_{ij}}
\end{equation}

On meshes where the perpendicular component is non-zero another scheme must be applied on top of the SG scheme. A centered-scheme has been chosen for this normal part:

\begin{equation}
    \vb{F}_{ij} \cdot \vb{S}_{ij}^\perp = \vb{V}_{ij} \frac{u_i + u_j}{2} \cdot \vb{S}_{ij}^\perp - D_{ij} \frac{\nabla u_i + \nabla u_j}{2} \cdot \vb{S}_{ij}^\perp
\end{equation}

The improved Scharfetter Gummel scheme (ISG) virtual node reconstruction is carried on the edge exactly as explained in the 1D case since locally the tangential flux reconstruction reduces to a 1D problem in Eq.~\eqref{eq:dd_projected}.

\subsubsection{Limited Lax-Wendroff scheme}

We are only interested in the convective part of the scheme for now so $u_t + \nabla \cdot (\vb{V} u) = 0$ is considered. The starting point of the Lax-Wendroff scheme is a Taylor expansion in time of the variable $u$ \cite{lamarquethesis, hirsch}

\begin{equation}
    u^{n+1} = u^n + \upD t  u_t + \frac{\upD t^2}{2} u_{tt} + \mc{O}(\upD t^3)
\end{equation}

\noindent Keeping only the first three terms and using the advection diffusion equation the Taylor expansion is rewritten as

\begin{equation}
    u^{n+1} = u^n - \upD t \nabla \cdot (\vb{V} u) + \frac{\upD t^2}{2} \nabla \cdot (\vb{V} \nabla \cdot (\vb{V} u))
\end{equation}

\noindent Integrating on the nodal volume $V_i$ following \eqref{eq:vertex_centered_flux} yields

\begin{equation}
    u_i^{n+1} = u_i^n - \frac{\upD t}{V_i} \sum_{\tau \in E(i)} \sum_{f \in \tau \cap \ppartial V_i}\qty[\int_f u \vb{V} \cdot \vb{n} \, \md S - \frac{\upD t}{2} \int_f \nabla \cdot (\vb{V} u) \vb{V} \cdot \vb{n} \, \md S]
\end{equation}

\noindent Assuming a constant velocity on the edge, the Lax-Wendroff schemes finally writes

\begin{equation}
    u_i^{n+1} = u_i^n - \frac{\upD t}{V_i} \sum_{\tau \in E(i)} \sum_{f \in \tau \cap \ppartial V_i}\qty[u_{ij} - \frac{\upD t}{2} \nabla u_{ij} \cdot \vb{V}_{ij}]\vb{V}_{ij}\cdot \vb{S}_{ij}
\end{equation}

\noindent where $ij$ denotes the mean of the values at the edge, \textit{e.g.} $u_{ij} = (u_i + u_j) / 2$.

\begin{figure}[htbp]
    \centering
    \begin{subfigure}[b]{0.45\textwidth}
        \centering
        \includegraphics[width=\textwidth]{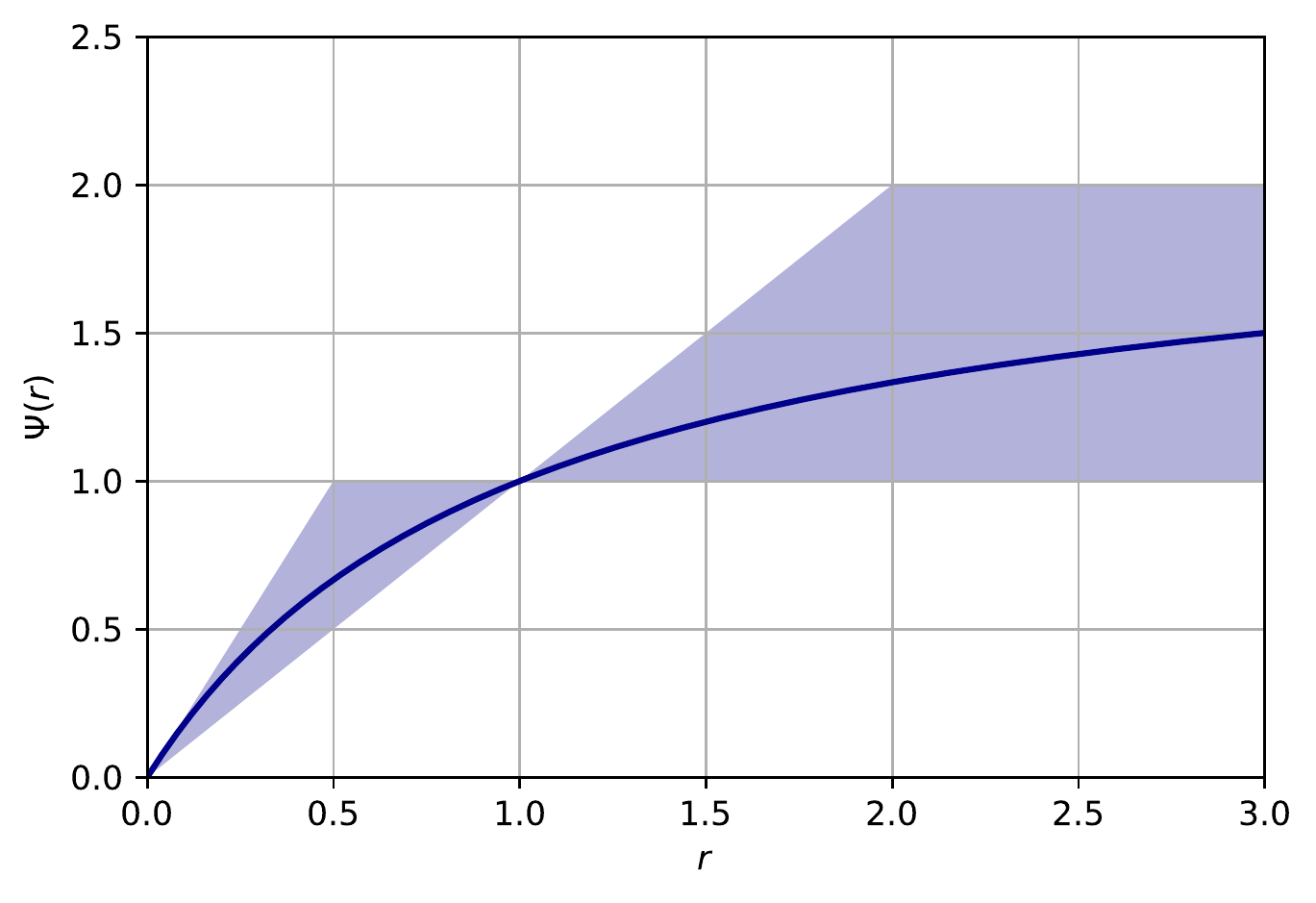}
        \caption{Van Leer limiter}
        \label{fig:vanleer_limiter}
    \end{subfigure}
    \centering
    \begin{subfigure}[b]{0.45\textwidth}
        \centering
        \includegraphics[width=\textwidth]{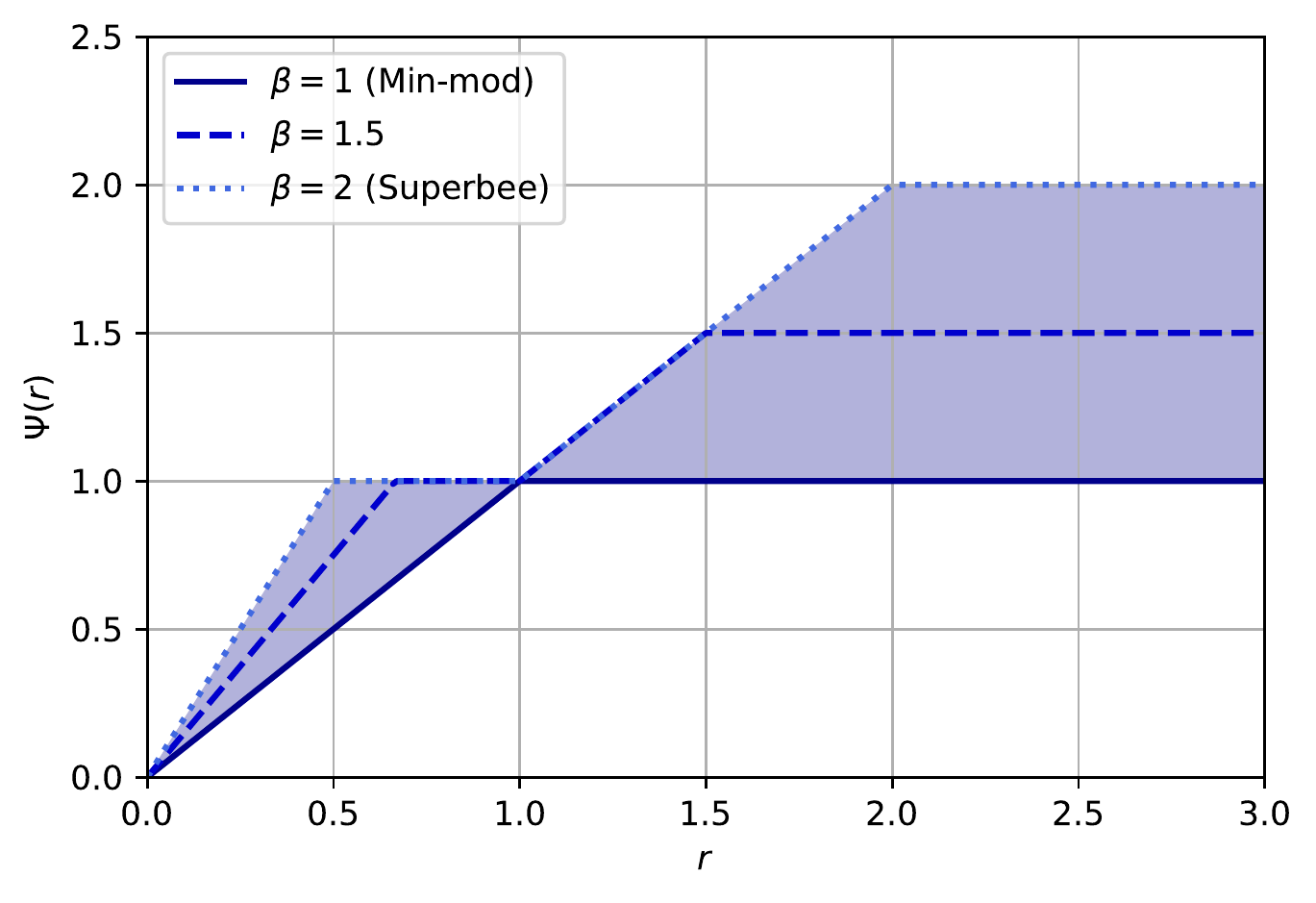}
        \caption{Sweby limiters}
        \label{fig:sweby_limiters}
    \end{subfigure}
    \caption{Limiters in Sweby diagrams.}
    \label{fig:limiters}
\end{figure}

We now add a limiter on the Lax-Wendroff scheme to make it TVD following \cite[Chap. 8]{hirsch}. In 1D finite volume formulation the limited Lax-Wendroff scheme then yields

\begin{equation}
    u_{i+1/2} = u_i + \frac{1-\sigma}{2}(u_{i + 1} - u_i) \Uppsi(R_i)
\end{equation}

\noindent where the limiting function is either the Van Leer \cite{vanleer_1974} or Sweby \cite{sweby_1984} limiter which are shown in Sweby diagrams in Fig.~\ref{fig:limiters}.

This cartesian limiting scheme needs to be adapted to unstructured meshes. In a cell of an unstrutured mesh, there is no real upwind node so that only a virtual upwind node $k$ can be defined at an edge $ij$. Only the upwind gradient $u_k - u_i$ is needed and is computed by leveraging the centered gradient of AVIP so that approximately we have

\begin{equation}
    \nabla u_i \cdot (2 \vb{ij}) = u_j - u_k
\end{equation}

\noindent The ratio of gradients is computed as

\begin{equation}
    R_{ij} = \frac{u_j - u_i - \nabla u_i \cdot (2 \vb{ij})}{u_j - u_i}
\end{equation}

In the end the limited Lax-Wendroff scheme in Finite Volume Vertex Centered formulation is given by

\begin{equation}
    u_i^{n+1} = u_i^n - \frac{\upD t}{V_i} \sum_{\tau \in E(i)} \sum_{f \in \tau \cap \ppartial V_i}\qty[u_i + \qty(\frac{u_j - u_i}{2} - \frac{\upD t}{2} \nabla u_{ij} \cdot \vb{V}_{ij}) \Uppsi(R_{ij})]\vb{V}_{ij}\cdot \vb{S}_{ij}
\end{equation}

\subsubsection{Central difference diffusive flux integration}

When using upwind or limited LW scheme for the advection term, a central difference scheme is applied on the diffusive flux $\vb{F}_D = -D \nabla u$ as

\begin{equation}
    \int_f \vb{F}_D \cdot \vb{n} \, \md S = - D_{ij} \frac{\nabla u_i + \nabla u_j}{2} \cdot \vb{S}_{ij}
\end{equation}

\subsection{Boundary conditions in AVIP}

\label{subsec:bcs_scalar}

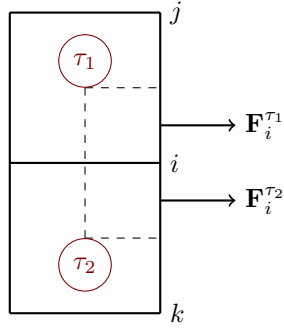
\begin{figure}[htbp]
    \centering
    \begin{tikzpicture}

    \draw[thick] (0, -2) -- (0, 2);
    \draw[thick] (-2, -2) -- (-2, 2);
    \draw[thick] (-2, 2) -- (0, 2) node[anchor=west] {$j$};
    \draw[thick] (-2, 0) -- (0, 0) node[anchor=west] {$i$};
    \draw[thick] (-2, -2) -- (0, -2) node[anchor=west] {$k$};
    \draw[dashed] (-1, -1) -- (-1, 1);
    \draw[dashed] (-1, 1) -- (0, 1);
    \draw[dashed] (-1, -1) -- (0, -1);

    \node[circle, draw=red_sketch, red_sketch, anchor=south] at (-1, 1) {$\tau_1$};
    \node[circle, draw=red_sketch, red_sketch, anchor=north] at (-1, -1) {$\tau_2$};

    \draw[thick, ->] (0, 0.5) -- (1, 0.5) node[anchor=west]{$\vb{F}_i^{\tau_1}$};
    \draw[thick, ->] (0, -0.5) -- (1, -0.5) node[anchor=west]{$\vb{F}_i^{\tau_2}$};

\end{tikzpicture}
    \caption{Fluxes at the boundary conditions.}
    \label{fig:flux_bcs}
\end{figure}

The boundary fluxes for quandrangular elements are shown in Fig.~\ref{fig:flux_bcs} for a boundary node $i$. Two choices for the flux computation are possible

\begin{itemize}
    \item The two fluxes are taken equal to the node flux
    \begin{equation}
        \vb{F}_i^{\tau_1} = \vb{F}_i^{\tau_2} = \vb{F}_i
    \end{equation}
    \item The fluxes are interpolated, which in two-dimensions reduce to
    \begin{gather}
        \vb{F}_i^{\tau_1} = 0.75 \vb{F}_i + 0.25 \vb{F}_j \\
        \vb{F}_i^{\tau_2} = 0.75 \vb{F}_i + 0.25 \vb{F}_k
    \end{gather}
\end{itemize}

Two types of boundary conditions are implemented:

\begin{itemize}
    \item Supersonic boundary conditions: $\vb{F} = \vb{F}_C + \vb{F}_D$
    \item Neumann boundary conditions: $\vb{F} = \vb{F}_C$ where $\nabla n \cdot \vb{n} = 0$ is thus implied
\end{itemize}

\subsubsection{Integrating the Euler equations}

Various cell-vertex numerical schemes are already present in AVBP \cite{lamarquethesis}. All schemes are however high-order, centered methods that do not possess the Total Variation Diminushing property. Artificial viscosity is applied to capture strong gradients and reduce wiggles.

To fill this gap, Riemann solvers have been developed based on \cite{toro} to capture shocks and strong discontinuities that are present in both plasma Euler equations and also in the mixture equations since streamer discharges can create strong shocks.
\subsection{Implementation in AVIP}

The implementation of the HLLC and MUSCL-Hancock procedure in AVIP are now detailed where the generalization of the previous relations to multidimensional and multicomponent flows in an unstructured mesh is given.  As with the scalar transport equations, we solve these equations in a Finite Volume Vertex-Centered fashion.

\subsubsection{The HLLC solver}

We start from the Euler equations in tensor form valid in all types of coordinates frames:

\begin{align}
\pdv{\U}{t} + \div \vb{F} = 0
\end{align}
where
\begin{align}
\U = \begin{bmatrix}
\rho \\ \rho \uv \\ \rho E
\end{bmatrix}
\qq{and}
\vb{F} = \begin{bmatrix}
\rho \uv \\ \rho \uv \otimes \uv + p \vb{I} \\ (\rho E + p)\uv
\end{bmatrix}
\end{align}

Integrating the system of equations on $V = V_i$, \textit{i.e.} we integrate on the nodal $i$ volume of the mesh and splitting of the fluxes across the relevant faces as done in Section.~\ref{subsec:scalar_implementation} yields the following discretization

\begin{equation}
    \dv{t} \vU_i + \frac{1}{V_i} \sum_{\tau \in E(i)} \sum_{f \in \ppartial V_i \cap \tau} \int_f \vF \cdot \vb{n} \, \md S = 0
\end{equation}

\noindent This discretization is exactly the same as the one detailed in Eq.~\eqref{eq:vertex_centered_flux} and Fig.~\ref{fig:2D_cells} of the previous chapter for scalar transport variables. Two ways of computing the face integral are possible: either by splitting across each dimension and solving each Riemann problem or by rotating the flux into the local basis of the edge \cite[Chap. 16]{toro}. The latter has been implemented in AVIP as it is less computationally expensive. Hence the face integral is computed in three steps:

\begin{enumerate}
    \item Projection of the solution in the local basis $\vb{T} \vU$
    \item Solve the Riemann problem in the local basis $\vb{F}_1 = \vb{F}_1(\vb{T}\vU)$
    \item Project back into the original basis $\vb{T}^{-1} \vb{F}_1(\vb{T}\vU)$
\end{enumerate}

In three dimensions and multi-component flows (where each species $k$ is described by its mass fractions $Y_k$) the conservative variables, fluxes are now given by:

\begin{equation}
    \vb{U}=
    \begin{bmatrix}
        \rho\\\rho u\\\rho v\\\rho w\\\rho E \\ \rho Y_1 \\ \vdots \\ \rho Y_N
    \end{bmatrix}
    \quad\vb{F}=
    \begin{bmatrix}
        \rho u \\
        \rho u^2 + p \\
        \rho v u \\
        \rho w u \\
        (\rho E + p) u \\
        \rho Y_1 u \\
        \vdots \\
        \rho Y_N u
    \end{bmatrix}
\end{equation}

The resulting star region state for multicomponent flows is now as follows:

\begin{equation}
    \vU_{*K} = \rho_K \qty(\frac{S_K - u_K}{S_K - S_*}) \begin{bmatrix}
        1 \\ S_* \\ v_K \\ w_K \\ E_K + (S_* - u_K)\qty[S_* + \frac{p_K}{\rho_K(S_K - u_K)}] \\ (\rho Y_1)_K \\ \vdots \\ (\rho Y_N)_K
    \end{bmatrix}
    \label{eq:hllc_formulation_mutlicomponent}
\end{equation}

\subsubsection{The MUSCL procedure implementation in AVIP}

\begin{figure}[htbp]
    \centering
    \begin{tikzpicture}

    \draw[thick] (-2.5, 0) node[anchor=north east, text=red_sketch] {$l$} -- (0, 0) node[anchor=north west] {$i$};
    \draw[thick] (0, 0) -- (2.5, 0);
    \draw[thick] (2.5, 0) node[anchor=north east] {$j$} -- (5.0, 0) node[anchor=north west, text=red_sketch] {$k$};
    \draw[thick] (-2.5, 2.5) -- (5.0, 2.5);
    \draw[thick] (-2.5, -2.5) -- (5.0, -2.5);

    \foreach \x in {-2.5, 0, 2.5, 5}
        \draw (\x, -2.5) -- (\x, 2.5);

    \draw[thick, dotted] (0, 1.25) -- (2.5, 1.25);
    \draw[thick, dotted] (1.25, 0) -- (1.25, 1.25);

    \draw[very thick, ->] (1.25, 0.625) -- (2.0, 0.625) node[anchor=south] {$\vb{S}_{ij}^\tau$};

    \node[anchor=center, text=red_sketch, draw=red_sketch, circle] at (-1.25, -1.25) {$\tau_u^2$};
    \node[anchor=center, text=red_sketch, draw=red_sketch, circle] at (-1.25, 1.25) {$\tau_u^1$};

\end{tikzpicture}
    \caption{Reconstruction of neighboring nodes inside AVIP in quadrangular cells.}
    \label{fig:nodes_quad_reconstruct}
\end{figure}
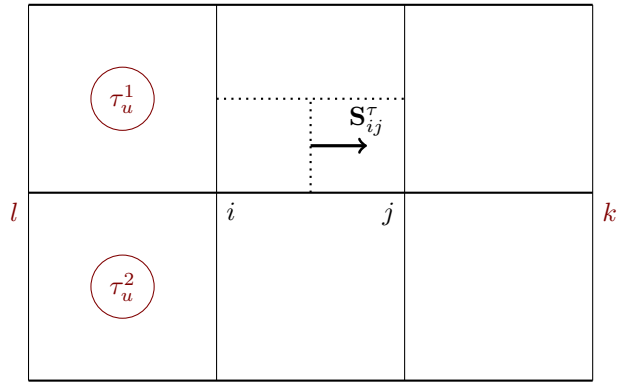

\begin{figure}[htbp]
    \centering
    \begin{tikzpicture}

    \draw[thick] (-3, 0) node[anchor=north east, text=red_sketch] {$l$} -- (0, 0) node[anchor=south] {$i$};
    \draw[thick] (0, 0) -- (3, 0);
    \draw[thick] (3, 0) node[anchor=south] {$j$} -- (6.0, 0) node[anchor=north west, text=red_sketch] {$k$};
    \draw[thick] (-1.5, 2.6) -- (4.5, 2.6);
    \draw[thick] (-1.5, -2.6) -- (4.5, -2.6);

    \foreach \x in {-3, 0, 3}{
        \draw (\x, 0) -- (\x + 1.5, 2.6);
        \draw (\x, 0) -- (\x + 1.5, -2.6);
    }
    \foreach \x in {-3, 0, 3}{
        \draw (\x + 3, 0) -- (\x + 1.5, 2.6);
        \draw (\x + 3, 0) -- (\x + 1.5, -2.6);
    }

    \draw[thick, dotted] (1.5, 0) -- (1.5, 0.86);
    \draw[thick, dotted] (0.75, 1.3) -- (1.5, 0.86);
    \draw[thick, dotted] (2.25, 1.3) -- (1.5, 0.86);

    \draw[very thick, ->] (1.5, 0.4) -- (2.0, 0.4) node[anchor=south] {$\vb{S}_{ij}^\tau$};

    \node[anchor=center, text=red_sketch, draw=red_sketch, circle] at (-1.5, -1) {$\tau_u^2$};
    \node[anchor=center, text=red_sketch, draw=red_sketch, circle] at (-1.5, 1) {$\tau_u^1$};

\end{tikzpicture}
    \caption{Reconstruction of neighboring nodes inside AVIP in triangular cells.}
    \label{fig:nodes_tri_reconstruct}
\end{figure}
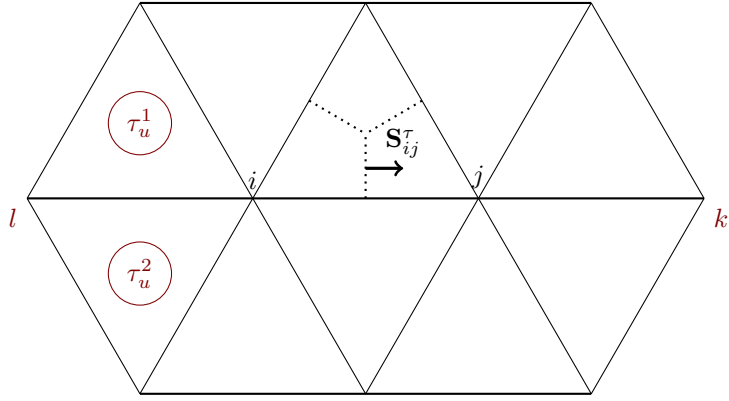

The equivalents of $\upD_{i-1/2}$ and $\upD_{i+1/2}$ in unstructured meshes are needed for MUSCL implementation. They correspond respectively to $\upD_{li}$ and $\upD_{ij}$ for $i$ and $\upD_{ij}$ and $\upD_{jk}$ for $j$ which are depicted in Figs.~\ref{fig:nodes_quad_reconstruct} and \ref{fig:nodes_tri_reconstruct}:

\begin{equation}
    \upD_{li} = \vU_i - \vU_l \quad \upD_{ij} = \vU_j - \vU_i \quad \upD_{jk} = \vU_k - \vU_j
\end{equation}

Inside the domain of computation the gradients computed in AVIP are centered gradients so that the following reconstruction inside a cell can be made:

\begin{equation}
    \vU_j - \vU_l = (\nabla \vU)_i \cdot (2\vb{ij})
    \label{eq:virtual_node_reconstruction}
\end{equation}

\begin{figure}[htbp]
    \centering
    \def\x{1068}
\begin{tikzpicture}
    \draw[thick, red_sketch, dashed] (-3, 0) node[anchor=east, text=red_sketch, draw=red_sketch, circle, dashed] {$l$} -- (0, 0) node[anchor=south] {$i$};
    \draw[thick] (0, 0) -- (3, 0);
    \draw[thick, red_sketch, dashed] (3, 0) node[anchor=south] {$j$} -- (6.0, 0) node[anchor=west, text=red_sketch, draw=red_sketch, circle, dashed] {$k$};

    \draw[thick] (0, 0) -- (-2.7, 2);
    \draw[thick] (1.5, 2.6) -- (-2.7, 2);
    \draw[thick] (1.5, 2.6) -- (0, 0);
    \draw[thick] (1.5, 2.6) -- (3, 0);
    \draw[thick] (-2.7, -1) -- (-2.7, 2);
    \draw[thick] (0, 0) -- (-2.7, -1);
    \draw[thick] (0, 0) -- (1.2, -2.2);
    \draw[thick] (-2.7, -1) -- (1.2, -2.2);
    \draw[thick] (3, 0) -- (1.2, -2.2);
    \draw[thick] (3, 0) -- (4.7, -1.7);
    \draw[thick] (3, 0) -- (4.7, 1.7);
    \draw[thick] (4.7, -1.7) -- (4.7, 1.7);
    \draw[thick] (4.7, -1.7) -- (1.2, -2.2);
    \draw[thick] (4.7, 1.7) -- (1.5, 2.6);

    \node[anchor=center, text=red_sketch, draw=red_sketch, circle] at (-2, 0.7) {$\tau_u^i$};
    \node[anchor=center, text=red_sketch, draw=red_sketch, circle] at (4.2, 0.6) {$\tau_u^j$};

\end{tikzpicture}
    \caption{Reconstruction of neighboring nodes using upwind cells.}
    \label{fig:nodes_tris_reconstruct_alauzet}
\end{figure}
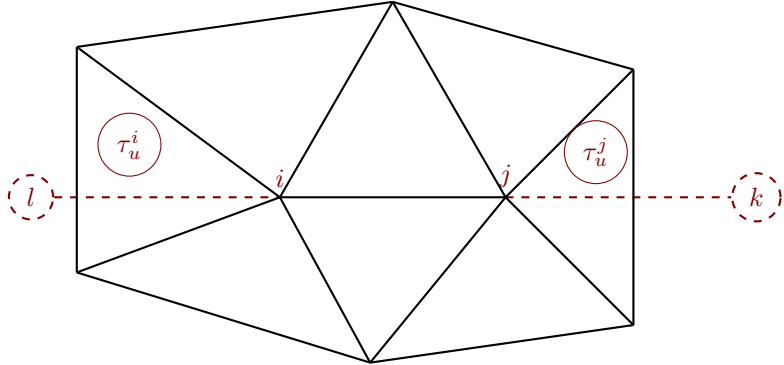

In the case of regular meshes such as Figs.~\ref{fig:nodes_quad_reconstruct} and \ref{fig:nodes_tri_reconstruct} the reconstructed nodes actually exist ($l$ and $k$) but in the case of irregular meshes in Fig.~\ref{fig:nodes_tris_reconstruct_alauzet} \textit{virtual} upwind nodes are retrieved ($l$ and $k$ in dashed circles) which are consistent with the structured cartesian formulation of limiters: $l$ and $k$ are at a distance $ij$ of $i$ and $j$, respectively.

This reconstruction of upwind nodes is different from the ones adopted in \cite{alauzet_2010} and \cite{Joncquieres2019} where upwind cells relative to the edge and node considered are sought shown in Fig.~\ref{fig:nodes_tris_reconstruct_alauzet}. This cell is defined as the one whose center is the most aligned with the edge $ij$. In the case of irregular meshes whose edges are not aligned this cell is well-defined ($\tau_u^i$ and $\tau_u^j$ in Fig.~\ref{fig:nodes_tris_reconstruct_alauzet}) however in the case of aligned edges such as those shown in Figs.~\ref{fig:nodes_quad_reconstruct} and \ref{fig:nodes_tri_reconstruct} two equivalent upwind cells can be defined $\tau_u^1$ and $\tau_u^2$ for the $i$ node. For rather regular elements a mean of the two upwind cells would be better than just picking one as this could lead otherwise to odd-even decoupling. This bad behavior in the case of regular elements vanishes for the virtual node reconstruction considered in Eq.~\eqref{eq:virtual_node_reconstruction}. Moreover finding the associated upwind cells for each edge is computationally expensive and communications are needed across partitions whereas this implementation is local to the cell.

\paragraph{Effect of multicomponent mixture}

For multicomponent gas mixture equations the MUSCL-Hancock procedure needs to be slightly modified compared to what is presented in \cite[Chap. 14]{toro}. In the HLLC solver, primitive variables are sought from conservative variables and so after interpolation with the gradients the pressure is retrieved from

\begin{equation}
    p = (\gamma - 1) \rho e_s
    \label{eq:pressure_sensible_energy}
\end{equation}

\noindent where $e_s$ is the internal or sensible energy. For multicomponent mixtures with temperature dependent $c_v$ this relation does not hold anymore. In AVIP the pressure is computed from the temperature which is deduced from an iterative procedure on the sensible energy $e_s$ so that

\begin{equation}
    T = e_s^{-1} e_s(T) \implies p = \rho r T
\end{equation}

Hence to circumvent the impossibility to use Eq.~\eqref{eq:pressure_sensible_energy} the pressure is interpolated along with the conserved variables at the beginning of the MUSCL reconstruction:

\begin{gather}
    \vU_i^L = \vU_i^n - \frac{1}{2} \bar{\upD}_i \quad \vU_i^R = \vU_i^n + \frac{1}{2} \bar{\upD}_i \\
    p_i^L = p_i^n - \frac{1}{2} \bar{\upD} p_i \quad p_i^R = p_i^n + \frac{1}{2} \bar{\upD} p_i
\end{gather}

This is an approximation compared to the original scheme: since the pressure is a non-linear function of the conserved variables a linear variation of the conservariables does not result in a linear variation of the pressure.

\subsubsection{Boundary conditions}

Flux closure is performed exactly as for the scalar transport equations detailed in Sec.~\ref{subsec:bcs_scalar}. Hence an equivalent of supersonic outlet is applied at the end of the numerical scheme yielding a result close to the cell-vertex schemes. Characteristic boundary conditions \cite{poinsot1992} implemented in AVBP in \cite{moureau2005} are then available where a detailed account can be found in \cite[Chap. 4.12.2]{lamarquethesis}. Values predicted by the numerical scheme is corrected by a wave decomposition of the Euler equations which is close to Riemann solvers.

A more consistent approach would be to include the boundary conditions directly in the numerical scheme by solving Riemann problems at the boundaries with the HLLC MUSCL-Hancock solver. Virtual states depending on the type of boundary are constructed in this case and Riemann problems are solved at each boundary interface. A comparison between NSCBC and Riemann solver boundary could be an interesting future work.

\subsubsection{Time integration of the HLLC MUSCL-Hancock solver}

The Riemann solvers are not spatio-temporal schemes like the classical LW and TTGC schemes of AVBP. Hence a time integration on top of if must be chosen. Many strategies can be adopted depending on the stability critertions we want to achieve \cite[Chap. 9]{hirsch}. Since we are dealing with unsteady flows the low-storage Runge-Kutta (RK) schemes for coupling with the HLLC MUSCL solver is chosen:

$$
\begin{aligned}
&\vb{U}^{(1)} = \vb{U}^n \\
&\cdots \\
&\vb{U}^{(j)} = \vb{U}^n + \alpha_j \vb{R}^{(j-1)}\\
&\cdots \\
&\vb{U}^{(K)} = \vb{U}^n + \alpha_K \vb{R}^{(K-1)}\\
&\vb{U}^{n+1} = \vb{U}^n + \Delta t \sum_{k=1}^K \beta_k \vb{R}^{(k)}
\end{aligned}
$$

where $\vb{R}$ is the residual, $\sum_k \beta_k = 1$ for consistency and each $\vb{U}^{(j)}$ is called a RK stage and $K$ the number of stages. In AVIP the choice of $\beta_K = 1$, $\beta_j = 0 \quad j=1 \ldots k-1$ is made where the RK1 to RK4 methods are defined by the coefficients given in Tab.~\ref{tab:rk_coeffs}.

\begin{table}[htbp]
    \centering
    \begin{tabular}{| c | c | c | c | c |}\hline
     Time integration method  &  $\alpha_2$  &  $\alpha_3$  &  $\alpha_4$  &  $\beta_K$ \\
    \hline RK1                      &  0           &  0           &  0           &  1         \\
     RK2                      &  1/2         &  0           &  0           &  1         \\
     RK3                      &  1/3         &  1/2         &  0           &  1         \\
     RK4                      &  1/4         &  1/3         &  1/2         &  1         \\
    \hline
    \end{tabular}
    \caption{Runge-Kutta coefficients.}
    \label{tab:rk_coeffs}
\end{table}

In \cite[Chap. 14]{toro}, a simple RK1 time integration scheme is always chosen in structured grids and seems to be stable for all the shock cases studied. The more stable but more costly RK2 and RK3 time integrations are also possible in AVIP and can be switched on whenever necessary.

\subsection{Geometric source terms}

Cylindrical frames are considered in AVIP for the simulation of plasma discharges in interaction with combustion. Navier-Stokes equations in axisymmetric conditions have the following form \cite{Cheng2022}:

\begin{equation}
    \pdv{r\vb{U}}{t} + \grad_\mathrm{2D} \cdot (r\vb{F}^C_{rz} + r\vb{F}^D_{rz}) = \vb{S}^C_{rz} + \vb{S}^D_{rz}
\end{equation}

\noindent where $\vb{F}^C_{rz}$ and  $\vb{F}^D_{rz}$ are the classical convective and diffusive Navier-Stokes fluxes. Integration of these fluxes is done as for the scalar transport equations detailed earlier in Section.~\ref{subsec:scalar_implementation}. However contrary to the scalar case, convective and diffusive geometric source terms are also present:

\begin{equation}
    \vb{S}^C_{rz} = \begin{bmatrix}
    0 \\  p \\  0 \\ 0
    \end{bmatrix}
    \qquad
    \vb{S}^D_{rz} = \begin{bmatrix}
        0 \\ - 2 \eta \frac{u_r}{r} - \lambda \qty(\frac{u_r}{r} + \frac{\ppartial u_r}{\ppartial r} + \frac{\ppartial u_z}{\ppartial z}) \\ 0 \\ 0
        \end{bmatrix}
    \label{eq:ns_cyl_source_terms}
\end{equation}

These geometric source terms are added after the transport residuals and radial velocity is corrected to zero after application of these geometric source terms to ensure $u_r = 0$ at the axis.

\subsection{Time step computation}

The dynamical time step computation is detailed in this section for the different sets of equations presented. Whenever multiple sets of equations are present (such as in PAC simulations where plasma and gas mixture equations are integrated), the minimum over all sets of equations is taken.

\subsubsection{Drift-diffusion equations}

We use a cell-based approach to compute the timestep. Four time steps can be computed for each species at each cell $\tau$:

\begin{itemize}
    \item Convective time step with CFL constant
    \begin{equation}
        \upD t_\mrm{conv} = \frac{\mrm{CFL}\upD x_\tau}{||\vb{V}_\tau||}
        \label{eq:dt_conv}
    \end{equation}
    \item Diffusive time step with Fourier constant $\mrm{F}$
    \begin{equation}
        \upD t_\mrm{diff} = \frac{\mrm{F} \upD x_\tau^2}{D_\tau}
        \label{eq:dt_diff}
    \end{equation}
    \item Dielectric time step with contant $A_D$
    \begin{equation}
        \upD t_\mrm{diel} = A_D \frac{\veps_0}{\sum_i |q_i n_i \mu_i|}
    \end{equation}
    \item Chemical time step with constant $A_\tau$
    \begin{equation}
        \upD t_\mrm{chem} = A_\tau \max_k S_\mrm{chem}^k
    \end{equation}
\end{itemize}

The minimum over all cells is taken as the reference timestep. The convective and diffusive time steps are computed for each species modeled with drift-diffusion equations and the resulting time step is the minimum of all the considered time steps.

\subsubsection{Gas mixture equations}

Convective and diffusive time steps are computed as in AVBP for the gas mixture equations following \cite[Chap. 4.10]{lamarquethesis} very smilar to Eqs.~\eqref{eq:dt_conv} and Eqs.~\eqref{eq:dt_diff}.

\section{Plasma discharges validation cases}
\label{sec:plasma_validation}

The simulation of plasma discharges is considered in this section. A stiff scalar advection case to validate the robustness of the two numerical schemes laid out in the previous section, ISG and LLW, is first presented. Discharge simulation capabilities are then validated on a streamer code benchmark \cite{bagheri_benchmark}. The last validation case is a pin-pin configuration taken from \cite{tholin2012} more representative of plasma discharges for combustion.

\subsection{Stiff scalar advection}

\begin{figure}[htbp]
    \centering
    \begin{subfigure}[b]{\textwidth}
        \centering
        \begin{tikzpicture}

    \draw[thick, ->] (-4.5, 0) -- (4.5, 0) node[anchor=north west] {$x$};

    \draw[thick] (-4, 2) -- (4, 2);
    \foreach \x in {-4, -2, 0, 2, 4}
    {
        \draw[thick] (\x, 0) -- (\x, 2);
    }
    \foreach \x in {-4, -2, 0, 2}
    {
        \draw[thick] (\x, 0) -- (\x+2, 2);
    }

\end{tikzpicture}
        \caption{Triangular elements}
        \label{fig:tri_1d}
    \end{subfigure}
    \centering
    \begin{subfigure}[b]{\textwidth}
        \centering
        \begin{tikzpicture}

    \draw[thick, ->] (-4.5, 0) -- (4.5, 0) node[anchor=north west] {$x$};

    \draw[thick] (-4, 2) -- (4, 2);
    \foreach \x in {-4, -2, 0, 2, 4}
        \draw[thick] (\x, 0) -- (\x, 2);

\end{tikzpicture}
        \caption{Quadrangular elements}
        \label{fig:quad_1d}
    \end{subfigure}
    \caption{1D domain with quadrangular and triangular meshes.}
    \label{fig:elements_1d}
\end{figure}
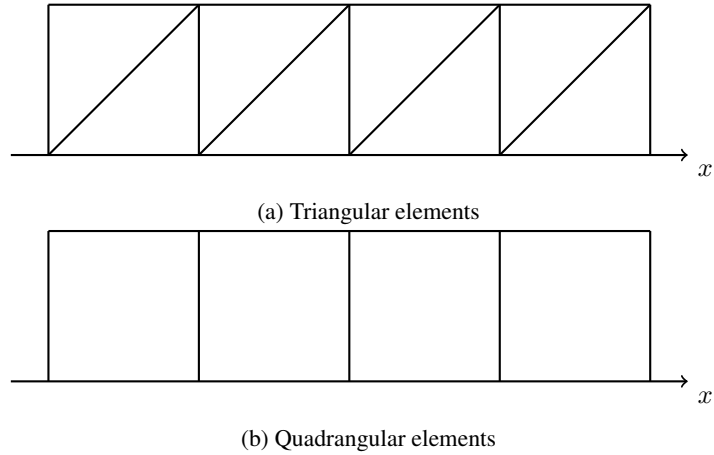

This test case is taken from \cite{Kulikovsky1995} where a drift-diffusion model in a 1D domain $[0, 1]$ is considered with a linear electric field $E(x) = Ax$. In AVIP triangular and quadrangular elements are tested for this 1D case with elements shown in Fig.~\ref{fig:elements_1d}. Mobility is set to -1 and the diffusion coefficient to 1 so that the studied equation is

\begin{equation}
    \pdv{n}{t} - \pdv{Axn}{x} - \pdv[2]{n}{x} = 0
\end{equation}

\begin{figure}[htbp]
    \centering
    \includegraphics[width=0.5\textwidth]{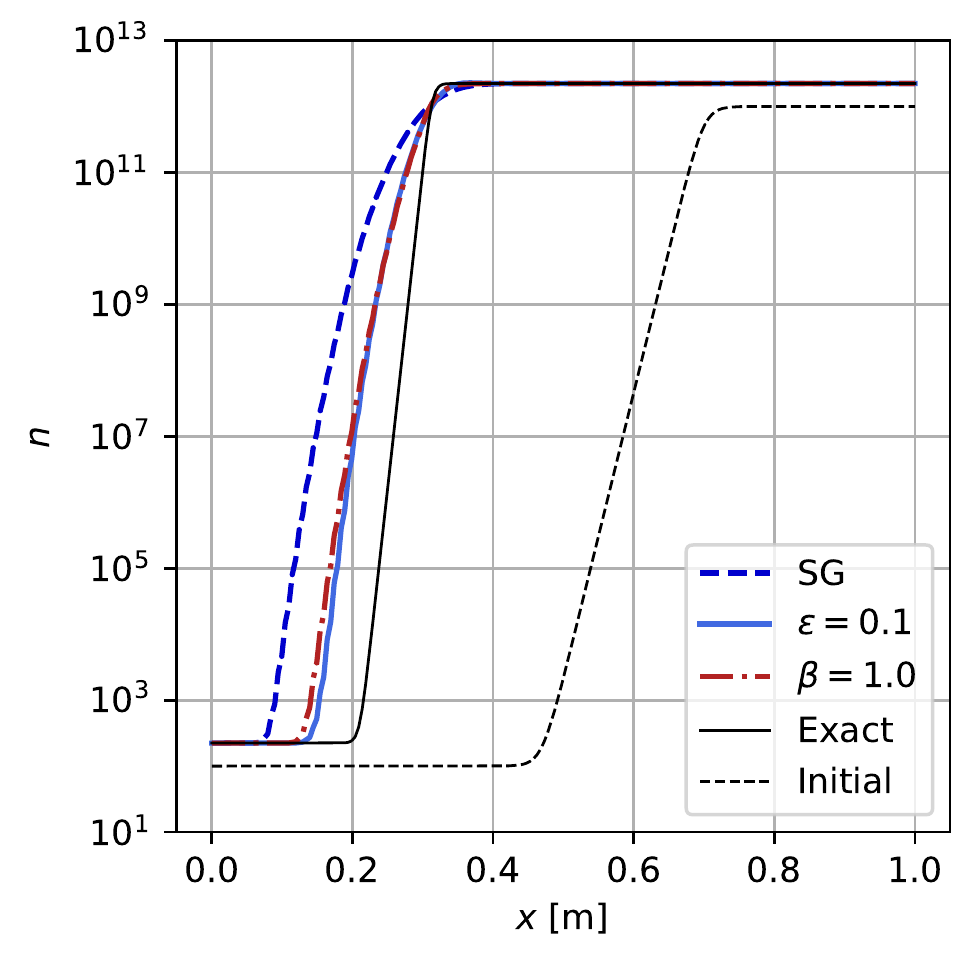}
    \caption{Four profiles for limited Lax-Wendroff scheme in 2D.}
    \label{fig:kuli_1d_schemes}
\end{figure}

The coefficient $A$ is set to $10^4$ so that advection is dominant and the effective equation can be considered without the diffusion term. The analytical solution is given by the method of characteristics to yield

\begin{equation}
    n(x, t) = n_0(x e^{At}) e^{At}
\end{equation}

\noindent where $n_0$ is the initial solution profile. The initial profile is advected to the left and compressed so that its standard deviation decreases. In plasma simulations low and high values of densities, sometimes several orders of magnitude apart, need to be well transported. Hence the following profile with amplitude $n_1$ and background $n_0$ is chosen:

\begin{equation}
    n_0(x) = n_1 + \frac{1}{2} \qty[1 + \tanh(\frac{x-x_0}{\sigma})]n_2
\end{equation}

where $n_1 = 10^{2}$ and $n_2 = 10^{12}$ so that 10 orders of magnitude separate the maximum and minimum values of the profile. This is a typical shape representative of a plasma discharge front. Results for SG, limited Lax-Wendroff with min-mod ($\beta=1.0$) and improved SG scheme with $\veps = 0.1$ are shown in Fig.~\ref{fig:kuli_1d_schemes} after 200 iterations at CFL = 0.4 in a 101 nodes mesh. Both schemes allow to retrieve better the discontinuity than the SG scheme which reduces to an upwind scheme for this advection dominated problem.

\begin{figure}[htbp]
    \centering
    \begin{subfigure}[b]{0.45\textwidth}
        \centering
        \includegraphics[width=\textwidth]{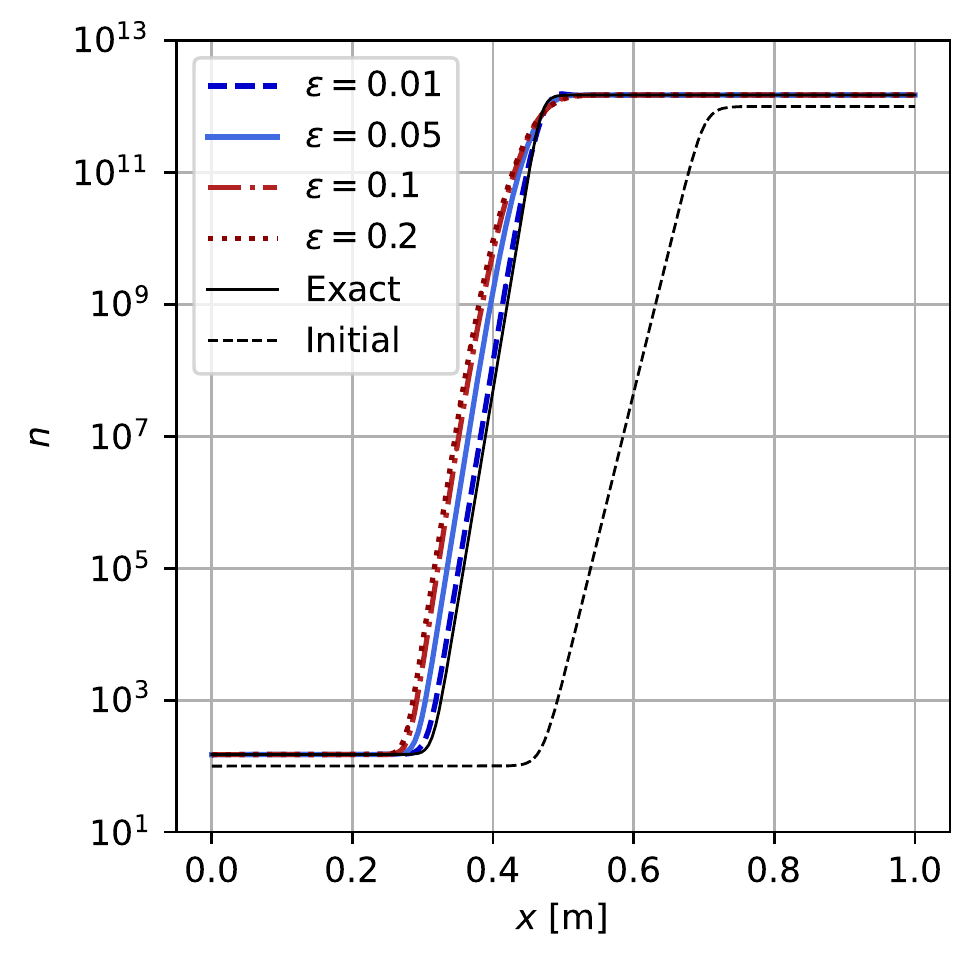}
        \caption{Improved SG}
        \label{fig:kuli_1d_sgs}
    \end{subfigure}
    \centering
    \begin{subfigure}[b]{0.45\textwidth}
        \centering
        \includegraphics[width=\textwidth]{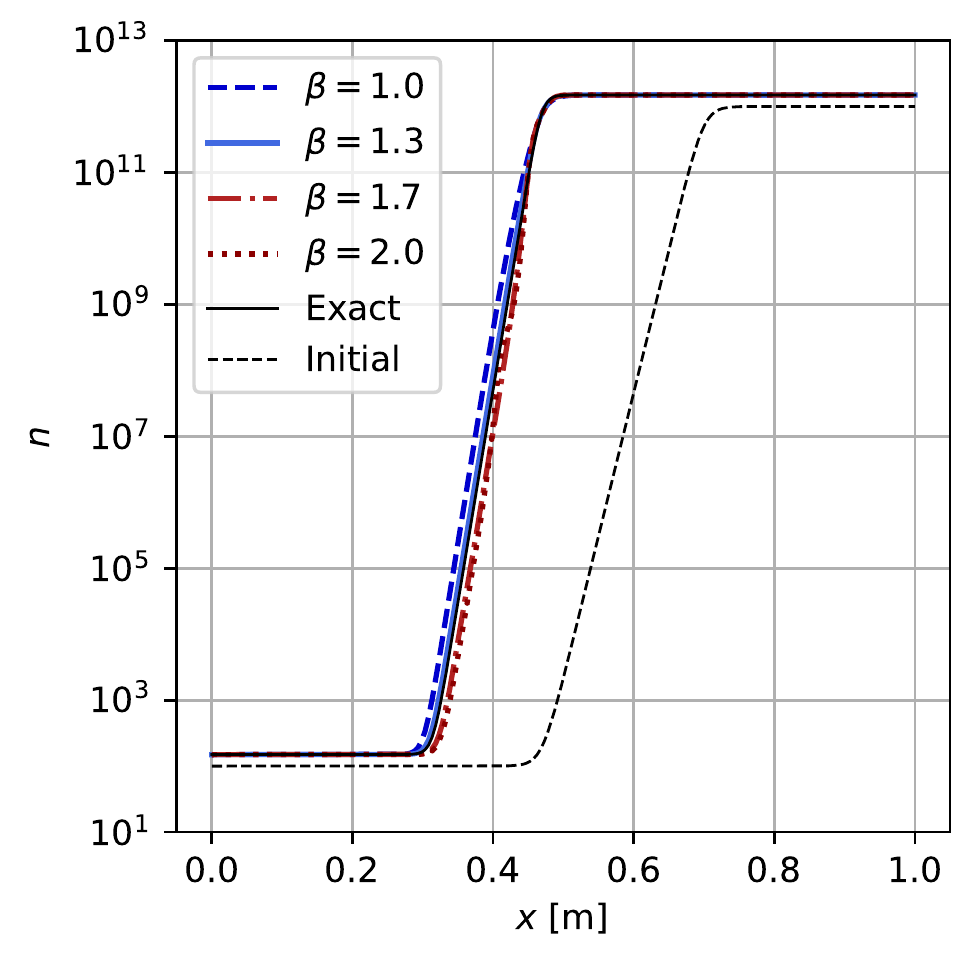}
        \caption{Limited Lax-Wendroff}
        \label{fig:kuli_1d_lw_lim}
    \end{subfigure}
    \caption{Influence of interface capturing parameters $\veps$ and $\beta$ for improved SG and limited LW schemes.}
    \label{fig:kuli_1d_sg_lw}
\end{figure}

Both schemes have parameters that allow to gain higher order and better capture discontinuities: $1 \leq \beta \leq 2$ for limited Lax-Wendroff and $0 \leq \veps \leq 1$ for improved SG. For Lax-Wendroff the higher the value of $\beta$ the better the scheme can capture discontinuities whereas for improved SG the lower the value of $\veps$ the better. The influence of those parameters on the hyperbolic tangent profile is shown in Fig.~\ref{fig:kuli_1d_sgs} and \ref{fig:kuli_1d_lw_lim}, respectively. Results are shown at CFL = 0.4 after 200 iterations on a finer grid than previously with 201 nodes in the $x$ direction. Both schemes tend to the exact solution as their parameters are closer to their limits: we can however observe that the improved SG scheme tends monotonically from left to right whereas the limited Lax-Wendroff is a bit left then a bit right of the profile.

\subsection{Bagheri benchmark}

AVIP is compared to the six streamer codes benchmark \cite{bagheri_benchmark}. The geometry is the same across all test cases: 2D axisymmetric square of $1.25 \times 1.25$ cm$^2$. Electrons $e$ and one species of positive ions $p$ are considered so that the governing equations are

\begin{gather}
\nabla^2 \phi = - \frac{e(n_p - n_e)}{\veps_0} \longrightarrow \vb{E} = - \nabla \phi \\
\pdv{n_e}{t} + \nabla \cdot (-n_e \mu_e \vb{E} - D_e \nabla n_e) = \bar{\alpha} \mu_e ||\vb{E}||n_e + S_{ph} \\
\pdv{n_p}{t}  = \bar{\alpha} \mu_e ||\vb{E}||n_e + S_{ph}
\end{gather}

\noindent where $\mu_e$ is the electron mobility, $D_e$ the electron diffusion coefficient, $\bar{\alpha}$ the effective ionization coefficient (ionization $\alpha$ minus attachment $\eta$), $\vb{E}$ the electric field and $S_{ph}$ the photoionization source term. The chemistry is composed of three reactions where we denote the neutral $\ce{A}$, the positive ion $\ce{A+}$ and the negative ion $\ce{A-}$ (not described):

\begin{align}
    &\qq{Ionization} \ce{e- + A -> 2e- + A+} \\
    &\qq{Attachment} \ce{e- + A -> A-} \\
    &\qq{Photionization} \ce{e- + $\gamma$ -> e- + A+}
\end{align}

\begin{figure}
    \centering
    \begin{subfigure}{0.34\textwidth}
        \centering
        \includegraphics[width=\textwidth]{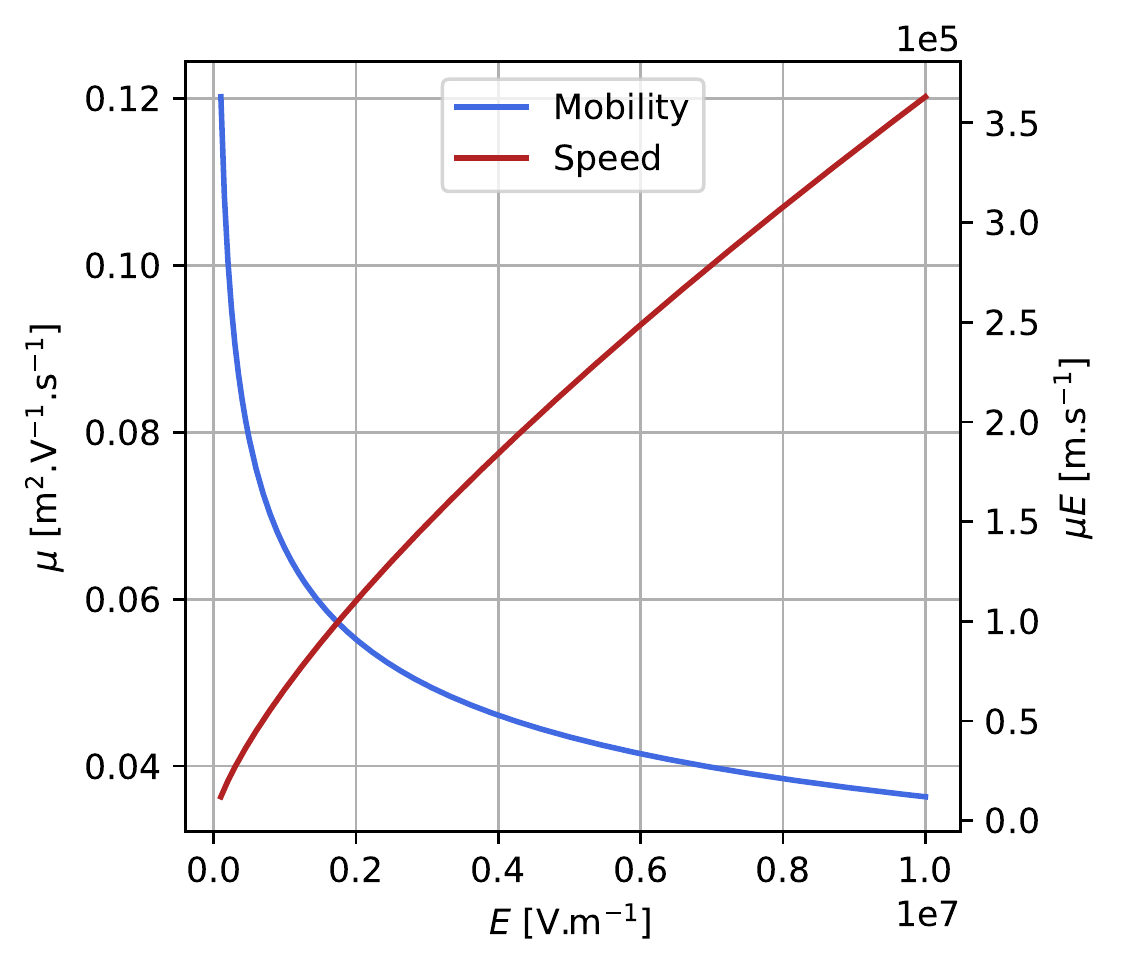}
        \caption{Mobility and speed}
    \end{subfigure}
    \begin{subfigure}{0.3\textwidth}
        \centering
        \includegraphics[width=\textwidth]{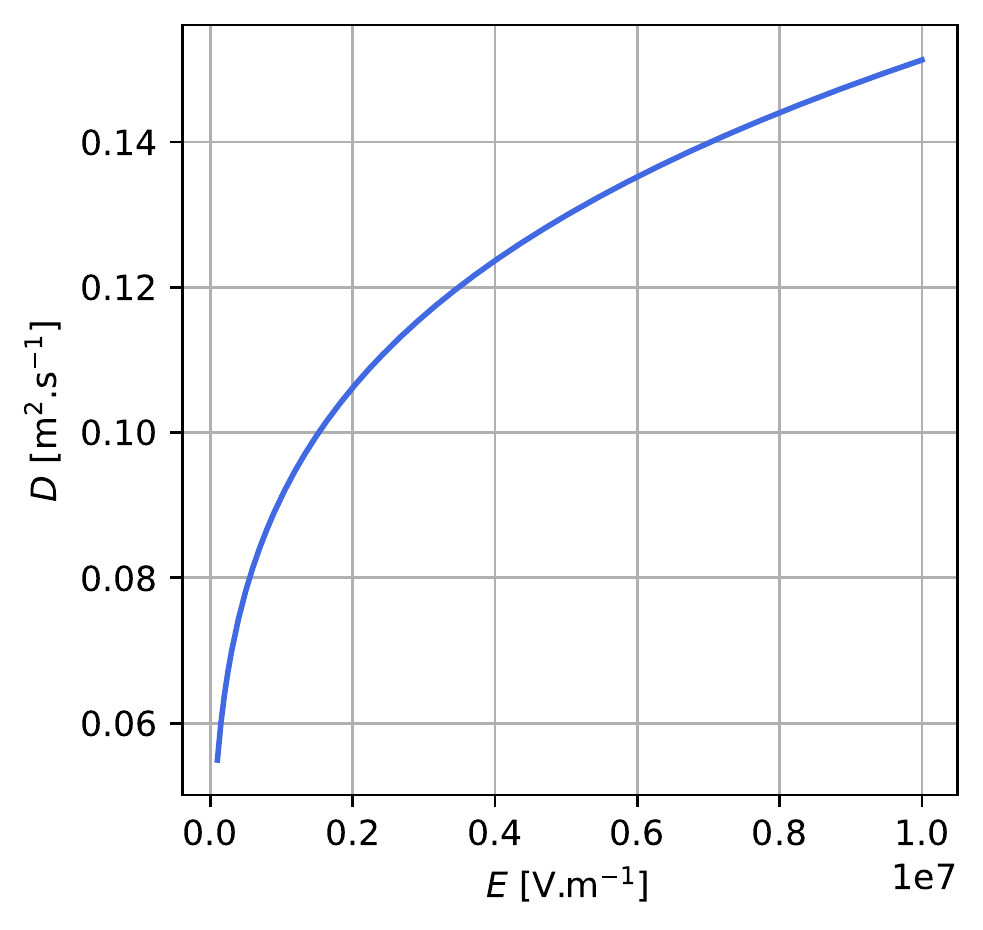}
        \caption{Diffusion coefficient}
    \end{subfigure}
    \begin{subfigure}{0.3\textwidth}
        \centering
        \includegraphics[width=\textwidth]{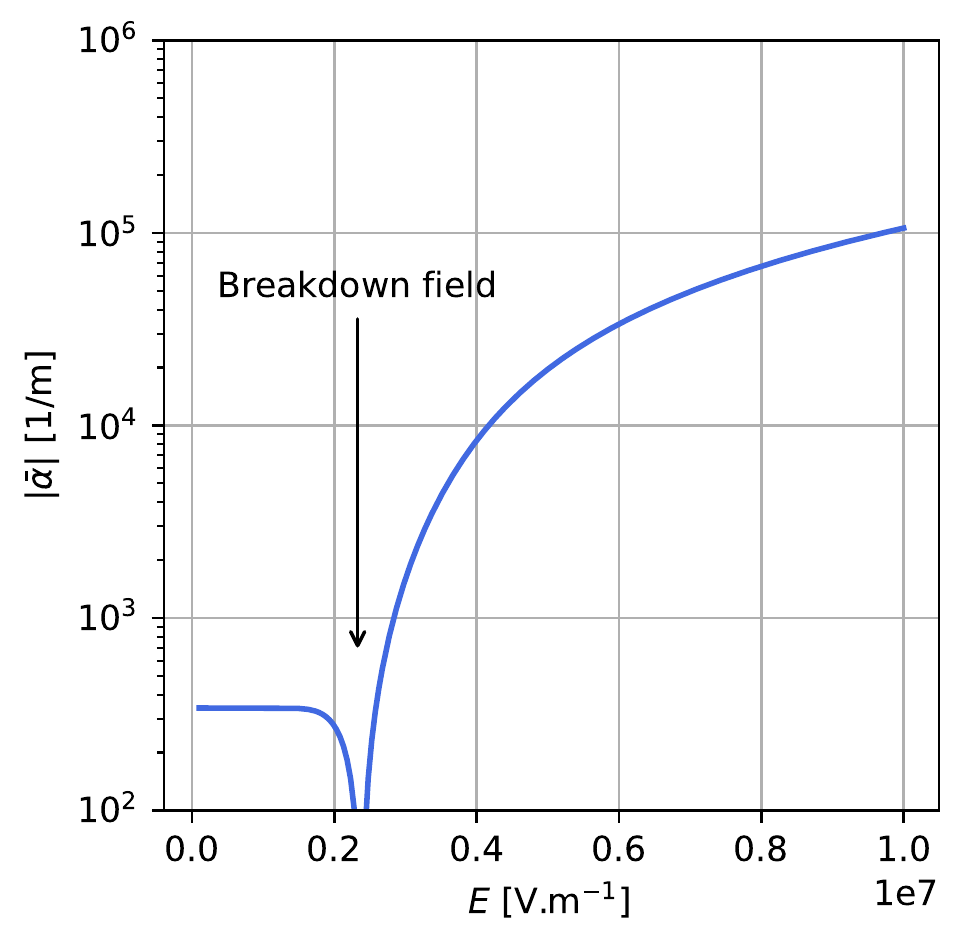}
        \caption{Effective ionization}
    \end{subfigure}
    \caption{Transport and chemistry coefficients as functions of the electric field norm for the \cite{bagheri_benchmark} validation case.}
    \label{fig:bagheri_coefs}
\end{figure}

We consider streamer discharges in dry air at $p=\SI{1}{\bar}$ and $T = \SI{300}{\kelvin}$ for which the coefficients are tabulated as a function of of the reduced electric field (local field approximation) from \cite{bagheri_benchmark}. The coefficients for the chosen conditions are plotted in Fig.~\ref{fig:bagheri_coefs}. These coefficients are only function of the electric field since the neutral density has been set for $p=\SI{1}{\bar}$ and $T = \SI{300}{\kelvin}$. Mobility decreases and diffusion increases as the electric field increases due to more collisions. Finally the effective ionization as a function of the electric field allows to retrieve the breakdown field $E_b$. This is a critical parameter as it controls the onset of the electron avalanche process described in \cite{plasmachemistry}: above $E_b$ discharges can multiply exponentially and a discharge can propagate. The value of the breakdown field is around \SI{2.2e6}{\mega\volt\per\metre} for this chemistry.

The positive streamer is initialized with a Gaussian profile on the positive species density only so that a strong charge density enhances locally the electric field to trigger a positive streamer propagation. A common background for both electrons and positive ions is also set and the densities are initialized as

\begin{gather}\label{eq:positive_ions_bagheri}
n_i(x, r) = n_0 \exp\left(-\frac{r^2 + (x - x_0)^2}{\sigma^2}\right) + n_\mrm{back}\\
n_e(x, r) = n_\mrm{back}
\end{gather}

\noindent where $n_0 = 5 \times 10^{18}$ m$^{-3}$ and $x_0 = 10^{-2}$ m are taken for all cases.

\subsubsection{Meshes}

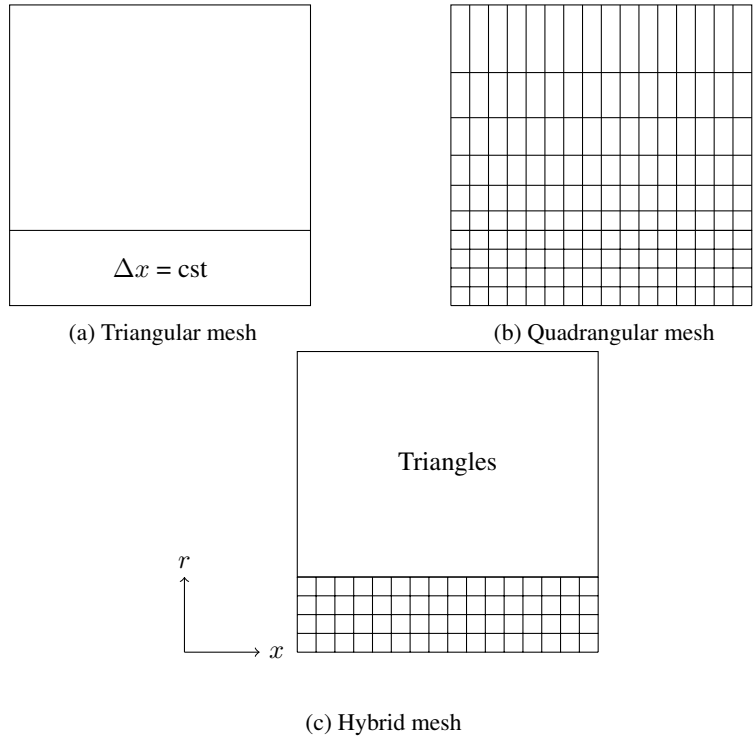
\begin{figure}[htbp]
    \centering
    \begin{subfigure}[b]{0.35\textwidth}
        \centering
        \begin{tikzpicture}
    \draw (0, 0) rectangle (4, 4);
    \draw (0, 1) -- (4, 1);
    \node[anchor=center] at (2, 0.5) {$\upD x$ = cst};
\end{tikzpicture}
        \caption{Triangular mesh}
        \label{fig:tri_mesh}
    \end{subfigure}
    \begin{subfigure}[b]{0.35\textwidth}
        \centering
        \begin{tikzpicture}
    \draw[step=0.25cm] (0, 0) grid (4, 1);
    \draw (0, 1) rectangle (4, 4);
    \foreach \i in {1, ..., 15}
        \draw (\i * 0.25, 1) -- (\i * 0.25, 4);
    \foreach \y in {1.26, 1.6, 2.0, 2.5, 3.1}
        \draw (0, \y) -- (4, \y);
\end{tikzpicture}
        \caption{Quadrangular mesh}
        \label{fig:quad_mesh}
    \end{subfigure}
    \begin{subfigure}[b]{0.35\textwidth}
        \centering
        \begin{tikzpicture}
    \draw[step=0.25cm] (0, 0) grid (4, 1);
    \draw (0, 1) rectangle (4, 4);
    \node[anchor=center] at (2, 2.5) {Triangles};
    \draw[->] (-1.5, 0) -- (-0.5, 0) node[anchor=west] {$x$};
    \draw[->] (-1.5, 0) -- (-1.5, 1) node[anchor=south] {$r$};
\end{tikzpicture}
        \caption{Hybrid mesh}
        \label{fig:hybrid_mesh}
    \end{subfigure}
    \caption{Meshes for \cite{bagheri_benchmark} test case.}
    \label{fig:bagheri_meshes}
\end{figure}

Triangular and quadrangular elements are considered for the validation of the schemes since AVIP can work on both types of elements. Three kind of meshes are considered: triangular meshes, quadrangular meshes and finally hybrid meshes which are shown in Fig.~\ref{fig:bagheri_meshes}. The refinement zone is a rectangle at the bottom of the computational domain around the axis $r = 0$. In triangular meshes, only the limited LW scheme can be used since the improved SG scheme only works in topologically dual meshes which is not ensured easily for triangular meshes. In quadrangular meshes, both schemes can be applied and finally in hybrid meshes the improved SG scheme can be applied in quadrangles whereas the limited Lax-Wendroff scheme is used in triangles.

Different mesh sizes are considered for each type of elements and are summarized in Tab.~\ref{tab:bagheri_meshes}. Note that the typical mesh size used in the benchmark is $\SI{3}{\micro\metre}$.

\begin{table}
    \centering
    \begin{tabular}{| c | c | c | c | c |}\hline
        Name & Type & $\upD x_\mrm{min}$ & $N_\mrm{nodes}$ & $N_\mrm{cells}$ \\\hline
        \texttt{tri5micro} & Tri & \SI{5}{\micro\metre} &  $4.2 \times 10^{5}$  & $8.4 \times 10^{5}$ \\
        \texttt{tri3micro} & Tri & \SI{3}{\micro\metre} &  $1 \times 10^{6}$  & $2.1 \times 10^{6}$ \\
        \texttt{tri2micro} & Tri & \SI{2}{\micro\metre} &  $2.33 \times 10^{6}$  & $4.66 \times 10^{6}$ \\
        \texttt{quad5micro} & Quad & \SI{5}{\micro\metre} &  $7.5 \times 10^{5}$  & $7.5 \times 10^{5}$ \\
        \texttt{quad3micro} & Quad & \SI{3}{\micro\metre} &  $1.8 \times 10^{6}$  & $1.8 \times 10^{6}$ \\
        \texttt{quad2micro} & Quad & \SI{2}{\micro\metre} &  $3.8 \times 10^{6}$  & $3.8 \times 10^{6}$ \\
        \texttt{hybrid5micro} & Hybrid & \SI{5}{\micro\metre} &  $6.5 \times 10^{5}$  & $7.1 \times 10^{5}$ \\
        \texttt{hybrid25micro} & Hybrid & \SI{2.5}{\micro\metre} &  $2.5 \times 10^{6}$  & $2.6 \times 10^{6}$ \\
        \hline
    \end{tabular}
    \caption{Summary of the meshes used for the benchmark}
\label{tab:bagheri_meshes}
\end{table}

\subsubsection{Presentation of cases}

Three test cases are considered in this benchmark to assess different aspects of streamer propagation. A constant background electric field is set to \SI{1.5}{\mega\volt\per\metre} in the $x$ direction: this value is below the breakdown electric field of \SI{2.2}{\mega\volt\per\metre} so that it is not enough to propagate a streamer by itself: the initial gaussian seed of charge density Eq.~\eqref{eq:positive_ions_bagheri} imposed at the beginning allows to locally increase the electric field \textit{above the breakdown field} so that a streamer can propagate. The three cases have two varying parameters: background charged species density $n_\mrm{back}$ and photoionization which are summarized in Tab.~\ref{tab:bagheri_cases}. Each test case is now going to be described along with its results.

\begin{table}[hbtp]
\centering
\begin{tabular}{| c | c | c |}\hline
 Casename  &  $n_\mrm{back}$ [m$^{-3}$]  &  Photoionization \\
\hline Case 1    &  $10^{13}$                  &  No              \\
 Case 2    &  $10^{9}$                   &  No              \\
 Case 3    &  $10^{9}$                   &  Yes             \\
\hline
\end{tabular}
\caption{Summary of the three cases parameters}
\label{tab:bagheri_cases}
\end{table}

\subsubsection{Case 1 results}

The first case has a rather strong background density at \SI{1e13}{\per\metre\cubed} and no photoionization: it is a canonical streamer propagation where photoionization has been replaced by a strong background density.

\begin{figure}[htbp]
    \begin{subfigure}{0.48\textwidth}
        \centering
        \includegraphics[width=\textwidth]{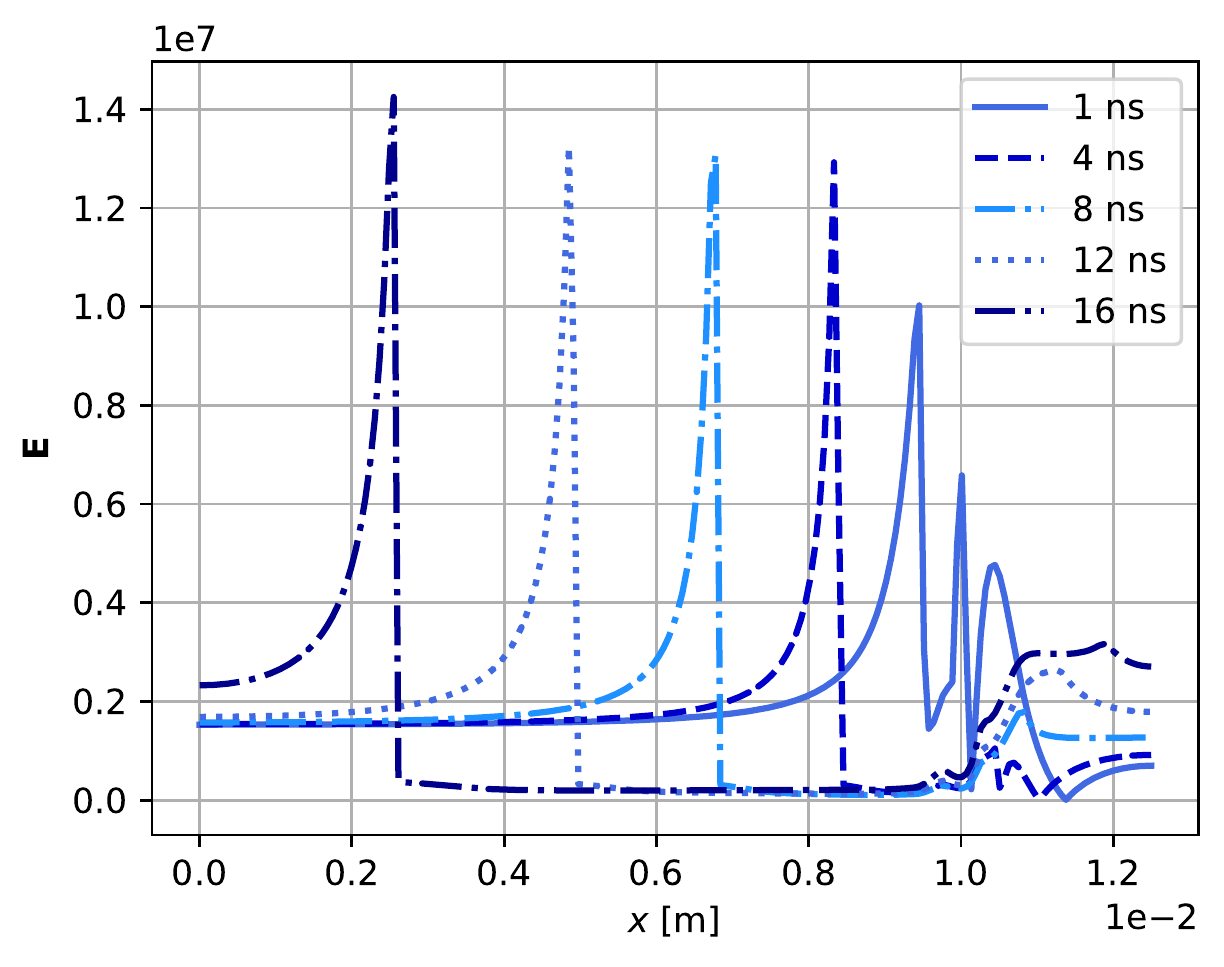}
    \end{subfigure}
    \begin{subfigure}{0.48\textwidth}
        \centering
        \includegraphics[width=\textwidth]{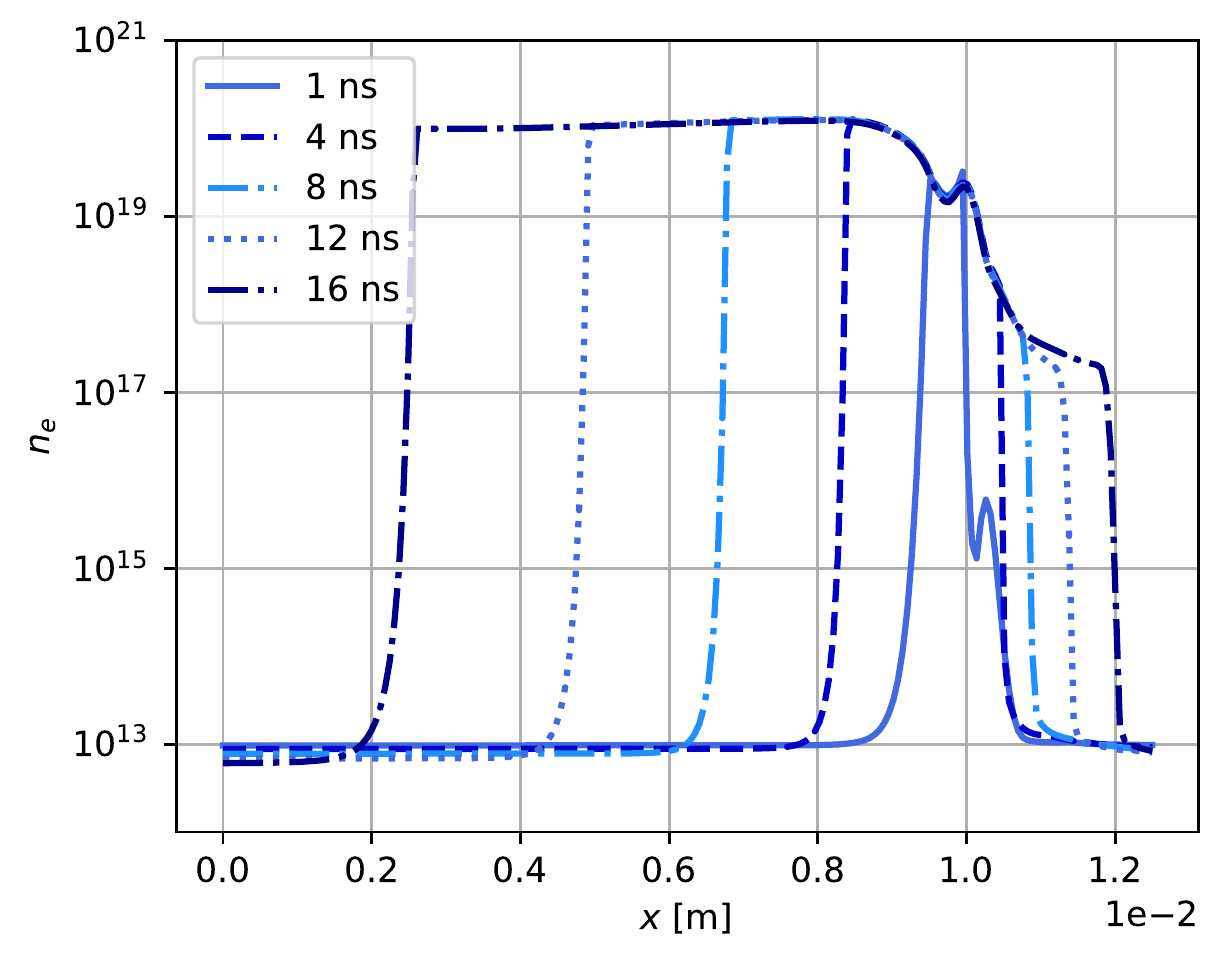}
    \end{subfigure}
    \caption{The norm of electric field (left) and the electron density (right) at different instants on the axis $r = 0$ for the ISG scheme with $\veps = 0.01$.}
    \label{fig:case1_instants}
\end{figure}

\begin{figure}[htbp]
    \centering
    \begin{subfigure}{0.8\textwidth}
        \centering
        \includegraphics[width=\textwidth]{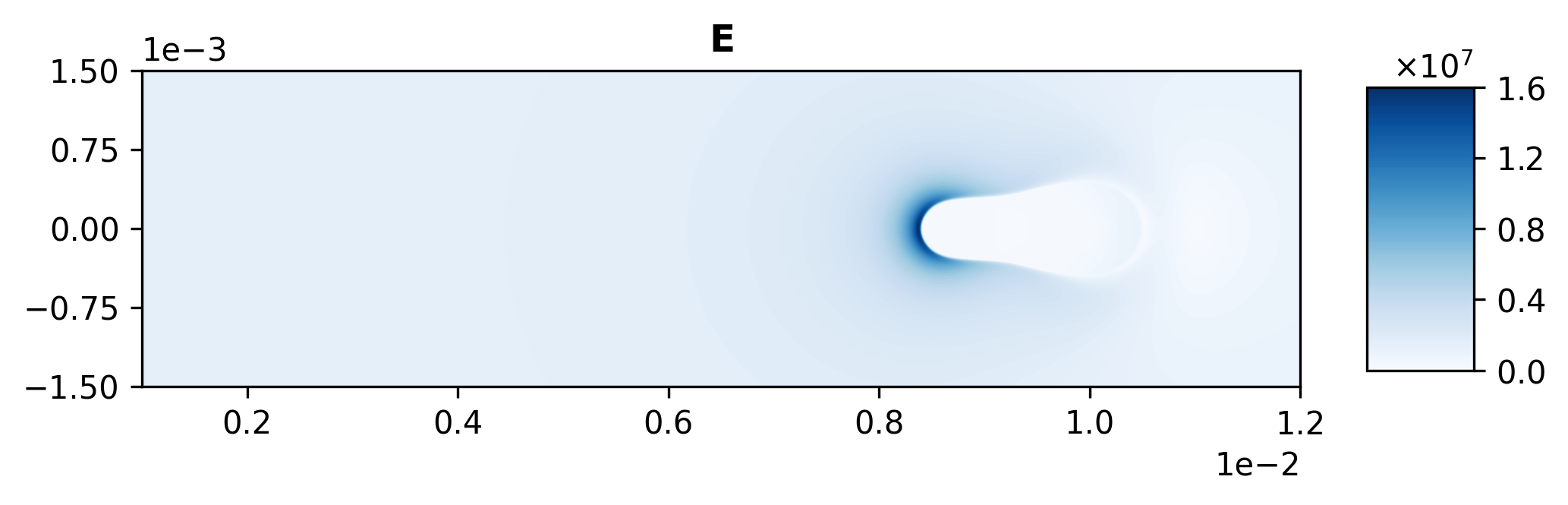}
        \caption{4 ns}
    \end{subfigure}
    \begin{subfigure}{0.8\textwidth}
        \centering
        \includegraphics[width=\textwidth]{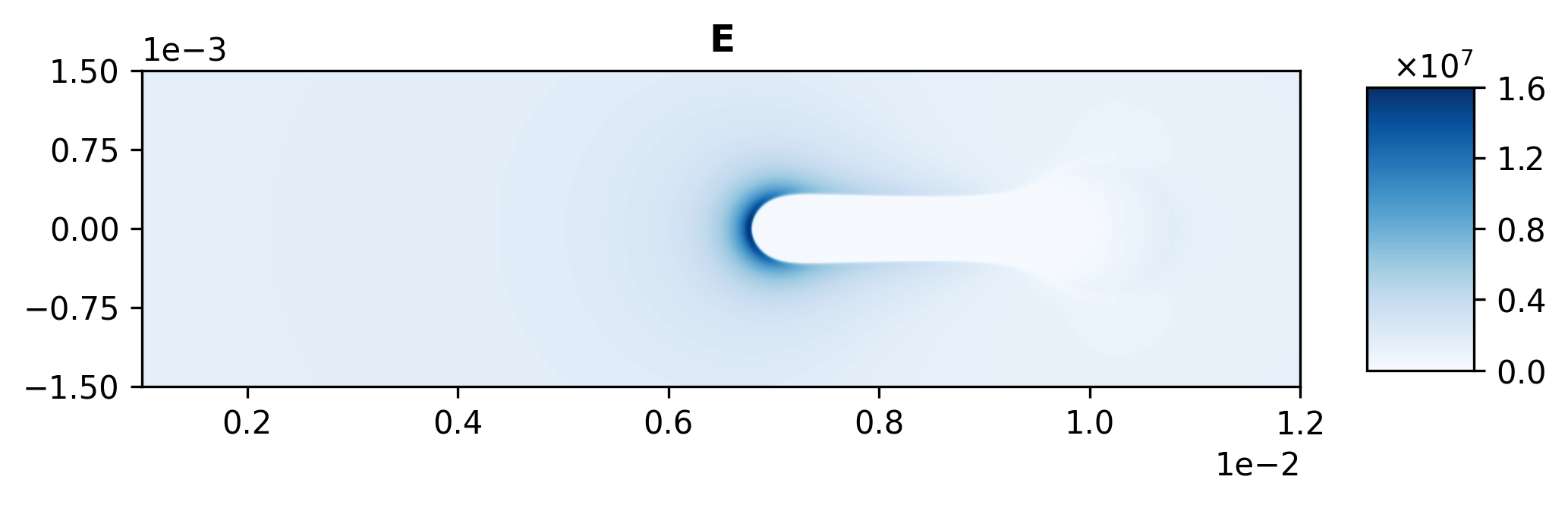}
        \caption{8 ns}
    \end{subfigure}
    \begin{subfigure}{0.8\textwidth}
        \centering
        \includegraphics[width=\textwidth]{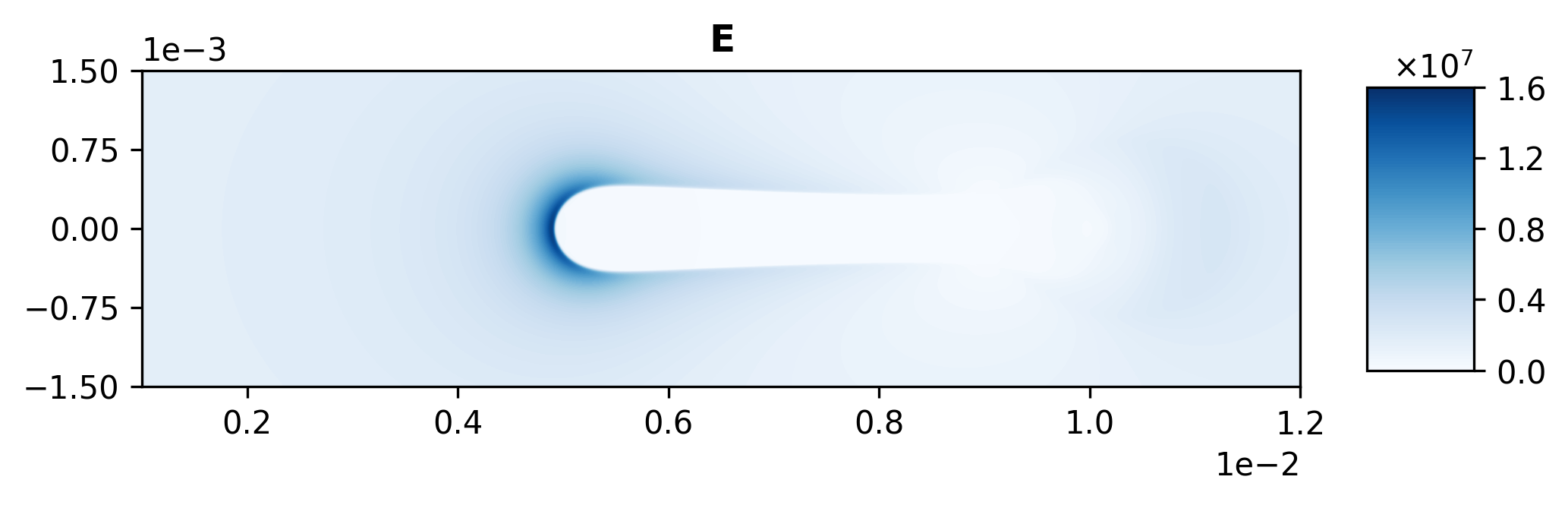}
        \caption{12 ns}
    \end{subfigure}
    \begin{subfigure}{0.8\textwidth}
        \centering
        \includegraphics[width=\textwidth]{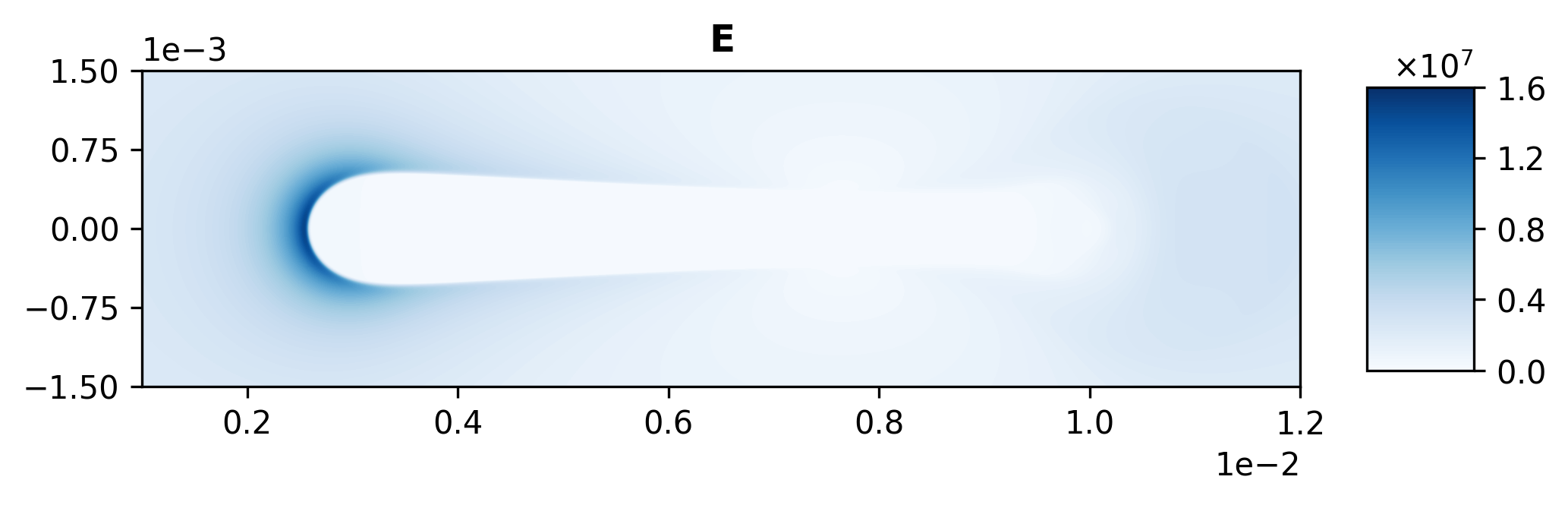}
        \caption{16 ns}
    \end{subfigure}
    \caption{Electric field of the streamer at different instants for ISG at $\upD x = \SI{3}{\micro\meter}$.}
    \label{fig:bagheri_case1_isg3micro}
\end{figure}

The electron density and electric field for the ISG scheme is shown in Fig.~\ref{fig:bagheri_case1_isg3micro} for the full quadrangular mesh at $\upD x = \SI{3}{\micro\metre}$. The propagation of the positive streamer from right to left is observed and the norm electric field at the head is around \SI{15}{\mega\volt\per\metre} which is above the breakdown field of the chemistry, causing the electron avalanche and the streamer propagation. Profiles of electron density and electric field norm along the axis are shown in Fig.~\ref{fig:case1_instants} where the propagation of the streamer head is observed: the peak of electric field moves from right to left and causes ionization leaving behind a higher electron density.

\begin{figure}[htbp]
    \centering
    \begin{subfigure}{0.8\textwidth}
        \centering
        \includegraphics[width=\textwidth]{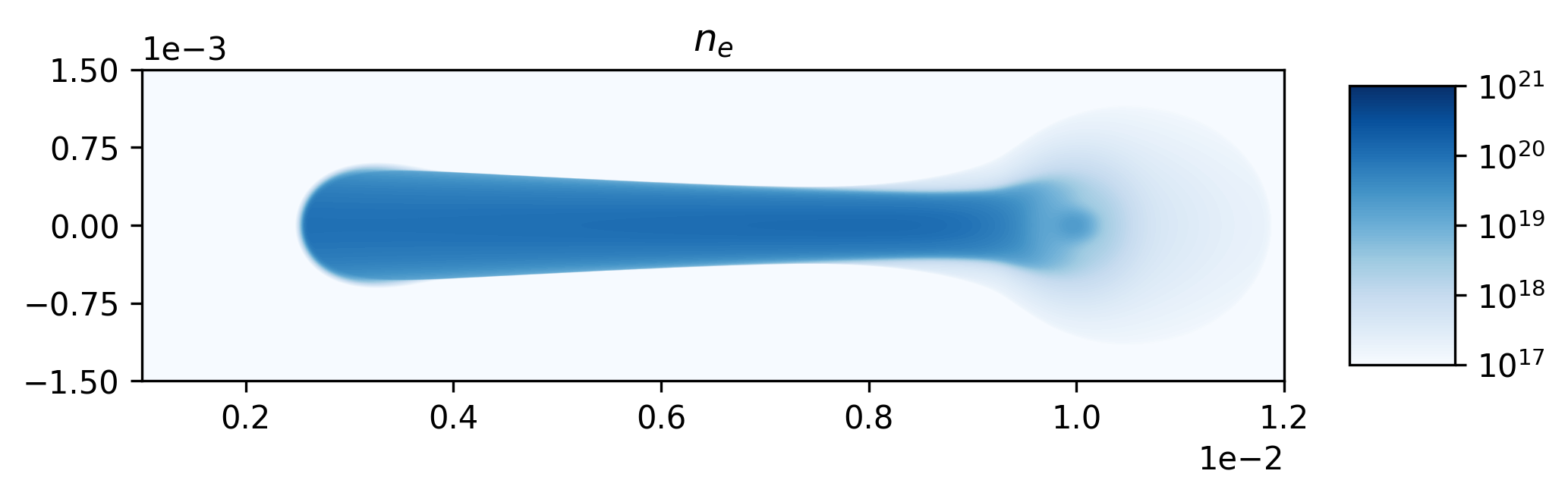}
        \caption{ISG $\upD x = \SI{3}{\micro\metre}$ - Quad}
        \label{fig:bagheri_ne_isg_3micro}
    \end{subfigure}
    \begin{subfigure}{0.8\textwidth}
        \centering
        \includegraphics[width=\textwidth]{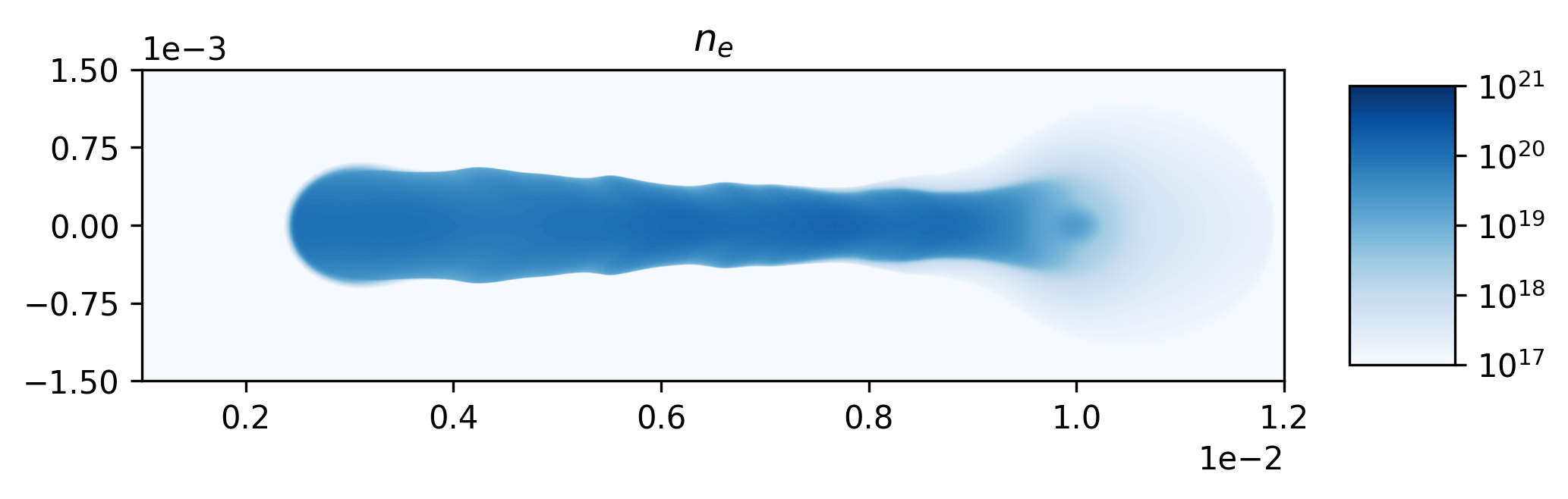}
        \caption{LLW $\upD x = \SI{3}{\micro\metre}$ - Tri}
        \label{fig:bagheri_ne_llw_3micro}
    \end{subfigure}
    \begin{subfigure}{0.8\textwidth}
        \centering
        \includegraphics[width=\textwidth]{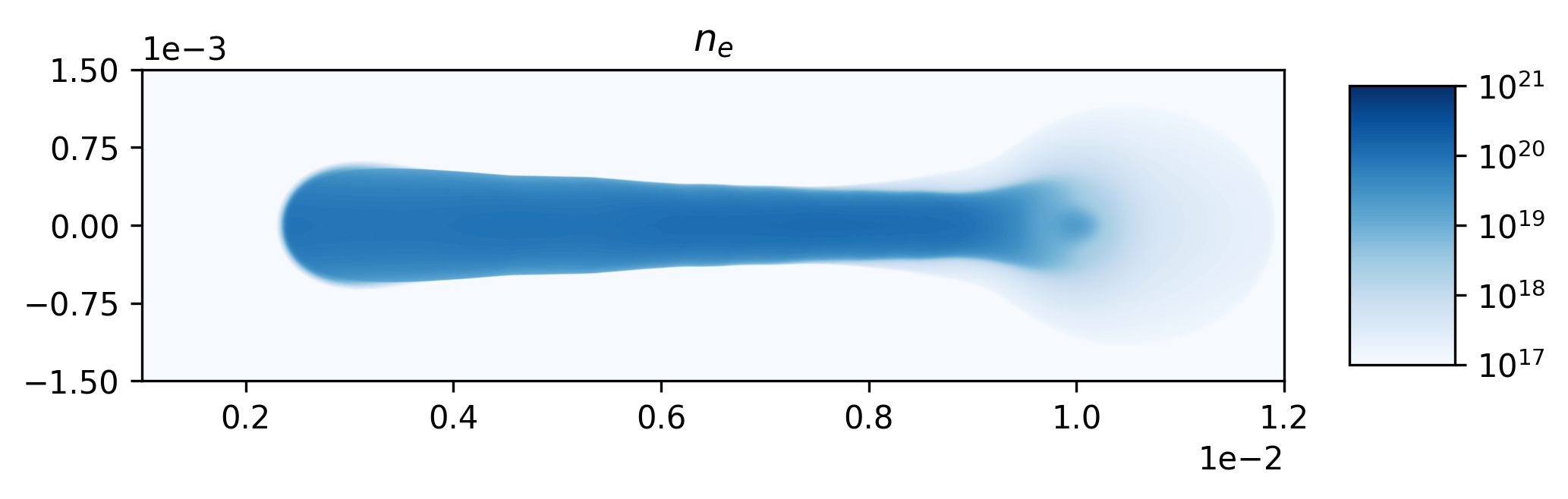}
        \caption{LLW $\upD x = \SI{2}{\micro\metre}$ - Tri}
        \label{fig:bagheri_ne_llw_2micro}
    \end{subfigure}
    \caption{Electron density at $t = \SI{16}{\nano\second}$ for different schemes and resolutions.}
    \label{fig:bagheri_case1_ndensity_comp}
\end{figure}

On the other hand the LLW scheme at the same resolution of $\upD x = \SI{3}{\micro\metre}$ using a triangular mesh produces oscillations. A comparison of the electron density at $t = \SI{16}{\nano\second}$ in Fig.~\ref{fig:bagheri_case1_ndensity_comp} shows that the LLW scheme (Fig.~\ref{fig:bagheri_ne_llw_3micro}) using a triangular mesh produces oscillations. These oscillations, present in both the electron density and electric field profiles, indicate that the streamer was on the verge of branching. By increasing the resolution to $\upD = \SI{2}{\micro\metre}$ (Fig.~\ref{fig:bagheri_ne_llw_2micro}), these oscillations disappear and hence trianglular elements are less stable than quadrangular elements.

\begin{figure}[htbp]
    \begin{subfigure}{0.48\textwidth}
        \centering
        \includegraphics[width=\textwidth]{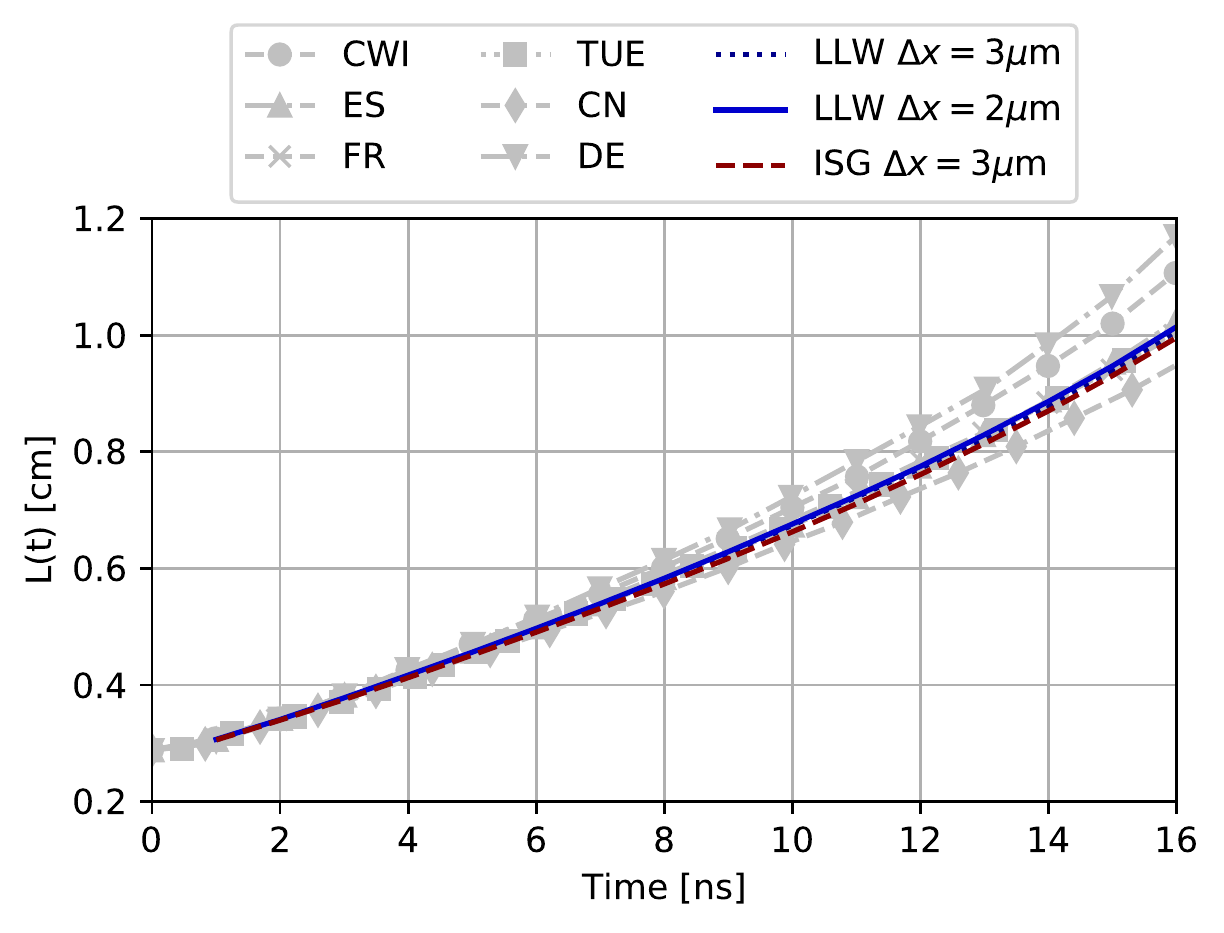}
    \end{subfigure}
    \begin{subfigure}{0.48\textwidth}
        \centering
        \includegraphics[width=\textwidth]{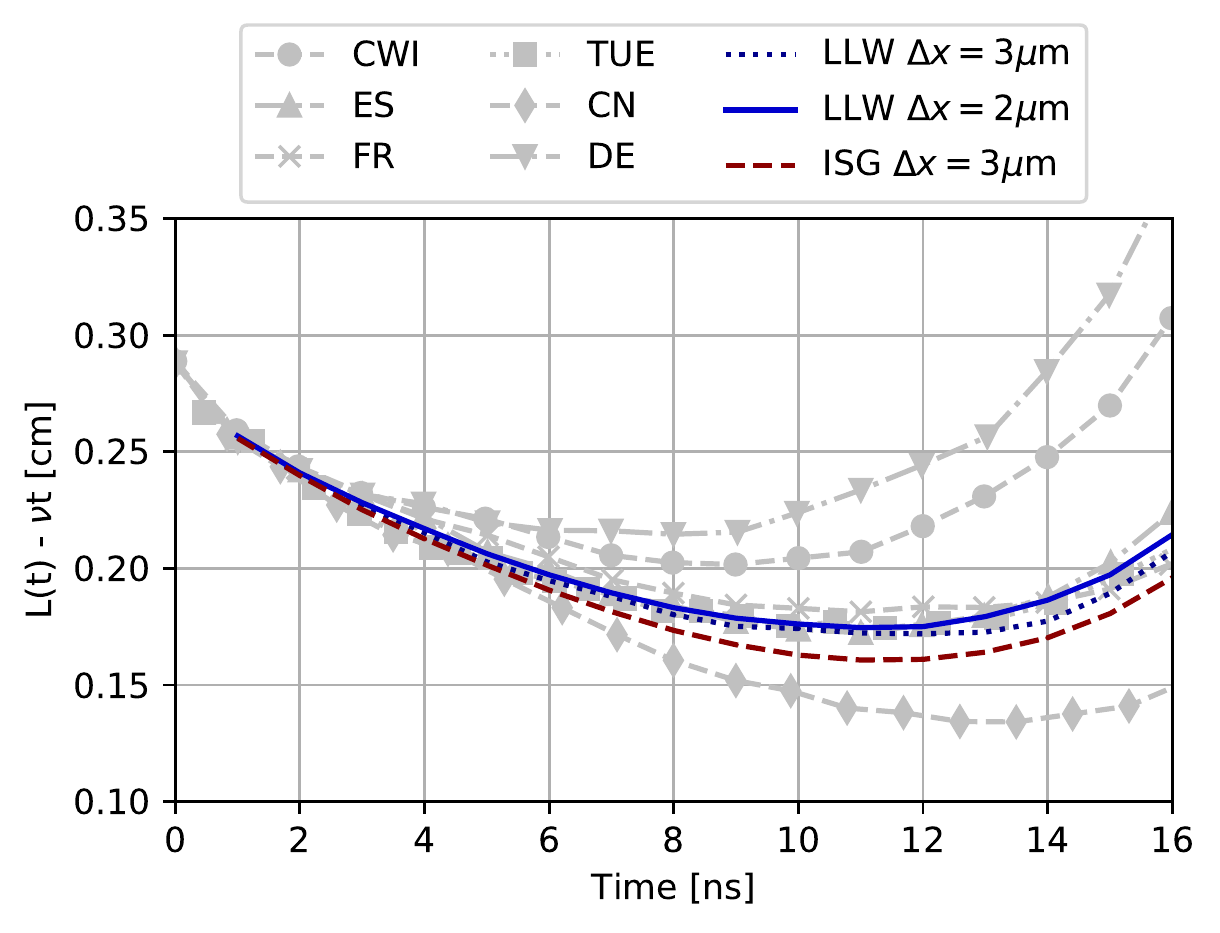}
    \end{subfigure}
    \caption{Streamer length as a function of time for all the streamer codes and AVIP.}
    \label{fig:figure_5_comp}
\end{figure}

\begin{figure}[htbp]
    \centering
    \includegraphics[width=0.6\textwidth]{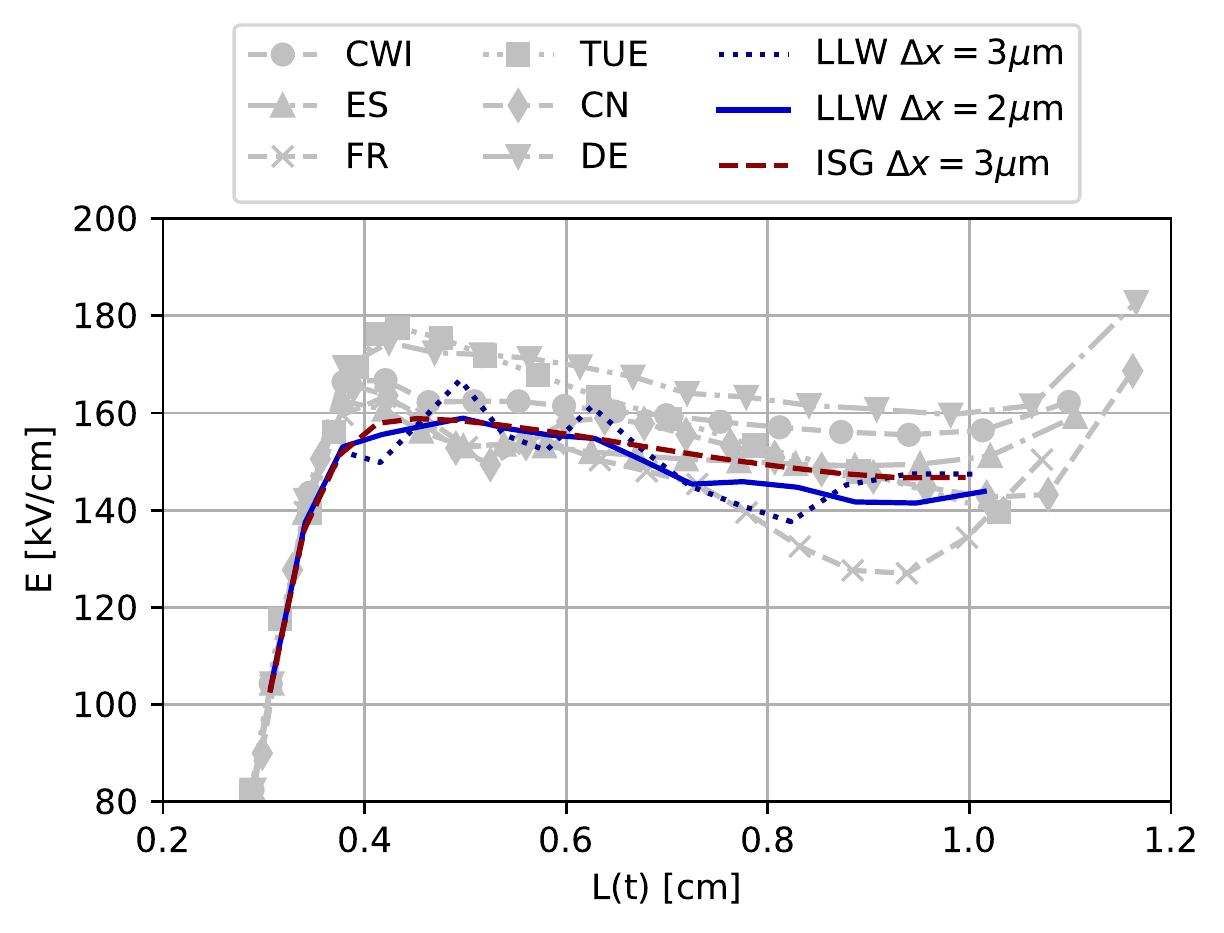}
    \caption{Electric field as a function of the streamer length for all the streamer codes and AVIP.}
    \label{fig:figure_6a_comp}
\end{figure}

A comparison of those three runs with the six codes of the benchmark is shown in Fig.~\ref{fig:figure_5_comp} for the streamer length which is defined as the distance between the initial gaussian seed and the maximum of the norm of the electric field. We can see that all three runs are satisfactory and lie between all the codes of the benchmark. However when looking at the value of the maximum electric field as a function of the streamer length in Fig.~\ref{fig:figure_6a_comp} we can clearly see the oscillations present in the LW $\upD x = \SI{3}{\micro\metre}$ run.

From this first case we can conclude that triangular meshes are able to propagate streamer but are less robust than quadrangular meshes. In all these axisymmetric simulations, the key is to keep the streamer on the axis because in this setting branching is not supposed to occur.

To combine the advantages of the stability of quadrangular elements with the flexibility of triangular elements we decide to use a hybrid approach whenever necessary which is shown in Fig.~\ref{fig:hybrid_mesh}: SG is applied in the regular and topologically dual quadrangular elements while the limited LW scheme is applied in the triangular elements. This new scheme is labeled ISG-LLW and it is implied that hybrid meshes are used whenever we use this scheme. This allows to lower the number of nodes and elements of the mesh while keeping the stability of the simulation.

\begin{figure}[htbp]
    \centering
    \begin{subfigure}{0.8\textwidth}
        \centering
        \includegraphics[width=\textwidth]{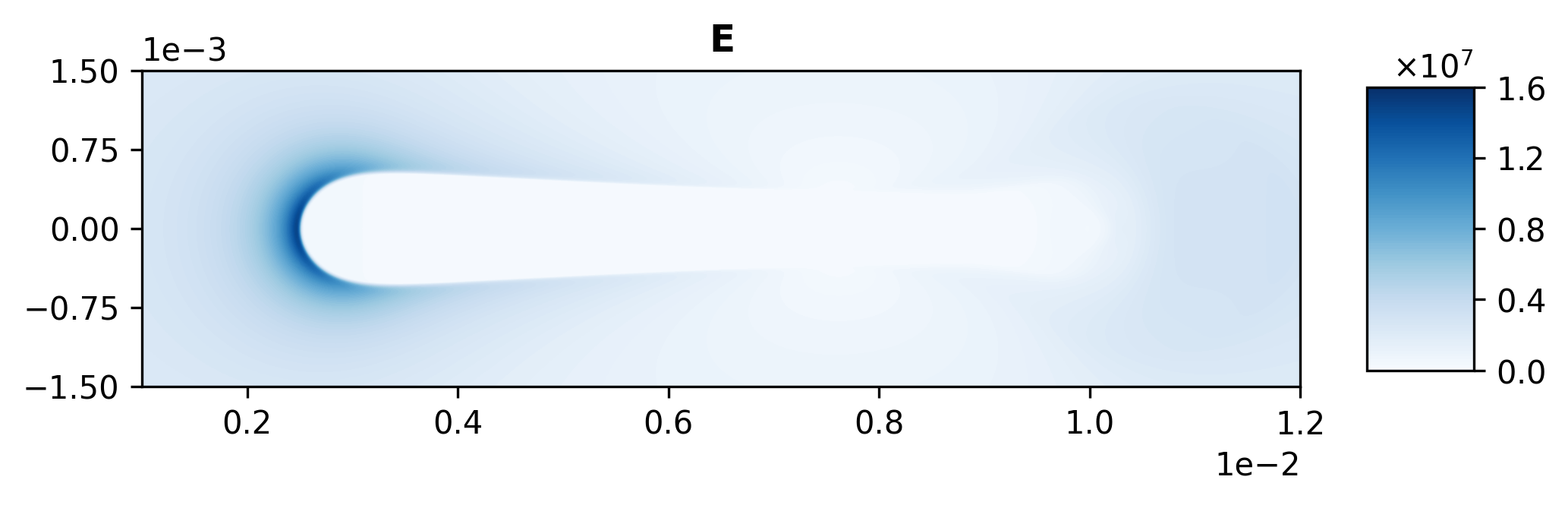}
        \caption{Norm of electric field}
        \label{fig:bagheri_Efield_hybrid}
    \end{subfigure}
    \begin{subfigure}{0.8\textwidth}
        \centering
        \includegraphics[width=\textwidth]{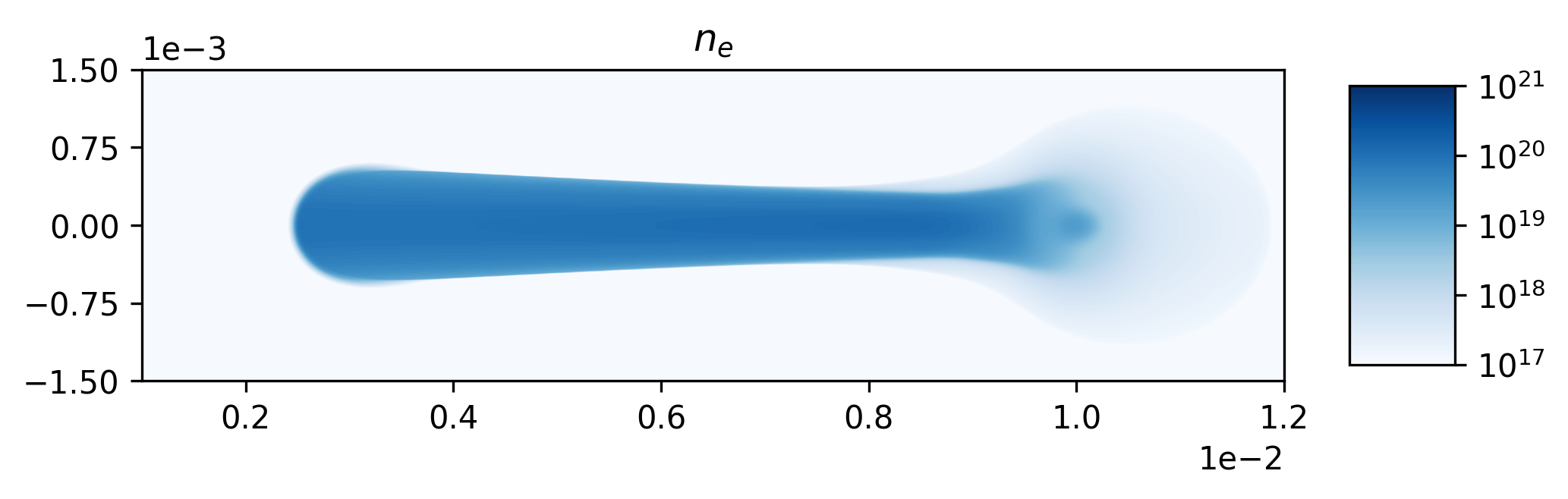}
        \caption{Electron density}
        \label{fig:bagheri_ne_hybrid}
    \end{subfigure}
    \caption{Electron density and norm of electric field at $t = \SI{16}{\nano\second}$ using the hybrid ISG-LLW numerical scheme on \texttt{hybrid5micro} mesh.}
    \label{fig:bagheri_case1_hybrid}
\end{figure}

Using this hybrid approach a run with $\upD x = \SI{5}{\micro\metre}$ has been tested using the \texttt{hybrid5micro} mesh. The electron density and electric field at the last instant are shown in Fig.~\ref{fig:bagheri_case1_hybrid} and are very close to the results using the full quadrangular mesh of Fig.~\ref{fig:bagheri_case1_isg3micro} which validates the approach of the hybrid scheme and shows again the robustness of quadrangular elements compared to triangular elements which are stable only for a resolution of $\upD x = \SI{2}{\micro\metre}$.

Finally a few words about performance for this test case. The six codes used in the benchmark along with their mesh characteristics and run time are presented in Tab.~\ref{tab:bagheri_case1_codes}. The octree code from the CWI group is by far the fastest at around 20 min core time thanks to a very low number of cells due to very efficient adaptive mesh refinement. The other codes use at least five times more cells, increasing significanty their core time from 6 h to 90 h. The equivalent results using AVIP and the various schemes and meshes mentioned in the section are presented in Tab.~\ref{tab:bagheri_case1_avip}. We use for all runs a bi-socket Intel node with 2 x 18 core Xeon Gold 6140 (2.3 Ghz clock speed and 96 Gb memory). The hybrid option, due to its low number of cells, is the fastest option putting AVIP in second place in terms of performance compared with the other codes.

\begin{table}[htbp]
    \centering
    \begin{tabular}{| c | c | c | c | c | c | c |}
        \hline
        Code & CWI & ES & FR & CN & TUE & DE \\\hline
    Adaptive refinement & yes & yes & no & yes & no & no \\
    Min grid size & \SI{3}{\micro\metre} & \SI{3}{\micro\metre} & \SI{3}{\micro\metre} & \SI{3}{\micro\metre} & \SI{3}{\micro\metre} & \SI{3}{\micro\metre} \\
    Max grid size &  &  &  & \SI{8}{\micro\metre} & & \SI{5}{\micro\metre} \\
    $N_\mrm{cells}$ & $1.2 \times 10^5$ & $2.0 \times 10^6$ & $1.1 \times 10^6$ & $6.5 \times 10^5$ & $4.2 \times 10^6$ & $5.1 \times 10^5$ \\
    Time step & dyn. & \SI{1}{\pico\second} & dyn. & dyn. & dyn. & dyn. \\
    CPU cores & 4 & 1 & 1 & 4 & 1 & 6 \\
    Run time & 5 min & 20 h & 6 h & 18 h & 25 h & 15 h \\
    Core time & 20 min & 20 h & 6 h & 72 h & 25 h & 90 h \\\hline
    \end{tabular}
    \caption{A summary of simulation settings for case 1 of the different codes used in the benchmark \cite{bagheri_benchmark}, adapted from the benchmark paper.}
    \label{tab:bagheri_case1_codes}
\end{table}

\begin{table}[htbp]
    \centering
    \begin{tabular}{| c | c | c | c | c |}
        \hline
        Mesh & \texttt{tri3micro} & \texttt{tri2micro} & \texttt{quad3micro} & \texttt{hybrid5micro} \\\hline
    Adaptive refinement & no & no & no & no \\
    Min grid size & \SI{3}{\micro\metre} & \SI{2}{\micro\metre} & \SI{2}{\micro\metre} & \SI{5}{\micro\metre} \\
    Max grid size & \SI{100}{\micro\metre} & \SI{100}{\micro\metre} & \SI{100}{\micro\metre} & \SI{100}{\micro\metre} \\
    $N_\mrm{cells}$ & $2.1 \times 10^6$ & $4.66 \times 10^6$ & $1.8 \times 10^6$ & $7.1 \times 10^5$ \\
    Time step & dyn. & dyn. & dyn. & dyn. \\
    CPU cores & 36 & 36 & 36 & 36 \\
    Run time & 370 s & 1337 s & 1280 s & 220 s \\
    Core time & 3.7 h & 13.4 h & 12.8 h & 2.2 h \\\hline
    \end{tabular}
    \caption{A summary of simulation settings for case 1 using AVIP.}
    \label{tab:bagheri_case1_avip}
\end{table}

\subsubsection{Case 2}

The second test case is the stiffest one: the background density is lowered by four orders of magnitude compared to the first case to \SI{1e9}{\per\metre\cubed}. This case is rather artificial since photoionization would not be negligible at these densities but it evaluates the robustness of the schemes implemented.

\begin{figure}[htbp]
    \centering
    \begin{subfigure}{0.8\textwidth}
        \centering
        \includegraphics[width=\textwidth]{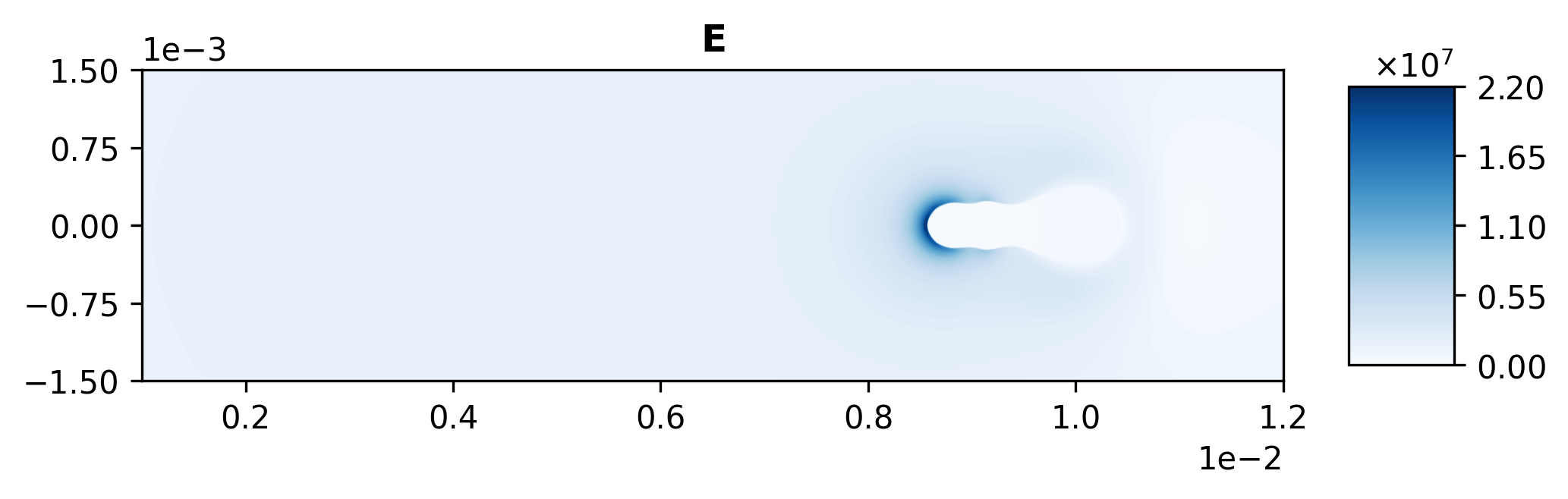}
        \caption{5 ns}
    \end{subfigure}
    \begin{subfigure}{0.8\textwidth}
        \centering
        \includegraphics[width=\textwidth]{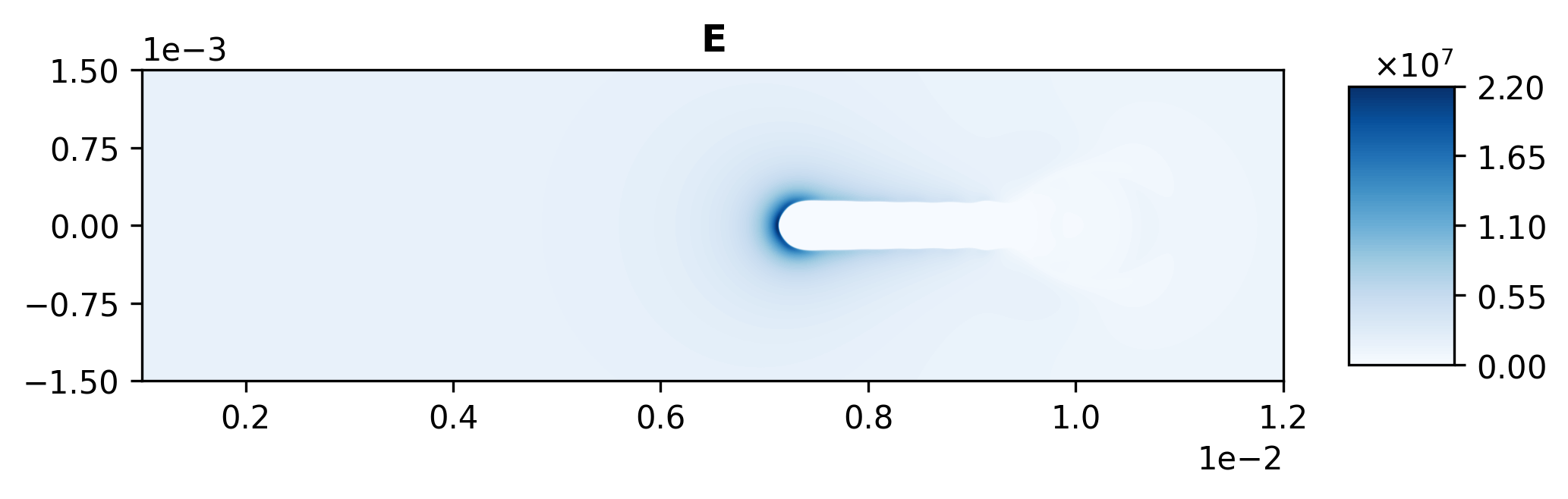}
        \caption{10 ns}
    \end{subfigure}
    \begin{subfigure}{0.8\textwidth}
        \centering
        \includegraphics[width=\textwidth]{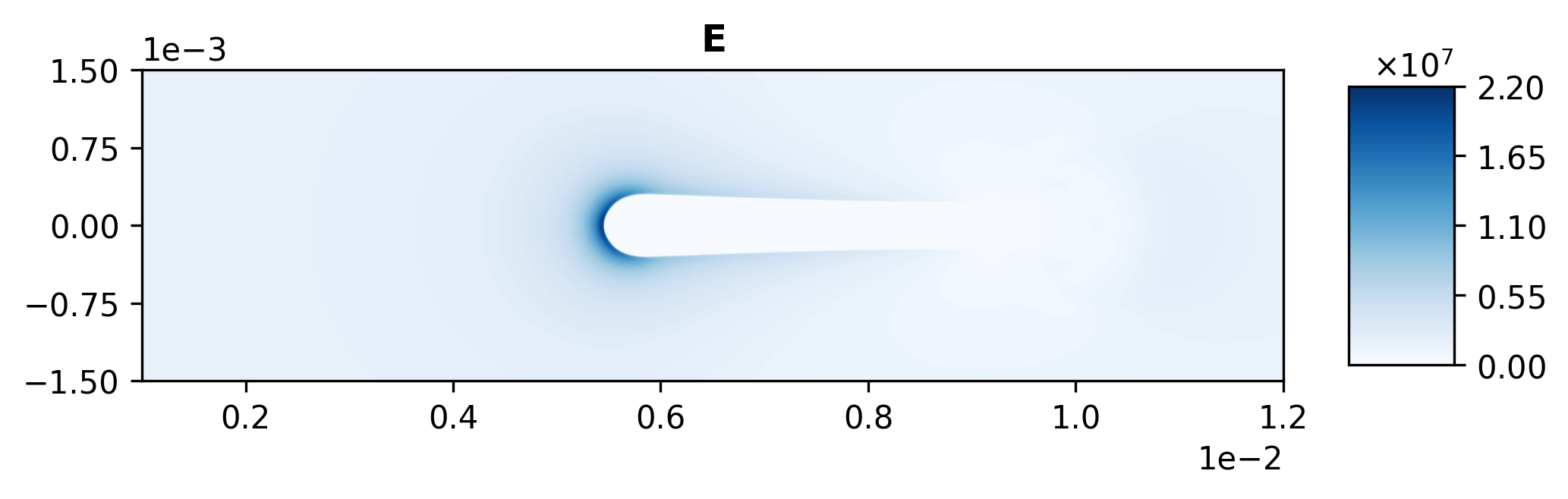}
        \caption{15 ns}
    \end{subfigure}
    \begin{subfigure}{0.8\textwidth}
        \centering
        \includegraphics[width=\textwidth]{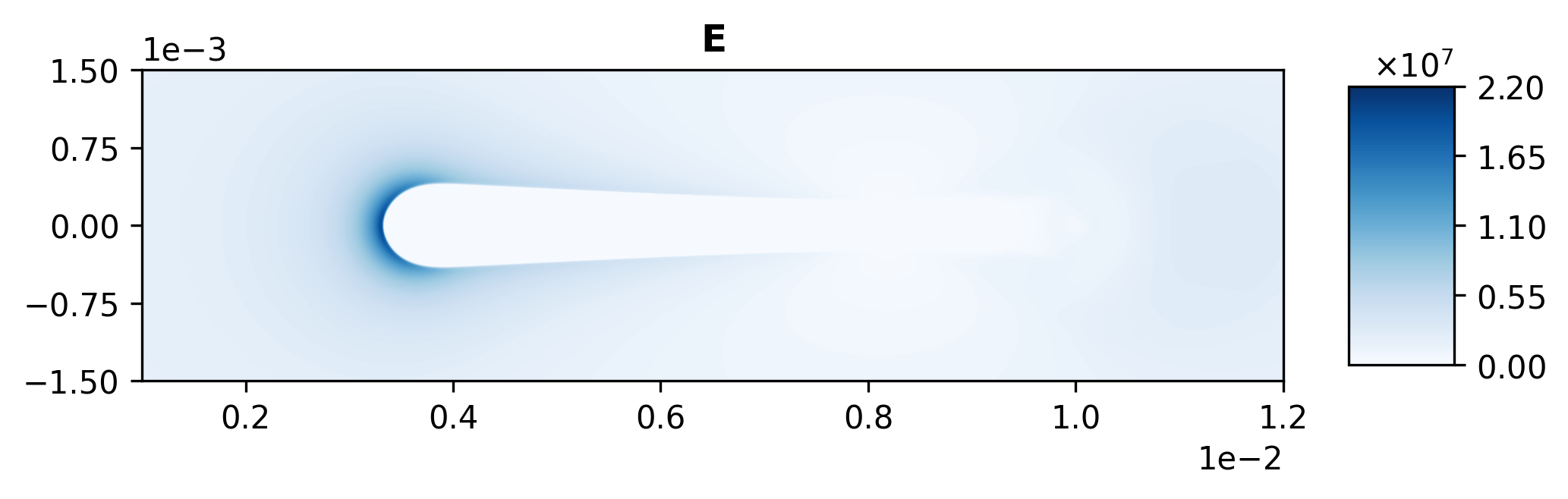}
        \caption{20 ns}
    \end{subfigure}
    \caption{Electric field of the streamer at different instants for hybrid ISG-LLW at $\upD x = \SI{2.5}{\micro\meter}$.}
    \label{fig:bagheri_case2_isgllw25micro}
\end{figure}

The resolutions used in the previous case using any scheme are not able to propagate the streamer without branching and here only below a resolution of $\upD x = \SI{2.5}{\micro\metre}$ the streamer propagates correctly as shown in Fig.~\ref{fig:bagheri_case2_isgllw25micro} for the ISG-LLW scheme using the \texttt{hybrid25micro} mesh. Since the background density is lower there is less charged species to collide with and therefore the speed of the streamer is lower in this case by a factor of 2. The peak electric field of the streamer head is also around 30\% higher for this case: gradients of charge density are higher due to the lower background causing a higher potential drop and thus higher electric field magnitude.

\begin{figure}[htbp]
    \begin{subfigure}{0.48\textwidth}
        \centering
        \includegraphics[width=\textwidth]{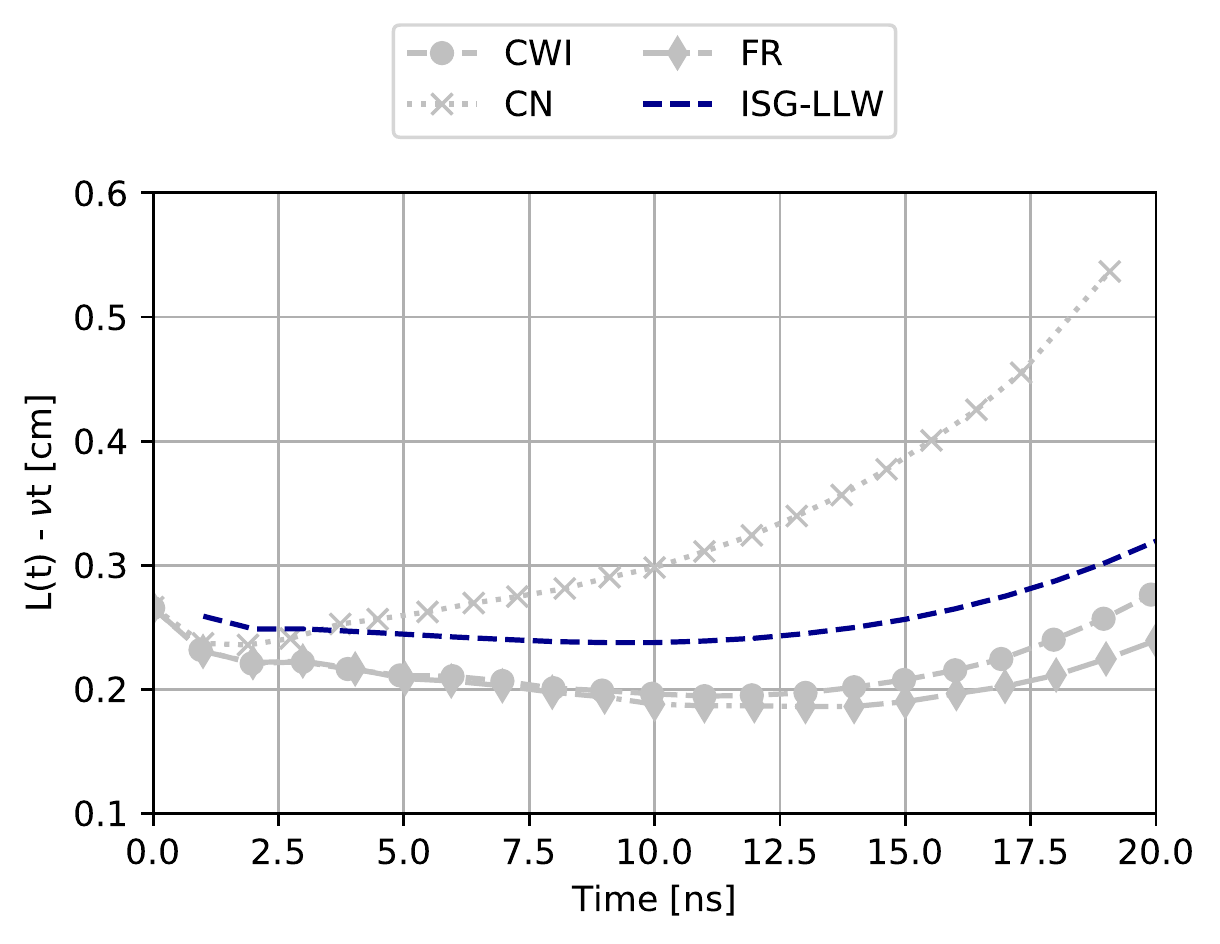}
    \end{subfigure}
    \begin{subfigure}{0.48\textwidth}
        \centering
        \includegraphics[width=\textwidth]{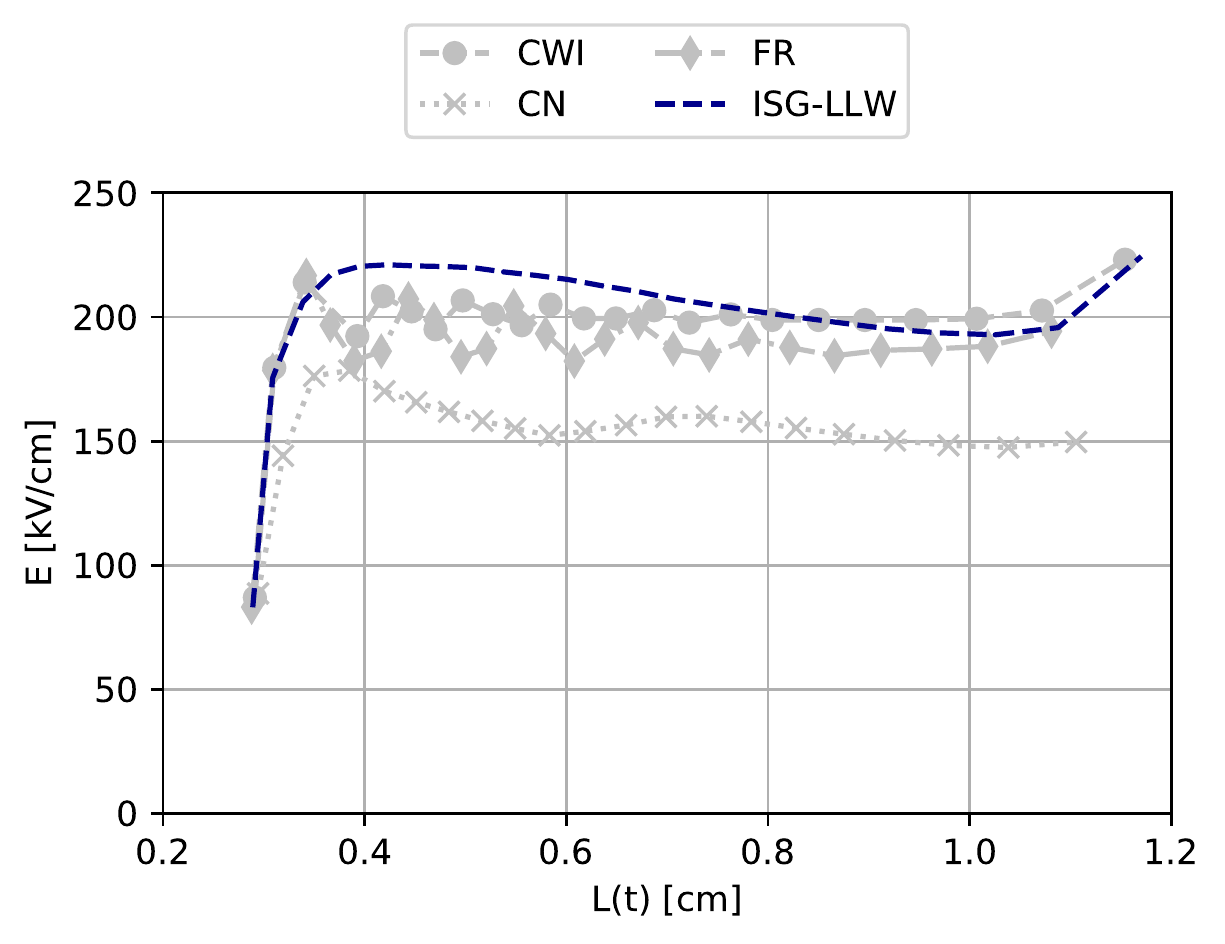}
    \end{subfigure}
    \caption{Streamer length as a function of time for the streamer codes and AVIP for case 2.}
    \label{fig:bagheri_case2_comp}
\end{figure}

The comparisons of the streamer length as a function of time and peak electric field as a function of the streamer length are shown in Fig.~\ref{fig:bagheri_case2_comp}. Only three out of the six streamer codes are able to correctly propagate the streamer in the benchmark paper \cite{bagheri_benchmark} for this case and only those results are shown. As for the first case, good agreement is found between AVIP and the other codes validating the hybrid approach for very stiff cases.

\subsubsection{Case 3}

\begin{figure}[htbp]
    \centering
    \begin{subfigure}{0.8\textwidth}
        \centering
        \includegraphics[width=\textwidth]{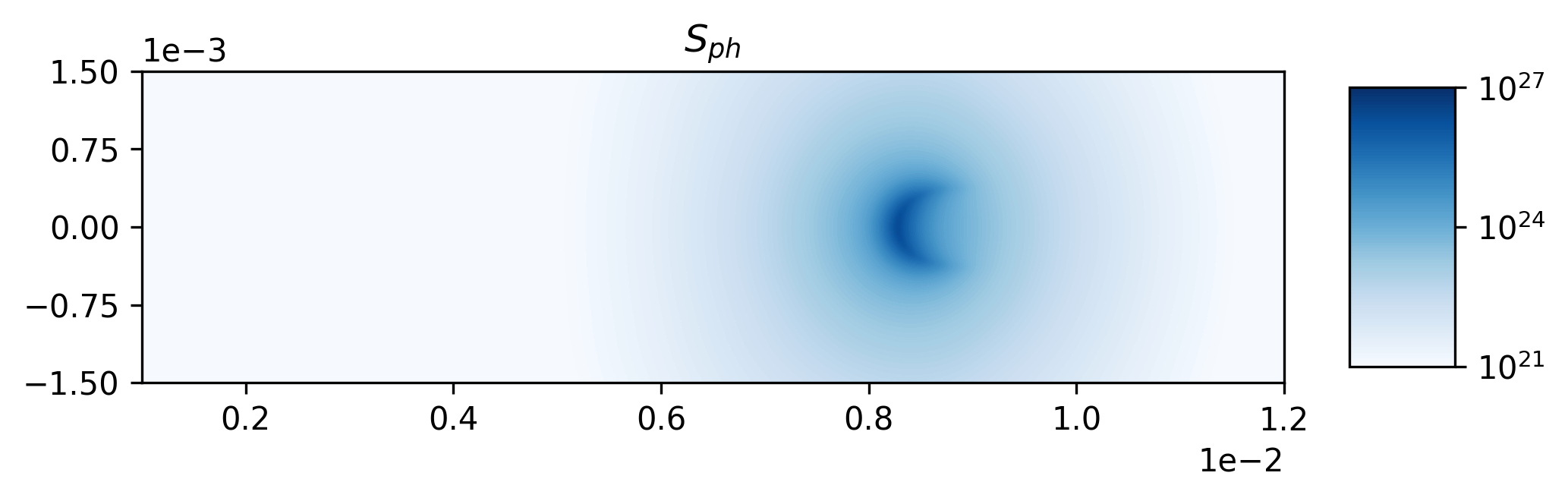}
        \caption{4 ns}
    \end{subfigure}
    \begin{subfigure}{0.8\textwidth}
        \centering
        \includegraphics[width=\textwidth]{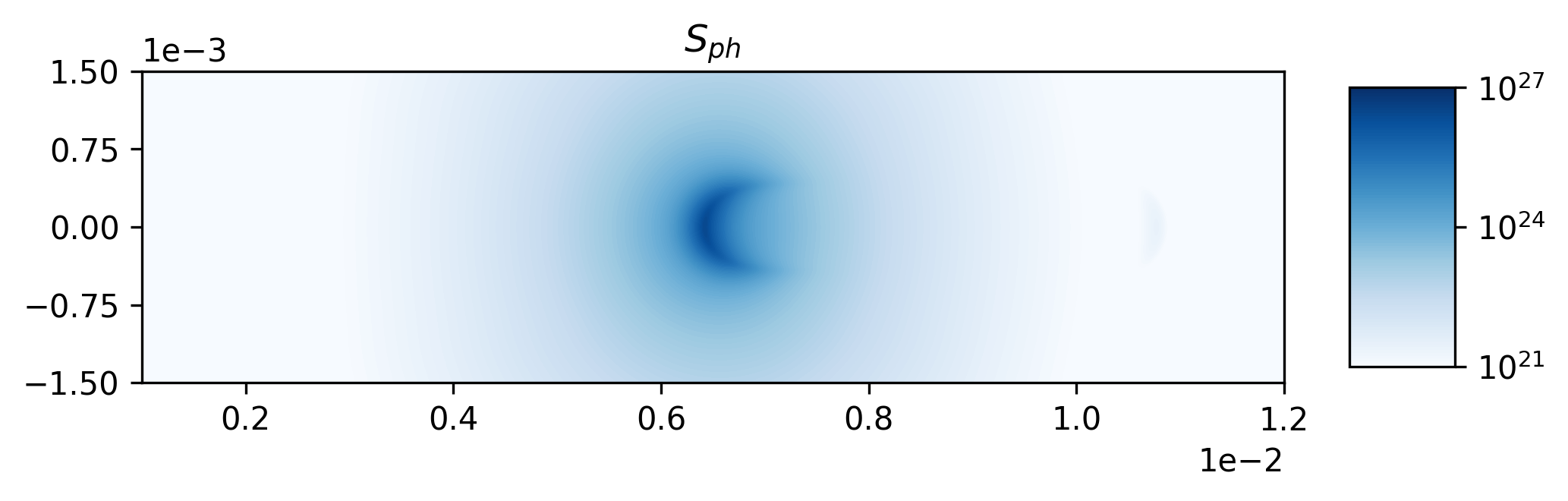}
        \caption{8 ns}
    \end{subfigure}
    \begin{subfigure}{0.8\textwidth}
        \centering
        \includegraphics[width=\textwidth]{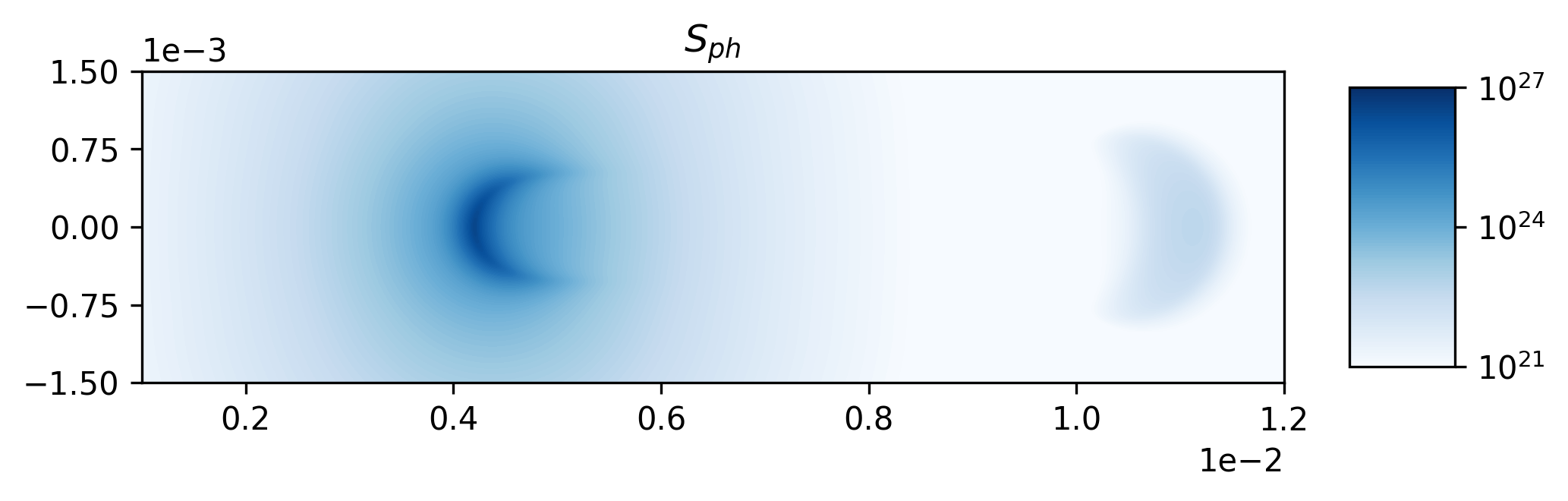}
        \caption{12 ns}
    \end{subfigure}
    \begin{subfigure}{0.8\textwidth}
        \centering
        \includegraphics[width=\textwidth]{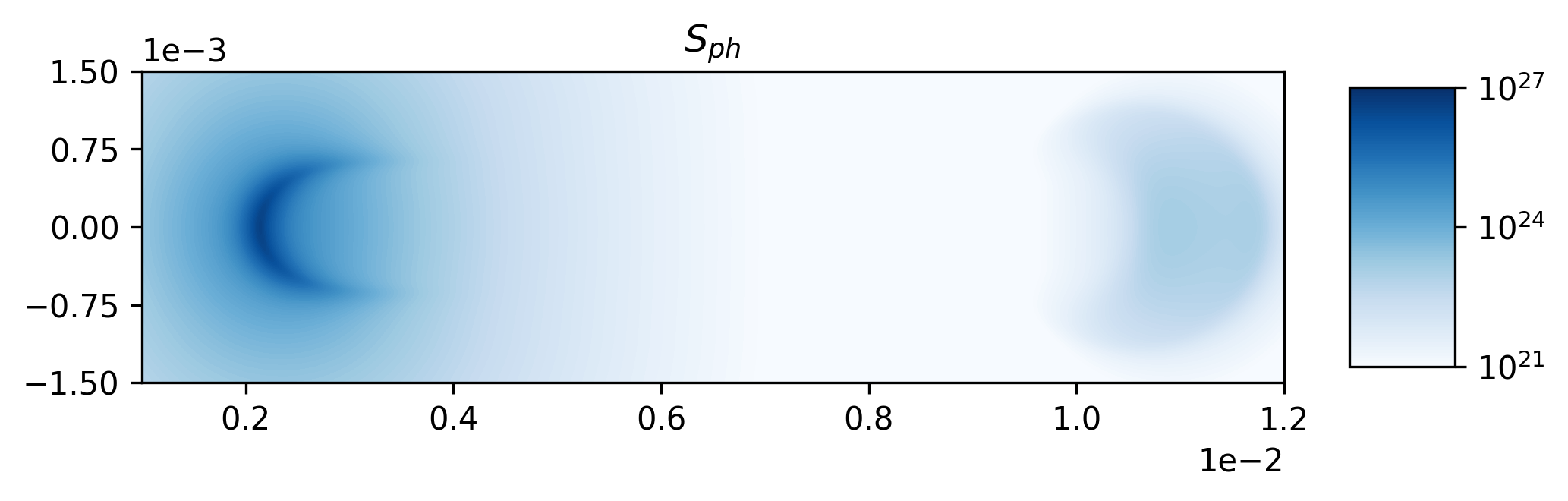}
        \caption{15 ns}
    \end{subfigure}
    \caption{Photoionization source term at different instants for hybrid ISG-LLW at $\upD x = \SI{5}{\micro\meter}$.}
    \label{fig:bagheri_case3_isgllw5micro}
\end{figure}

The third test case is the same as the second one but with photoionization activated. Photoionization in AVIP is implemented using the three-term exponential fitting from \cite{bourdon2007} which is one of the three models used in \cite{bagheri_benchmark}. This case is the less stiff among the three cases as this photoionization source term allows to stabilize the streamer: in front of the peak electric field the photoionization feeds the streamer head with charged species. The photoionization source term at different instants is shown in Fig.~\ref{fig:bagheri_case3_isgllw5micro}. We can see that it provides the streamer a diffusive stream of charged species which ionize the whole area around the streamer head.

\begin{figure}[htbp]
    \begin{subfigure}{0.48\textwidth}
        \centering
        \includegraphics[width=\textwidth]{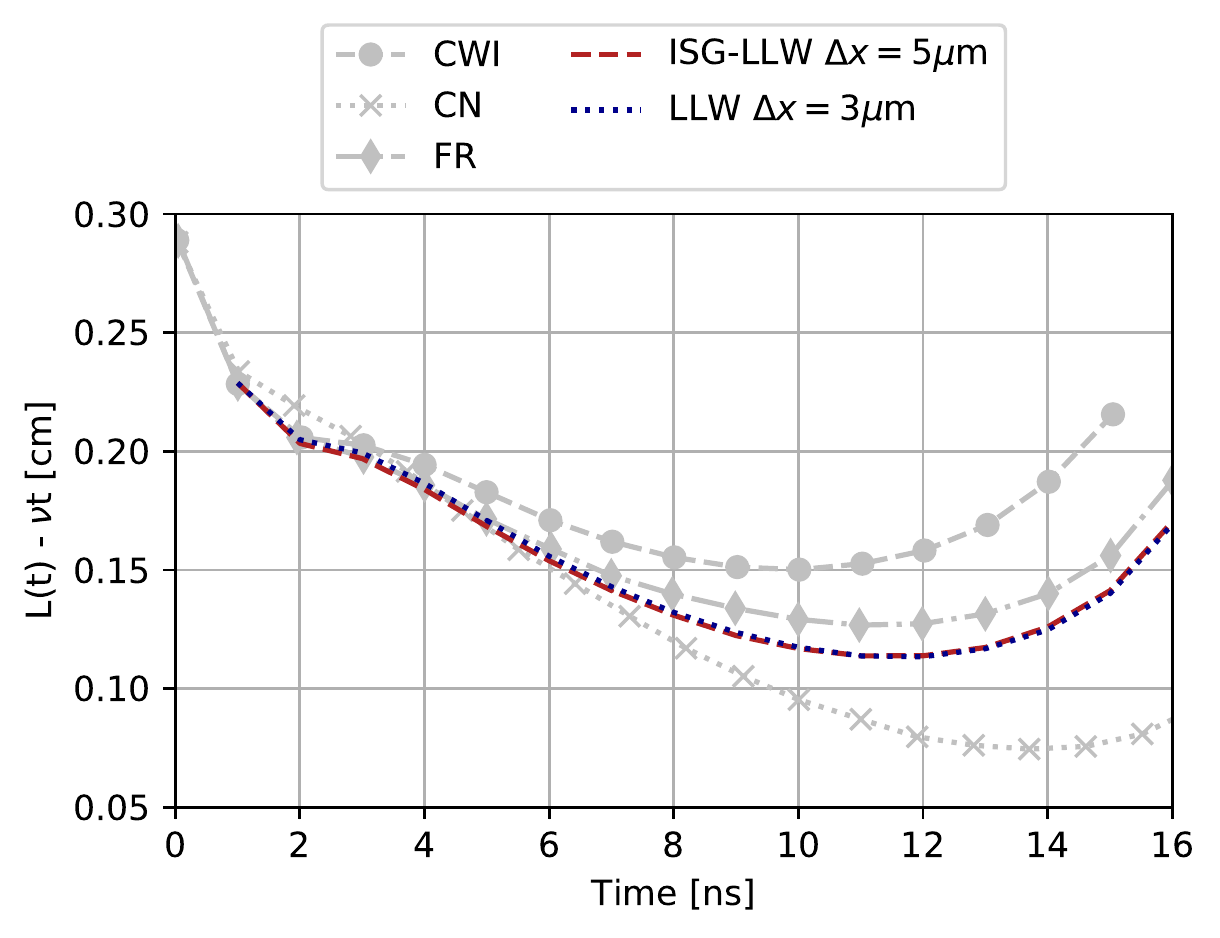}
    \end{subfigure}
    \begin{subfigure}{0.48\textwidth}
        \centering
        \includegraphics[width=\textwidth]{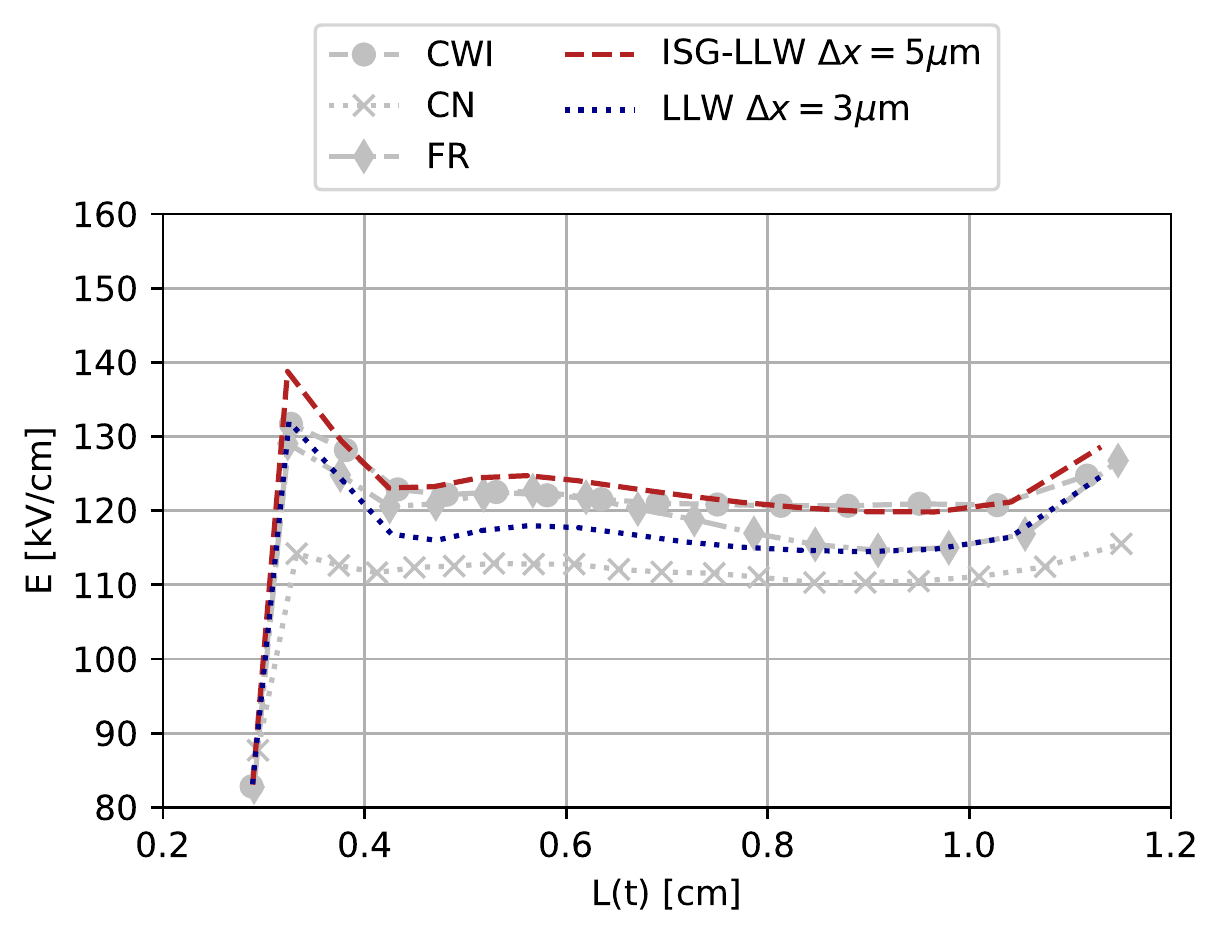}
    \end{subfigure}
    \caption{Streamer length as a function of time for all the streamer codes and AVIP.}
    \label{fig:bagheri_case3_comp}
\end{figure}

All the streamer codes presented in the benchmark are able to run this case with rather coarse resolutions compared to the previous cases. The comparison of AVIP with the three codes already present in Case 2 is shown in Fig.~\ref{fig:bagheri_case3_comp}. Good agreeement is found with LLW on triangular meshes and ISG-LLW on hybrid meshes validating the photoionization implementation in AVIP in both triangular and quadrangular meshes.

\subsection{Hyperbole electrodes}

We now turn to a more complex case with hyperbolic shape electrodes. The chemistry is taken from \cite{Morrow1997} to be representative of air. It involves three types of species: electrons $e$, positive $p$ and negative ions $n$ for which the governing equations read:

\begin{gather}
    \nabla^2 \phi = - \frac{e(n_p - n_e - n_n)}{\veps_0} \longrightarrow \vb{E} = - \nabla \phi \\
    \frac{\partial n_e}{\partial t} + \nabla \cdot \qty(n_e \mathbf{W_e} - D_e \nabla n_e) = n_e \alpha |W_e| - n_e \eta |W_e| - n_e n_p \beta \\
    \frac{\partial n_p}{\partial t} = n_e \alpha |W_e|-n_e n_p \beta-n_n n_p \beta\\
    \frac{\partial n_n}{\partial t} =  n_e \eta |W_e| - n_n n_p \beta
\end{gather}

\noindent where $\alpha = \alpha(E/N)$ is the ionization coefficient, $\eta = \eta(E/N)$ the attachment coefficient, $N$ the neutral gas density, $\beta$ the recombination rate, $\vb{W}_e = - \mu_e \vb{E}$ the drift-velocity of the electrons and $\mu_e = \mu_e(E/N)$ the electron mobility. The chemistry thus involves six reactions which are summarized below:

\begin{align}
    &\qq{Ionization} \ce{e- + A -> 2e- + A+} \\
    &\qq{Attachment} \ce{e- + A -> A-} \\
    &\qq{Photionization} \ce{e- + $\gamma$ -> e- + A+} \\
    &\qq{Recombination e-p} \ce{e- + A+ -> A} \\
    &\qq{Recombination n-p} \ce{A- + A+ -> 2A}
\end{align}

\begin{figure}[htbp]
    \centering
    \begin{subfigure}{0.34\textwidth}
        \centering
        \includegraphics[width=\textwidth]{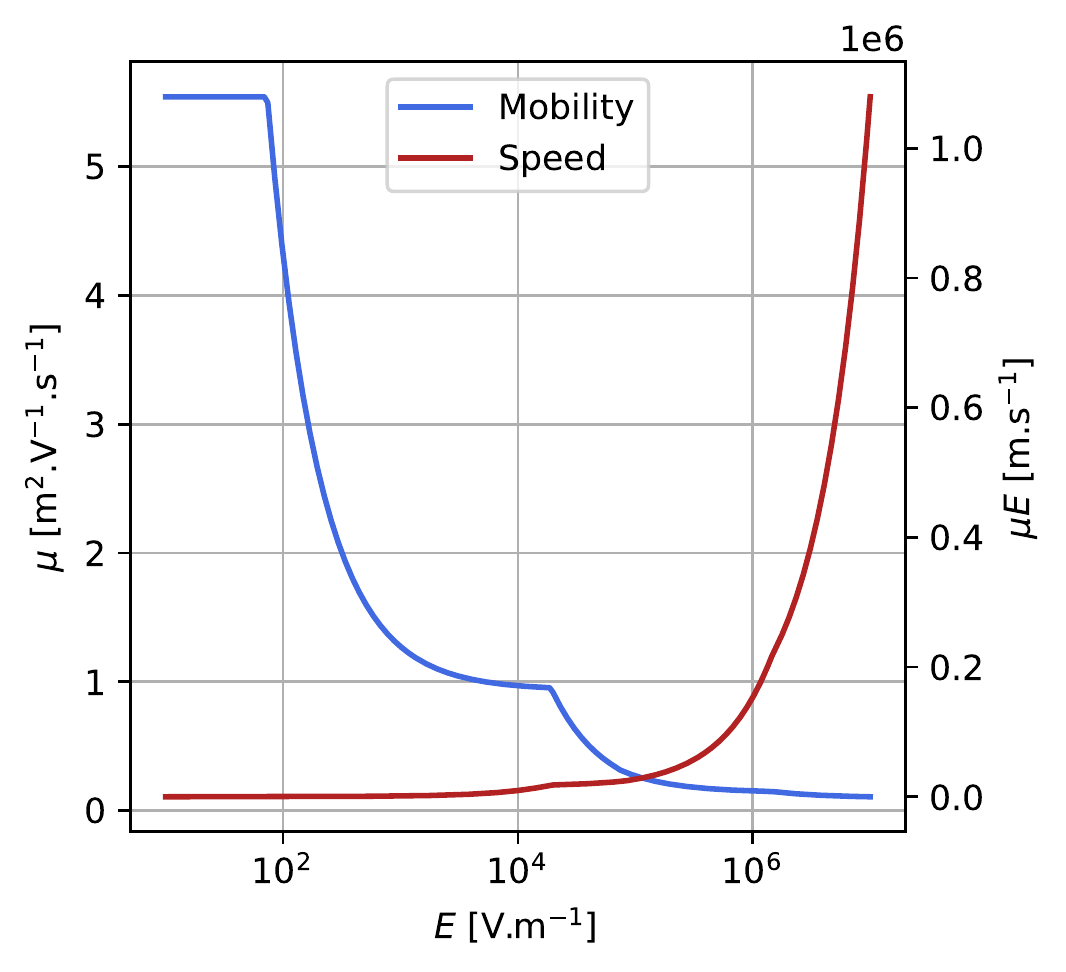}
        \caption{Mobility and speed}
    \end{subfigure}
    \begin{subfigure}{0.3\textwidth}
        \centering
        \includegraphics[width=\textwidth]{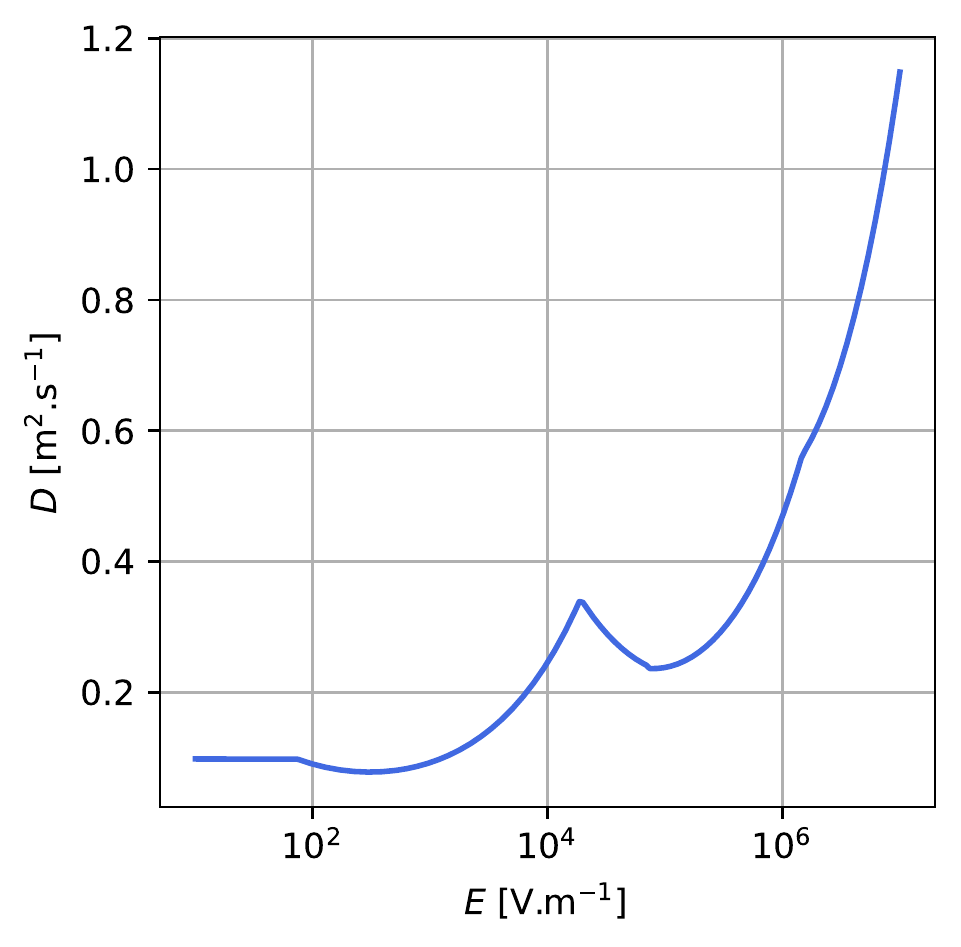}
        \caption{Diffusion coefficient}
    \end{subfigure}
    \begin{subfigure}{0.3\textwidth}
        \centering
        \includegraphics[width=\textwidth]{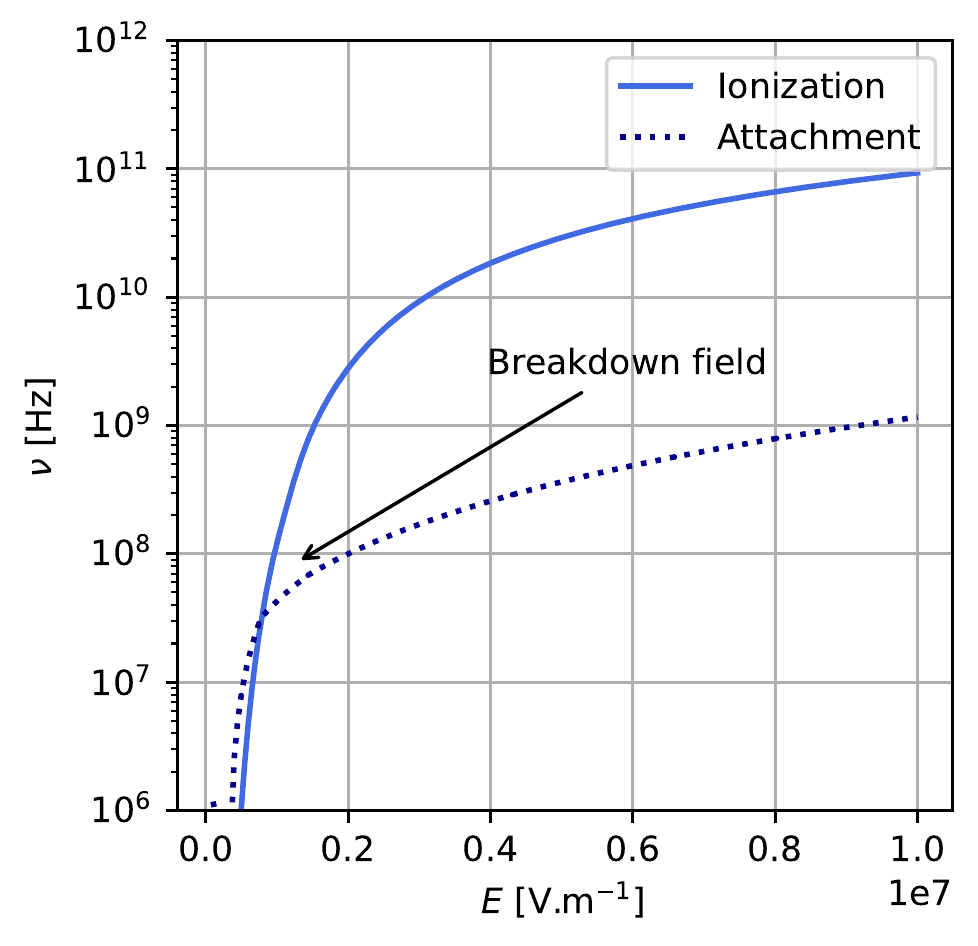}
        \caption{Effective ionization}
    \end{subfigure}
    \caption{Transport and chemistry coefficients as a function of the electric field norm for the \cite{Morrow1997} chemistry at 1000 K.}
    \label{fig:morrow_1000K}
\end{figure}

The local field approximation is assumed for this chemistry, \textit{i.e.} the transport coefficients depend on the reduced electric field $E/N$. The transport and chemistry coefficients are plotted at 1000 K and shown in Fig.~\ref{fig:morrow_1000K}. These coefficients essentially follow the same trends as the chemistry from the \cite{bagheri_benchmark} benchmark. At 1000 K the breakdown field is around \SI{8}{\mega\volt\per\metre}, 3 times lower than the value of \SI{22}{\mega\volt\per\metre} at 300 K in the benchmark cases. so that discharges propagate more easily at higher temperatures for a given voltage.

\begin{figure}[htbp]
    \centering
    \begin{tikzpicture}
  \draw[thick] (-1.5, 0) -- (1.5, 0);
    \draw[domain=0:1.64, smooth, variable=\t] plot ({1.5 * cosh(\t)}, {0.4 * sinh(\t)});
    \draw[domain=0:1.64, smooth, variable=\t] plot ({-1.5 * cosh(\t)}, {0.4 * sinh(\t)});
    \draw[domain=13.9:166, smooth, variable=\theta] plot ({4.12 * cos(\theta)}, {4.12 * sin(\theta)});

    \node[anchor=center, red_sketch] at (0, -1) {Neumann};
    \node[anchor=center, red_sketch] at (-3.4, -0.2) {Dirichlet $V_c = \SI{0}{\volt}$};
    \node[anchor=center, red_sketch] at (3, -0.2) {Dirichlet $V_a$};
    \node[anchor=center, red_sketch] at (0, 3.8) {Neumann};

    \node[anchor=center, blue_sketch] at (0, -0.5) {Symmetry};
    \node[anchor=center, blue_sketch] at (-3, 0.3) {Neumann};
    \node[anchor=center, blue_sketch] at (3, 0.3) {Neumann};
    \node[anchor=center, blue_sketch] at (0, 4.4) {Outlet};

    \draw[<->] (-1.5, 0.2) -- (1.5, 0.2);
    \node[anchor=center] at (0, 0.5) {5 mm};
  \end{tikzpicture}
    \caption{Hyperbolic shaped electrodes configuration with Poisson boundary conditions (red) and plasma species transport boundary conditions (blue).}
    \label{fig:pin-pin_circ_streamer}
\end{figure}
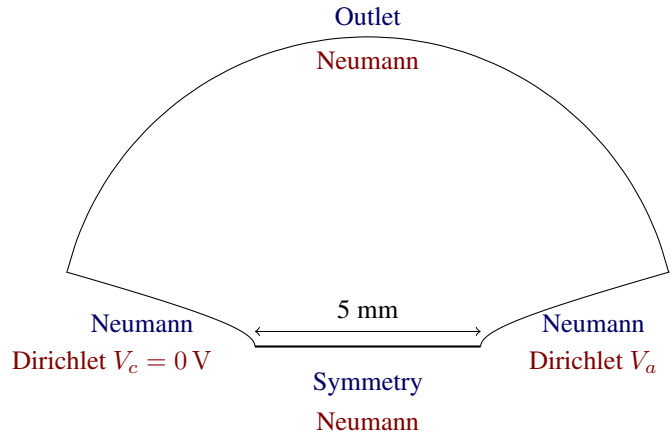

We choose a geometry shown in Fig.~\ref{fig:pin-pin_circ_streamer} with a 5 mm gap between the two hyperbole electrodes. To easily mesh the electrodes, triangular meshes with a smallest cell-size of about \SI{3}{\micro\metre} in the middle of the gap are used and prove to yield stable simulations.

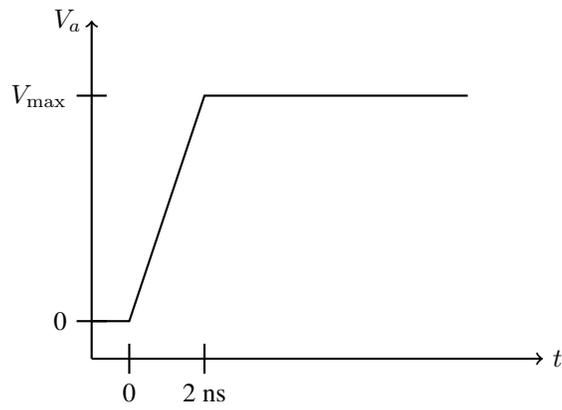
\begin{figure}[htbp]
    \centering
    \begin{tikzpicture}
    \draw[thick, ->] (-0.5, -0.5) -- (5.5, -0.5) node[anchor=west] {$t$};
    \draw[thick, ->] (-0.5, -0.5) -- (-0.5, 4) node[anchor=east] {$V_a$};

    \draw[thick] (-0.7, 0) node[anchor=east] {0}-- (-0.3, 0);
    \draw[thick] (-0.7, 3) node[anchor=east] {$V_\mathrm{max}$} -- (-0.3, 3);
    \draw[thick] (0, -0.7) node[anchor=north] {0} -- (0, -0.3);
    \draw[thick] (1, -0.7) node[anchor=north] {2 ns} -- (1, -0.3);
    \draw[thick] (-0.5, 0) -- (0, 0) -- (1, 3) -- (4.5, 3);

\end{tikzpicture}
    \caption{Anode potential time evolution.}
    \label{fig:pin-pin_anode_time}
\end{figure}

Dirichlet boundary conditions are applied at the electrodes, the left electrode is the grounded cathode $V_c = 0$ whereas the right electrode is the anode at potential $V_a$. A rise time of 2 ns from zero to the maximum value of the potential is set and the time evolution of the anode potential is shown in fig.~\ref{fig:pin-pin_anode_time}. The electric field is oriented from right to left and we expect two streamers to propagate across the gap. For the remaining Poisson BCs, Neumann conditions are enforced at the axis and in the farfield. For plasma transport species, Neumann boundary conditions are applied at the electrodes, a symmetry is applied at the axis whereas an outlet condition is imposed at the farfield.

For all simulations presented in this section, a constant neutral background of electrons and positive ions at \SI{1e15}{\per\metre\cubed} is set initially.

\begin{figure}[htbp]
    \centering
    \includegraphics[width=0.6\textwidth]{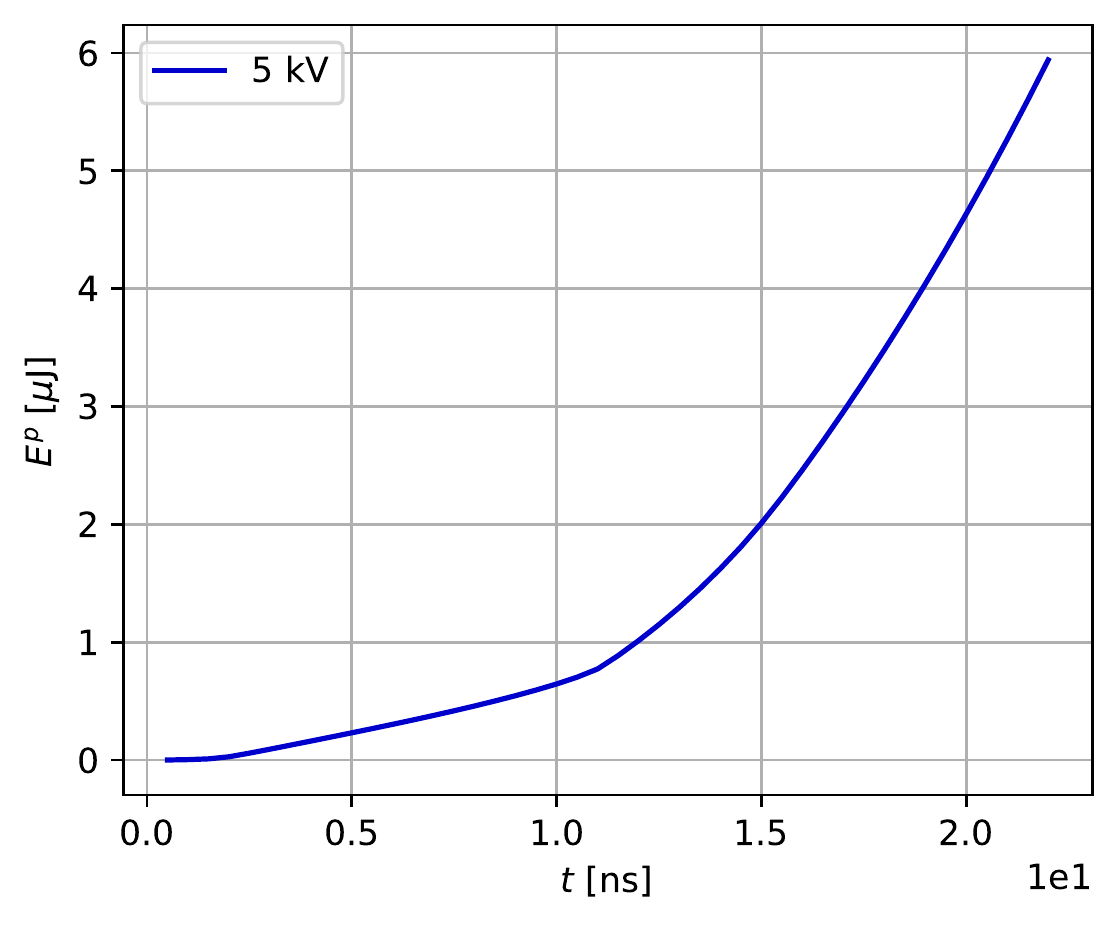}
    \caption{Discharge energy time evolution for the $R_c = \SI{200}{\micro\metre}$ at 1000 K and with peak anode potential of 5 kV.}
    \label{fig:discharge_energy_5kV}
\end{figure}

\begin{figure}[htbp]
    \centering
    \includegraphics[width=\textwidth]{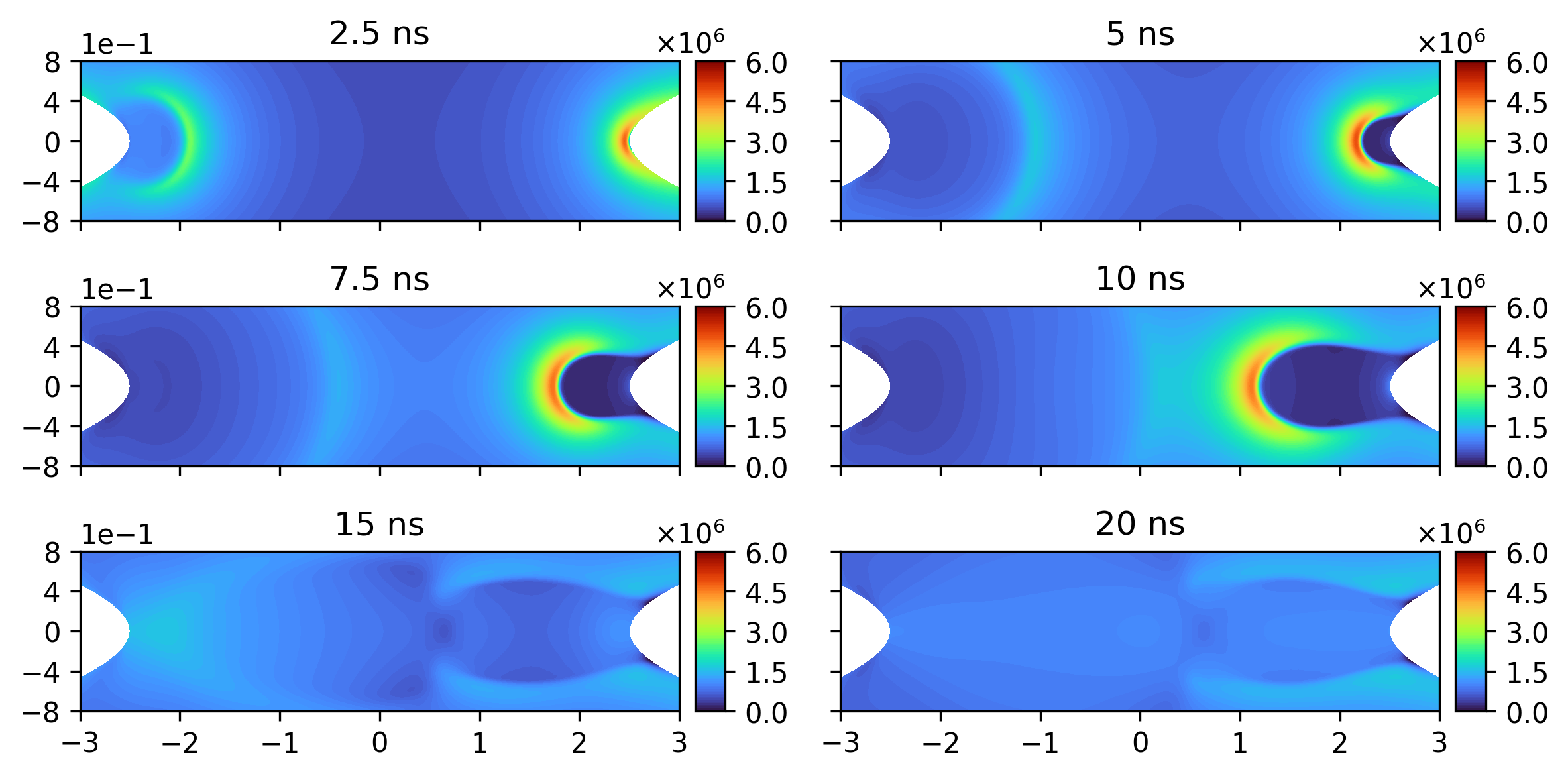}
    \caption{Electric field [V.m$^{-1}$] at different instants for two hyperbolic shape electrodes with $R_c = \SI{200}{\micro\metre}$ at 1000 K. Length units are in mm.}
    \label{fig:pin-pin-rc200_Efield}
\end{figure}

\begin{figure}[htbp]
    \centering
    \includegraphics[width=\textwidth]{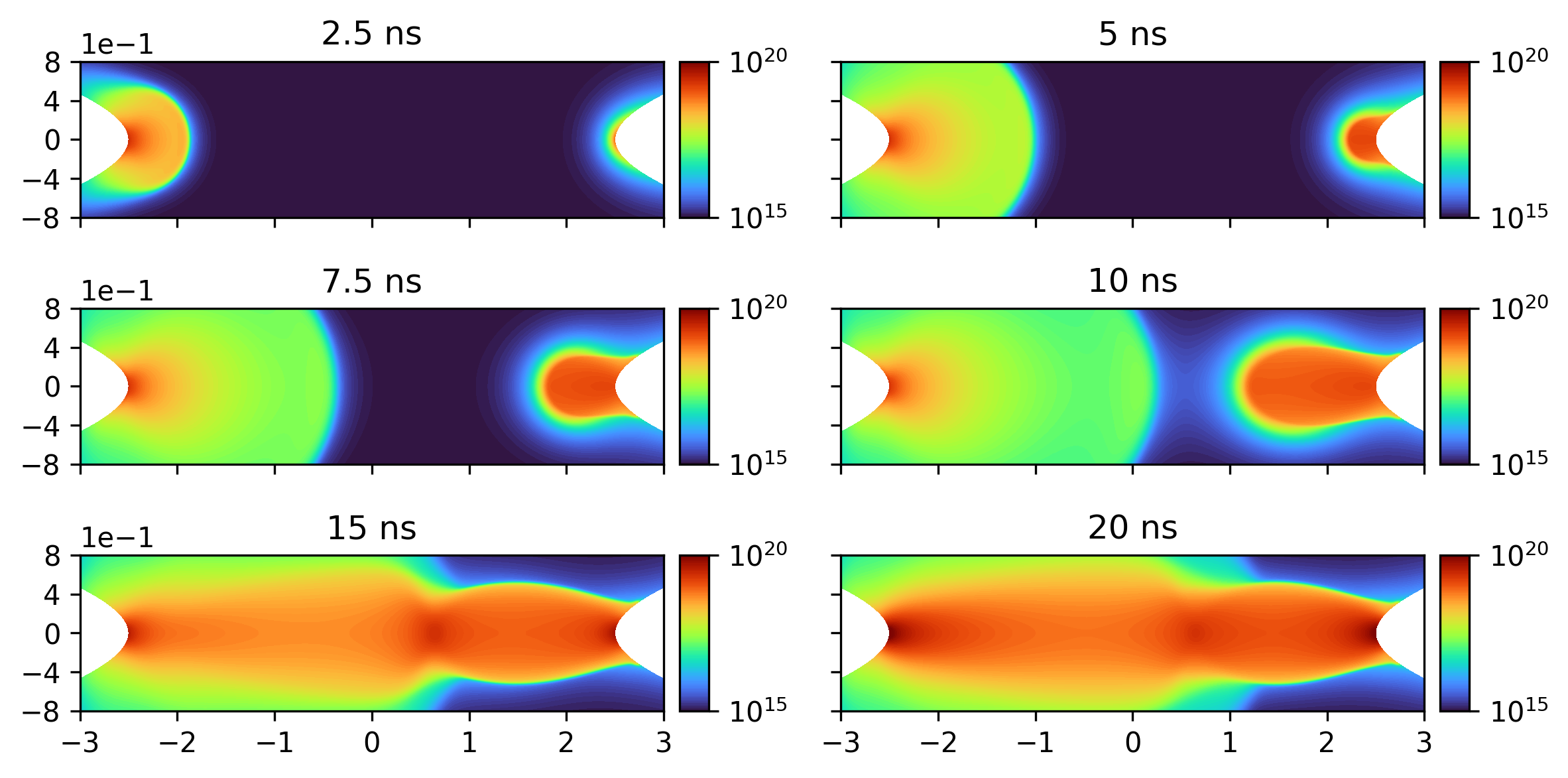}
    \caption{Electron density [m$^{-3}$] at different instants for two hyperbolic shape electrodes with $R_c = \SI{200}{\micro\metre}$ at 1000 K. Length units are in mm.}
    \label{fig:pin-pin-rc200_nElectron}
\end{figure}

\begin{figure}[htbp]
    \centering
    \includegraphics[width=\textwidth]{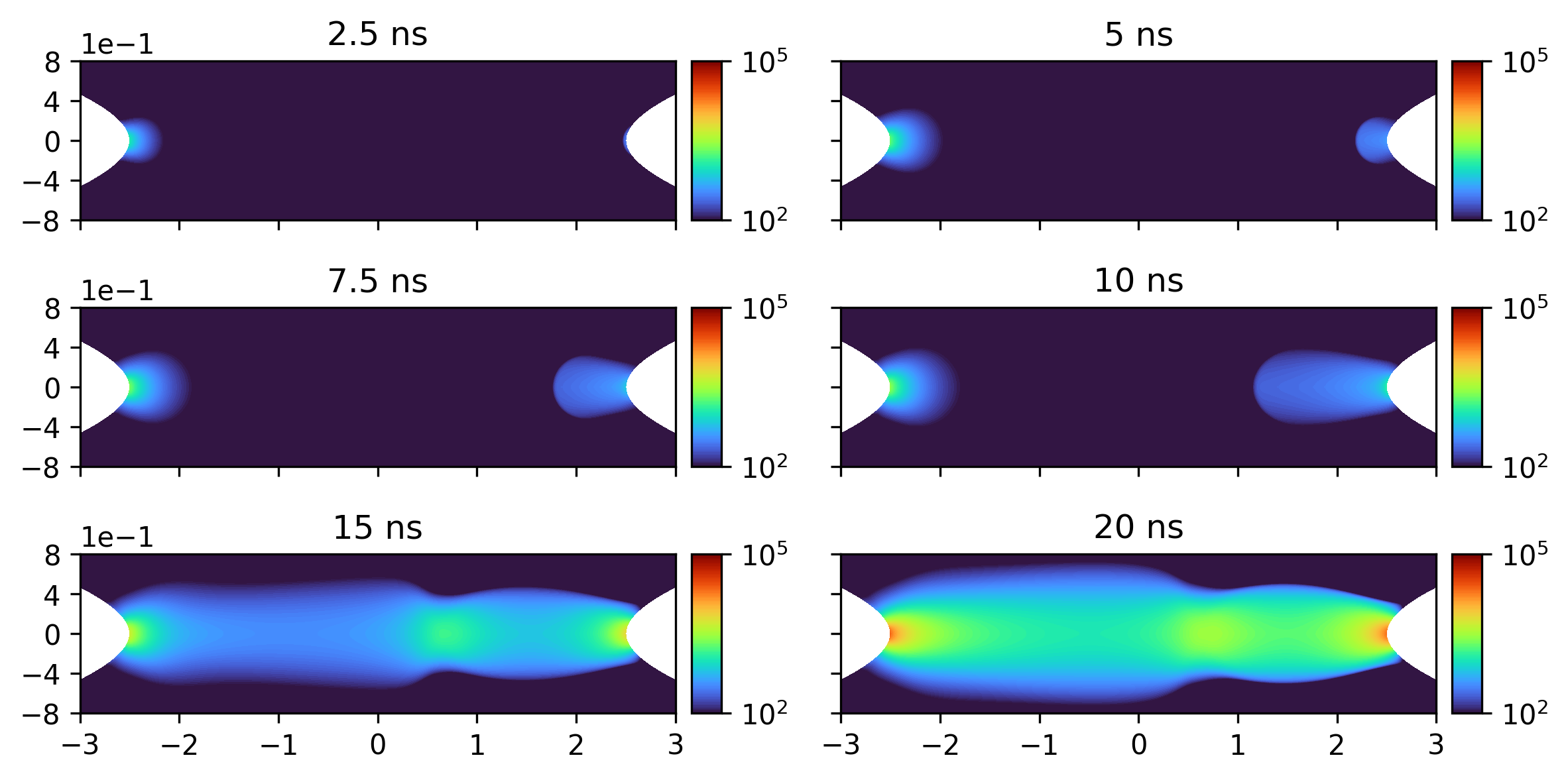}
    \caption{Discharge energy density [J.m$^{-3}$] at different instants for two hyperbolic shape electrodes with $R_c = \SI{200}{\micro\metre}$ at 1000 K. Length units are in mm.}
    \label{fig:pin-pin-rc200_dis_energy}
\end{figure}

Simulations at 1000 K using two hyperbole electrodes of radius $R_c = \SI{200}{\micro\metre}$ are shown in Figs.~\ref{fig:pin-pin-rc200_Efield}, \ref{fig:pin-pin-rc200_nElectron} and \ref{fig:pin-pin-rc200_dis_energy} for the norm of electric field, electron density and discharge energy density, respectively. The mesh contains $6.3 \times 10^5$ nodes and $1.3 \times 10^6$ triangular cells. Three phases are observed: a first phase where the streamer heads are created due to the voltage rise. Once the electric field at the heads (especially the positive streamer around the anode) is high enough, the streamers can propagate inside the gap. Both streamer eventually merge to create a conducting channel. Only after the creation of this conducting channel does the discharge energy density in Fig.~\ref{fig:pin-pin-rc200_dis_energy} significantly increases. We also note that the highest values of the discharge energy density are located close to the tips of the electrodes. The time evolution of the discharge energy is shown in Fig.~\ref{fig:discharge_energy_5kV} where two regimes can be observed: the first one is the slow increase of the discharge energy up to 12 ns. This time corresponds to the creation of the conducting channel and after 12 ns the discharge energy rises much faster.

\begin{figure}
    \centering
    \begin{tikzpicture}
    \draw[domain=-1.64:1.64, smooth, variable=\t] plot ({2 * cosh(\t)}, {0.6 * sinh(\t)});
    \draw[domain=-1.64:1.64, smooth, variable=\t] plot ({-2 * cosh(\t)}, {0.6 * sinh(\t)});

    \draw[domain=-1.64:1.64, smooth, variable=\t, light-gray] plot ({1.3 * cosh(\t)}, {1. * sinh(\t)});
    \draw[domain=-1.64:1.64, smooth, variable=\t, light-gray] plot ({-1.3 * cosh(\t)}, {1. * sinh(\t)});

    \draw[domain=-1:1, smooth, variable=\t, light-gray] plot ({0.6 * cosh(\t)}, {2 * sinh(\t)});
    \draw[domain=-1:1, smooth, variable=\t, light-gray] plot ({-0.6 * cosh(\t)}, {2 * sinh(\t)});

  \node[anchor=center] at (-4.2, 0) {Cathode $V_c = 0$};
  \node[anchor=center] at (4, 0) {Anode $V_a > 0$};

  \draw[<-, blue_sketch] (-1, 0) -- (-1.8, 0);
  \draw[<-, blue_sketch] (1.8, 0) -- (1, 0);
  \draw[<-, blue_sketch] (0.4, 0) -- (-0.4, 0);

  \draw[<-, blue_sketch] (-1, 1) -- (-1.8, 0.5);
  \draw[<-, blue_sketch] (1.8, 0.5) -- (1, 1);

  \draw[<-, blue_sketch] (-1, -1) -- (-1.8, -0.5);
  \draw[<-, blue_sketch] (1.8, -0.5) -- (1, -1);

  \node[anchor=center, blue_sketch] at (0, 0.5) {$-\mu_e \vb{E}$};

  \node[anchor=center, light-gray] at (0, 2.5) {$\phi = \mrm{cst}$};

  \end{tikzpicture}
    \caption{Sketch of streamer dynamics in pin-pin configurations.}
    \label{fig:pin-pin_dynamics}
\end{figure}
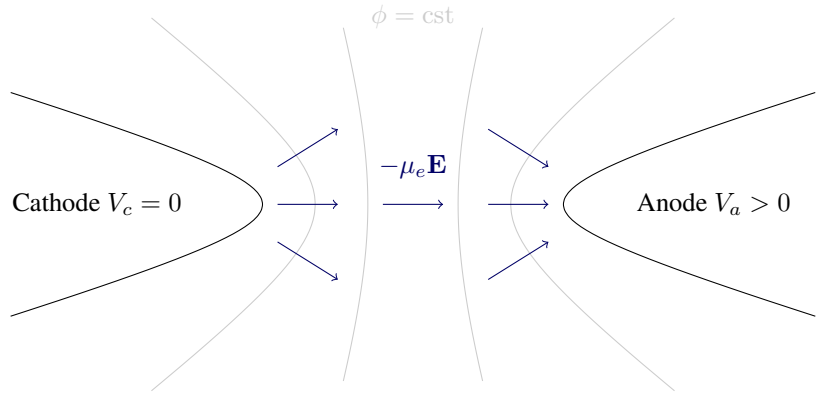

The streamer coming from the anode, the positive streamer, is much more concentrated than the negative streamer coming from the anode. This can be explained by the sketch shown in Fig.~\ref{fig:pin-pin_dynamics}. The electric field follows the isopotentials shown in gray in the sketch which are hyperboles as well. It can be seen that locally the electric field and hence the electron drift-velocity $-\mu_E \vb{E}$ have focusing directions around the anode and diffusive directions around the cathode.

\begin{figure}[htbp]
    \begin{subfigure}{0.48\textwidth}
        \centering
        \includegraphics[width=\textwidth]{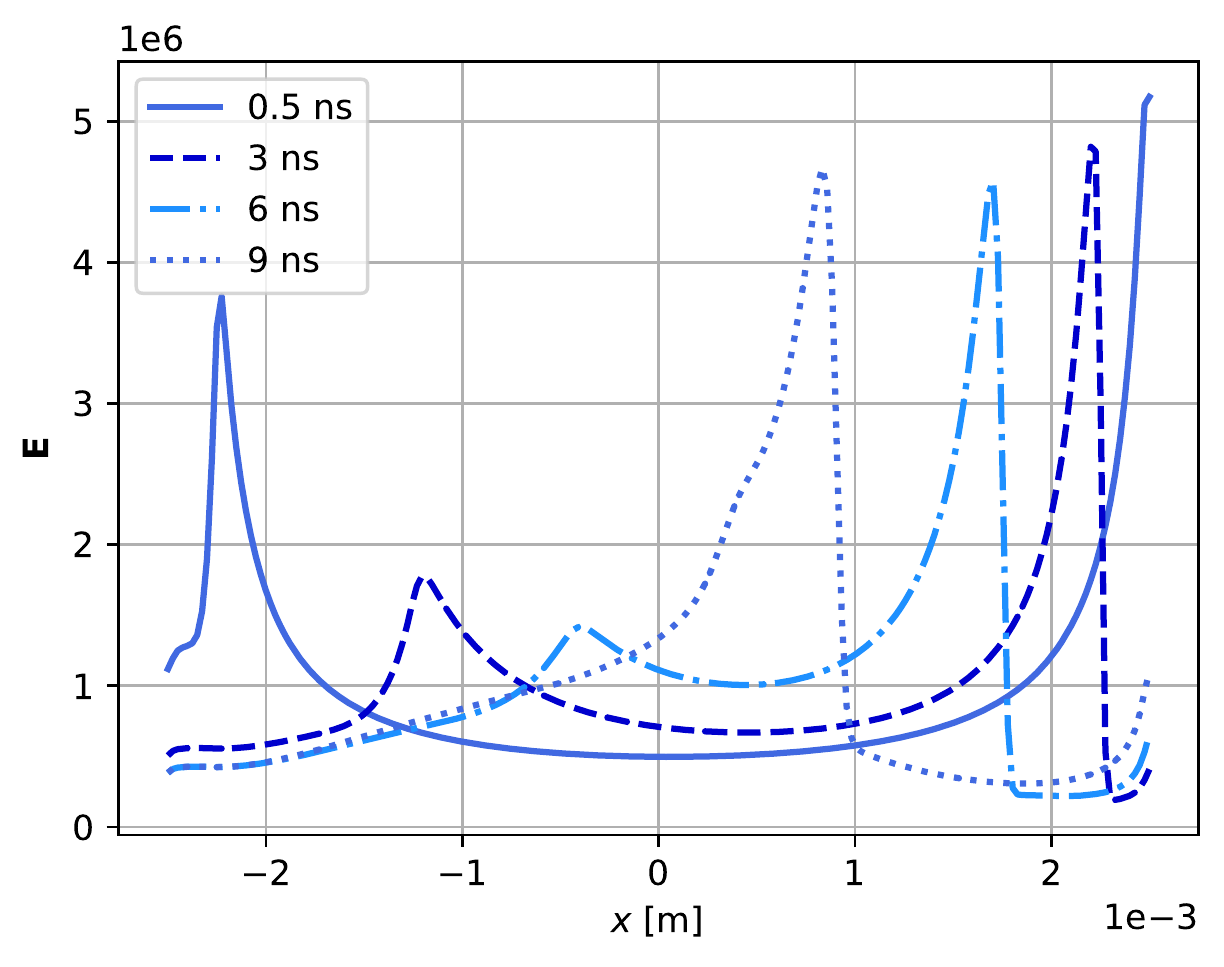}
    \end{subfigure}
    \begin{subfigure}{0.48\textwidth}
        \centering
        \includegraphics[width=\textwidth]{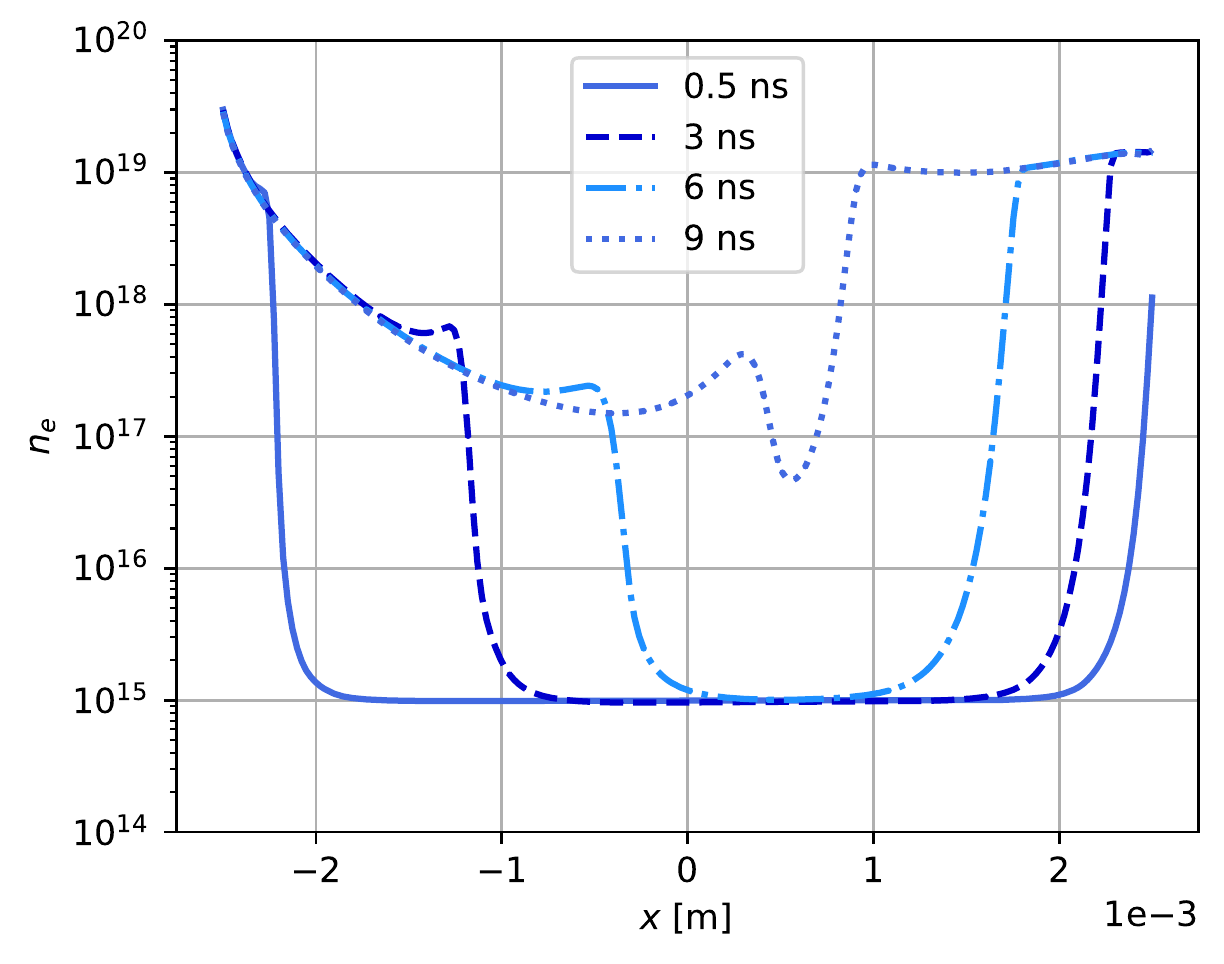}
    \end{subfigure}
    \caption{The norm of electric field (left) and the electron density (right) at different instants on the axis $r = 0$ for the LLW scheme with $\beta = 1.0$.}
    \label{fig:rc200_1000K_instants}
\end{figure}

One-dimensional profiles of electron density and electric field norm are shown in Fig.~\ref{fig:rc200_1000K_instants}. At $t=0.5$ ns the voltage is still rising and no propagation can be seen. At $t=2.5$ ns the onset of the negative streamer, with a local maximum electric field, is observed. This negative streamer has already moved around 0.5 mm while the positive streamer is still stuck aroudn the anode. We can note that the peak electric field is two times higher for the positive streamer than for the negative streamer at this time. At $t$ = 7.5 ns the negative streamer is at the middle of the gap and the peak electric field has considerably decreased while the positive streamer moved about half the distance and maintained its peak electric field. However, once the positive streamer is started it bridges the gap faster and overall streamer propagation speed is on the order of cm/ns approximately in these cases.

\section{Euler and Navier-Stokes validation cases}
\label{sec:ns_validation}

The validation of the HLLC-MUSCL scheme is carried out. The order of the schemes and more cases are shown in \cite{Cheng2022} and emphasis on the robustness of the scheme is made here. To do so shocks are first simulated in one-dimension with the canonical shock tube and two-dimensions with an explosion and a 2D Riemann problem. Finally the scheme is tested on a 1D flame as its final aim is to be used in plasma assisted combustion simulations.

\subsection{One-dimensional shock tube}

\label{subsec:1d_shock_cases}

The first shock presented in \cite[Chap. 10]{toro} is used to illustrate the TVD property of the Riemann solvers and more shocks cases can be found in \cite{Cheng2022}. The initial condition and solution of the shock are presented in Table \ref{tab:shock_1d}.

\begin{table}[htbp]
    \centering
    \begin{tabular}{| c | c | c | c | c |}\hline
        1   &  $\rho$  &  $u$    &  $p$    &  $a$   \\
       \hline $W_L$:    &  1.000   &  0.750  &  1.000  &  1.183 \\
        $W_L^*$:  &  0.580   &  1.361  &  0.466  &  1.061 \\
        $W_R^*$:  &  0.340   &  1.361  &  0.466  &  1.386 \\
        $W_R$:    &  0.125   &  0.000  &  0.100  &  1.058 \\
       \hline
    \end{tabular}
\caption{Summary of the shock test case.}
\label{tab:shock_1d}
\end{table}

This case is a modified version of the famous Sod problem \cite[Chap. 10.8]{toro} where there is a left rarefaction wave, a right travelling contact wave and a right shock. Both triangular and quadrangular elements types shown in Fig.~\ref{fig:elements_1d} are used and results are similar between them.

\begin{figure}[htbp]
    \centering
    \includegraphics[width=0.6\textwidth]{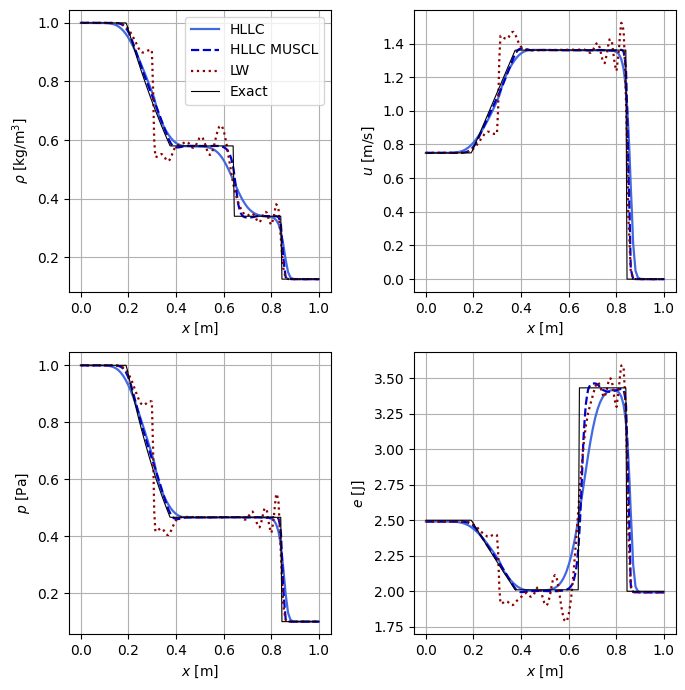}
    \caption{Case 1 with the three schemes for triangular mesh.}
    \label{fig:shock1_tri}
\end{figure}

Results for the most robust scheme used in AVBP \cite{lamarquethesis}, LW, along with HLLC and HLLC-MUSCL are shown in Fig.~\ref{fig:shock1_tri} for quadrangular elements. The LW scheme produces strong oscillations whereas HLLC and HLLC MUSCL yield stable simulations that correctly capture the shock and rarefaction wave. Moreover the MUSCL reconstruction proves to be effective as a higher accuracy is clearly observed.

\subsection{Multidimensional shocks}

Multidimensional shocks are now studied in this part where two generalizations of the one-dimensional shock tube are studied: cylindrical and spherical explosions \cite[Chap. 17]{toro} as well as two-dimensional Riemann problems \cite{deng2019}.

\subsubsection{Explosions}

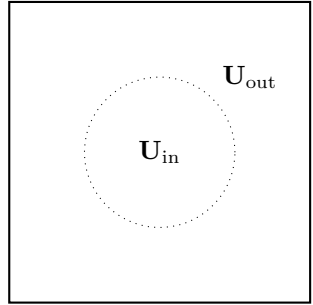
\begin{figure}[htbp]
    \centering
    \begin{tikzpicture}

    \draw[thick] (0, 0) rectangle (4, 4);
    \draw[dotted] (2, 2) circle (1);

    \node[anchor=center] at (2, 2) {$\mathbf{U}_\mathrm{in}$};
    \node[anchor=center] at (3.2, 3) {$\mathbf{U}_\mathrm{out}$};

\end{tikzpicture}
    \caption{Two dimensional explosion problem.}
    \label{fig:explosion_2d_sketch}
\end{figure}

\begin{figure}[htbp]
    \centering
    \begin{subfigure}[b]{0.3\textwidth}
        \centering
        \includegraphics[width=\textwidth]{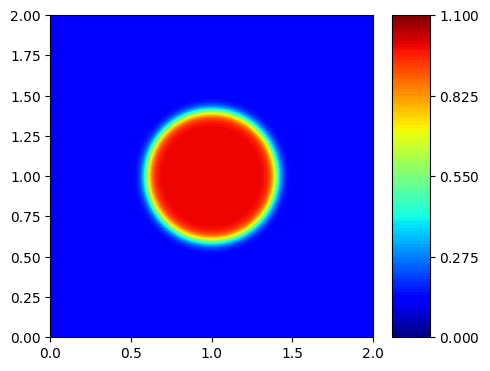}
        \caption{$t=0$}
        \label{fig:expl2d_instant1}
    \end{subfigure}
    \centering
    \begin{subfigure}[b]{0.3\textwidth}
        \centering
        \includegraphics[width=\textwidth]{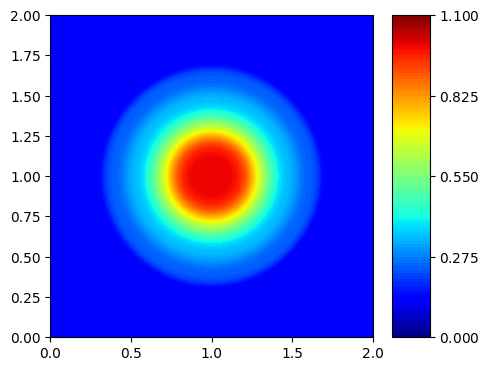}
        \caption{$t=0.15$}
        \label{fig:expl2d_instant2}
    \end{subfigure}
    \begin{subfigure}[b]{0.3\textwidth}
        \centering
        \includegraphics[width=\textwidth]{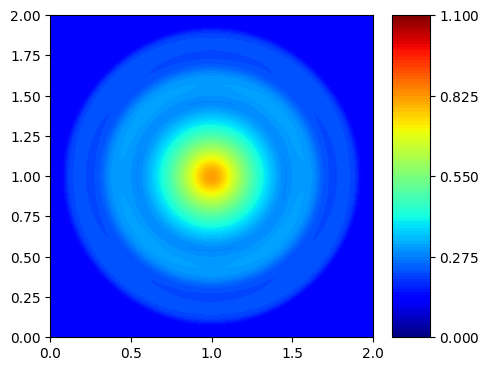}
        \caption{$t=0.3$}
        \label{fig:expl2d_instant3}
    \end{subfigure}
    \caption{Two-dimensional explosion at different instants.}
    \label{fig:expl2d_instants}
\end{figure}

The one-dimensional shock tube is extended as a circle in two dimensions and a sphere in three dimensions. An interior $\vU_\mrm{in}$ and exterior $\vU_\mrm{out}$ states are defined in this case as shown in Fig.~\ref{fig:explosion_2d_sketch}. A smoothing function is applied so that there is no staircase profile along the interface. The geometry and initial conditions of \cite[Chap. 17.1]{toro} are taken so that we simulate the two states

\begin{equation}
    \vU_\mrm{in} = \begin{bmatrix}
        1 \\ 0 \\ 0 \\ 1
    \end{bmatrix}
    \quad
    \vU_\mrm{out} = \begin{bmatrix}
        0.125 \\ 0 \\ 0 \\ 0.1
    \end{bmatrix}
\end{equation}

\noindent in a $[0, 2] \times [0, 2]$ square with $101 \times 101$ resolution. The results are shown at different instants in Fig.~\ref{fig:expl2d_instants} for a CFL of 0.5. An outgoing shock and inward rarefaction waves are present and well captured by the HLLC MUSCL RK2 scheme as shown in Fig.~\ref{fig:expl2d_instants} for the density field.

\begin{figure}[htbp]
    \centering
    \includegraphics[width=0.8\textwidth]{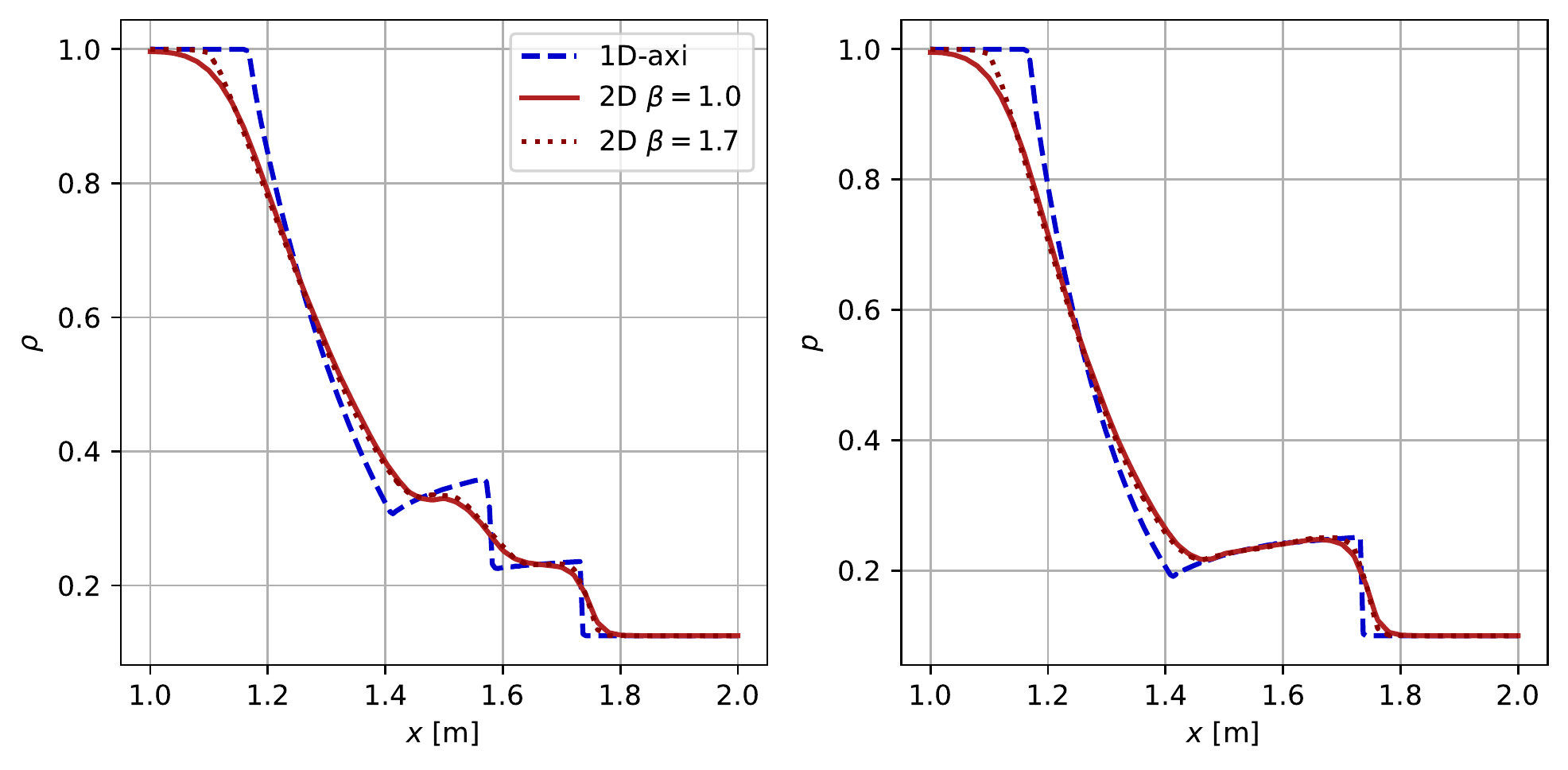}
    \caption{Comparison of 1D axisymmetric and 2D pressure and density profiles.}
    \label{fig:riemann_2d_vs_1daxi}
\end{figure}

This test case also allows to validate the pressure source term Eq.~\eqref{eq:ns_cyl_source_terms} in the axisymmetric formulation of the Euler equations as results should be consistent when comparing a 1D line in axisymmetric conditions with the 2D simulation. Comparisons of 2D simulations with HLLC-MUSCL RK2 at $\beta=1.0$ and $\beta=1.7$ with a 1D axisymmetric simulation with a much finer mesh of 1001 nodes along the $y$ direction are shown in Fig.~\ref{fig:riemann_2d_vs_1daxi}. Radial shocks have a tendency to create local minimas at the interfaces compared to the planar shock presented in Section.~\ref{subsec:1d_shock_cases}. The 1D-axisymmetric simulation and 2D simulations are consistent and yield the same results and differences are due to coarser resolution for the 2D test case. As done in \cite[Chap. 16]{toro}, the consistency of the solutions allows to consider the fine-mesh 1D-axisymmetric solution to be the exact one validating the HLLC-MUSCL scheme for multi-dimensional cylindrical shocks.

\subsubsection{2D Riemann problems}

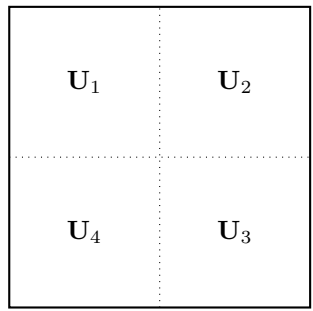
\begin{figure}[htbp]
    \centering
    \begin{tikzpicture}

    \draw[thick] (0, 0) rectangle (4, 4);
    \draw[dotted] (2, 0) -- (2, 4);
    \draw[dotted] (0, 2) -- (4, 2);

    \node[anchor=center] at (1, 3) {$\mathbf{U}_1$};
    \node[anchor=center] at (3, 3) {$\mathbf{U}_2$};
    \node[anchor=center] at (3, 1) {$\mathbf{U}_3$};
    \node[anchor=center] at (1, 1) {$\mathbf{U}_4$};

\end{tikzpicture}
    \caption{Two dimensional Riemann problem.}
    \label{fig:riemann_2d_sketch}
\end{figure}

\begin{figure}[htbp]
    \centering
    \begin{subfigure}[b]{0.45\textwidth}
        \centering
        \begin{tikzpicture}

    \draw[thick] (0, 0) rectangle (4, 4);
    \draw[dotted] (2, 0) -- (2, 4);
    \draw[dotted] (0, 2) -- (4, 2);

    \node[anchor=center] at (1, 3.5) {$\rho_1 = 2.0$};
    \node[anchor=center] at (3, 3.5) {$\rho_2 = 1.0$};
    \node[anchor=center] at (3, 1.5) {$\rho_3 = 3.0$};
    \node[anchor=center] at (1, 1.5) {$\rho_4 = 1.0$};

    \draw[thick, ->, blue_sketch] (0.5, 2.5) -- (1.5, 3.16);
    \node[anchor=center, blue_sketch] at (1.2, 2.5) {$\vb{u}_1$};
    \draw[thick, ->, blue_sketch] (2.5, 3.16) -- (3.5, 2.5);
    \node[anchor=center, blue_sketch] at (3.0, 2.5) {$\vb{u}_2$};
    \draw[thick, ->, blue_sketch] (3.5, 1.16) -- (2.5, 0.5);
    \node[anchor=center, blue_sketch] at (3.2, 0.5) {$\vb{u}_3$};
    \draw[thick, ->, blue_sketch] (1.5, 0.5) -- (0.5, 1.16);
    \node[anchor=center, blue_sketch] at (1.0, 0.5) {$\vb{u}_4$};

\end{tikzpicture}
        \caption{Case 1 2D-Riemann problem.}
        \label{fig:riemann_2d_sketch_case1}
    \end{subfigure}
    \centering
    \begin{subfigure}[b]{0.45\textwidth}
        \centering
        \begin{tikzpicture}

    \draw[thick] (0, 0) rectangle (4, 4);
    \draw[dotted] (2, 0) -- (2, 4);
    \draw[dotted] (0, 2) -- (4, 2);

    \node[anchor=center] at (1, 3.5) {$\rho_1 = 2.0$};
    \node[anchor=center] at (3, 3.5) {$\rho_2 = 1.0$};
    \node[anchor=center] at (3, 1.5) {$\rho_3 = 3.0$};
    \node[anchor=center] at (1, 1.5) {$\rho_4 = 1.0$};

    \draw[thick, ->, blue_sketch] (1.5, 2.5) -- (0.5, 3.16);
    \node[anchor=center, blue_sketch] at (1.0, 2.5) {$\vb{u}_1$};
    \draw[thick, ->, blue_sketch] (3.5, 3.16) -- (2.5, 2.5);
    \node[anchor=center, blue_sketch] at (3.2, 2.5) {$\vb{u}_2$};
    \draw[thick, ->, blue_sketch] (2.5, 1.16) -- (3.5, 0.5);
    \node[anchor=center, blue_sketch] at (3.0, 0.5) {$\vb{u}_3$};
    \draw[thick, ->, blue_sketch] (0.5, 0.5) -- (1.5, 1.16);
    \node[anchor=center, blue_sketch] at (1.2, 0.5) {$\vb{u}_4$};

\end{tikzpicture}
        \caption{Case 2 2D-Riemann problem.}
        \label{fig:riemann_2d_sketch_case2}
    \end{subfigure}
    \begin{subfigure}[b]{0.45\textwidth}
        \centering
        \begin{tikzpicture}

    \draw[thick] (0, 0) rectangle (4, 4);
    \draw[dotted] (2, 0) -- (2, 4);
    \draw[dotted] (0, 2) -- (4, 2);

    \node[anchor=center] at (1, 3.5) {$\rho_1 = 2.0$};
    \node[anchor=center] at (3, 3.5) {$\rho_2 = 1.0$};
    \node[anchor=center] at (3, 1.5) {$\rho_3 = 3.0$};
    \node[anchor=center] at (1, 1.5) {$\rho_4 = 1.0$};

    \draw[thick, ->, blue_sketch] (1.5, 2.5) -- (0.5, 3.16);
    \node[anchor=center, blue_sketch] at (1.0, 2.5) {$\vb{u}_1$};
    \draw[thick, ->, blue_sketch] (2.5, 2.5) -- (3.5, 3.16);
    \node[anchor=center, blue_sketch] at (3.2, 2.5) {$\vb{u}_2$};
    \draw[thick, ->, blue_sketch] (2.5, 1.16) -- (3.5, 0.5);
    \node[anchor=center, blue_sketch] at (3.0, 0.5) {$\vb{u}_3$};
    \draw[thick, ->, blue_sketch] (1.5, 1.16) -- (0.5, 0.5);
    \node[anchor=center, blue_sketch] at (1.2, 0.5) {$\vb{u}_4$};

\end{tikzpicture}
        \caption{Case 3 2D-Riemann problem.}
        \label{fig:riemann_2d_sketch_case3}
    \end{subfigure}
    \centering
    \begin{subfigure}[b]{0.45\textwidth}
        \centering
        \begin{tikzpicture}

    \draw[thick] (0, 0) rectangle (4, 4);
    \draw[dotted] (2, 0) -- (2, 4);
    \draw[dotted] (0, 2) -- (4, 2);

    \node[anchor=center] at (1, 3.5) {$\rho_1 = 2.0$};
    \node[anchor=center] at (3, 3.5) {$\rho_2 = 1.0$};
    \node[anchor=center] at (3, 1.5) {$\rho_3 = 3.0$};
    \node[anchor=center] at (1, 1.5) {$\rho_4 = 1.0$};

    \draw[thick, ->, blue_sketch] (0.5, 3.16) -- (1.5, 2.5);
    \node[anchor=center, blue_sketch] at (1.0, 2.5) {$\vb{u}_1$};
    \draw[thick, ->, blue_sketch] (3.5, 3.16) -- (2.5, 2.5);
    \node[anchor=center, blue_sketch] at (3.2, 2.5) {$\vb{u}_2$};
    \draw[thick, ->, blue_sketch]  (3.5, 0.5) -- (2.5, 1.16);
    \node[anchor=center, blue_sketch] at (3.0, 0.5) {$\vb{u}_3$};
    \draw[thick, ->, blue_sketch] (0.5, 0.5) -- (1.5, 1.16);
    \node[anchor=center, blue_sketch] at (1.2, 0.5) {$\vb{u}_4$};

\end{tikzpicture}
        \caption{Case 4 2D-Riemann problem.}
        \label{fig:riemann_2d_sketch_case4}
    \end{subfigure}
    \caption{The four 2D Riemann problems considered.}
    \label{fig:2d_riemann_cases}
\end{figure}
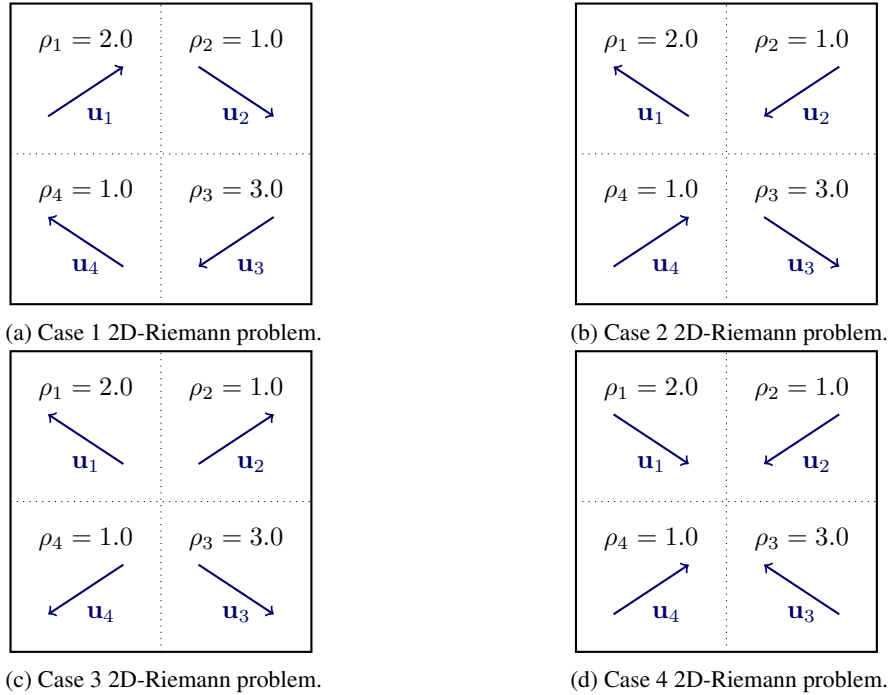

Another generalization of the 1D shock tube is to define four states in a square as shown in Fig.~\ref{fig:riemann_2d_sketch}. Four different cases are considered here: the initial values for the first two cases are taken from \cite{deng2019} and the third and fourth cases are variants of these. The geometry is a $[-0.5, -0.5] \times [0.5, 0.5]$ square with $602 \times 602$ resolution. All units are adimensionalized here: pressure is always constant at 1.0 in all cases while the absolute values of the speeds on the x and y axis are 0.75 and 0.5 respectively. The different studied configurations are shown in Fig.~\ref{fig:2d_riemann_cases} where the orientations of the speeds are given with arrows. The first case is a mixing interface where the four states turn clockwise. The second case consists of a shock in one diagonal and an expansion in the other diagonal. The third and fourth cases are respectively an all-in or all-out configuration in terms of speeds' directions.

\begin{figure}[htbp]
    \centering
    \begin{subfigure}[b]{0.45\textwidth}
        \centering
        \includegraphics[width=\textwidth]{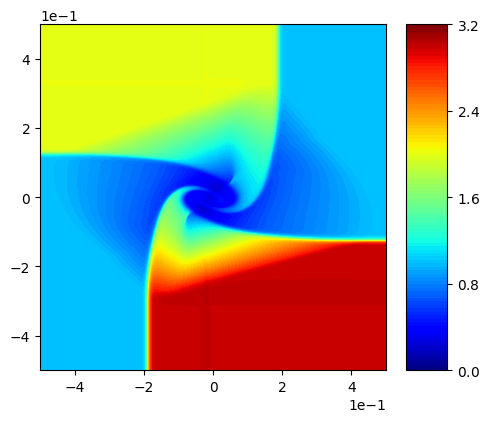}
        \caption{Case 1}
        \label{fig:riemann_2d_sol_case1}
    \end{subfigure}
    \centering
    \begin{subfigure}[b]{0.43\textwidth}
        \centering
        \includegraphics[width=\textwidth]{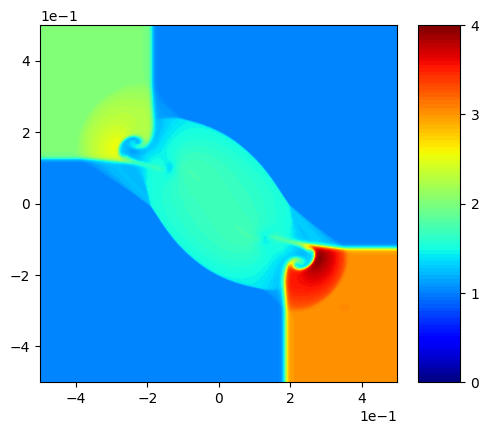}
        \caption{Case 2}
        \label{fig:riemann_2d_sol_case2}
    \end{subfigure}
    \begin{subfigure}[b]{0.45\textwidth}
        \centering
        \includegraphics[width=\textwidth]{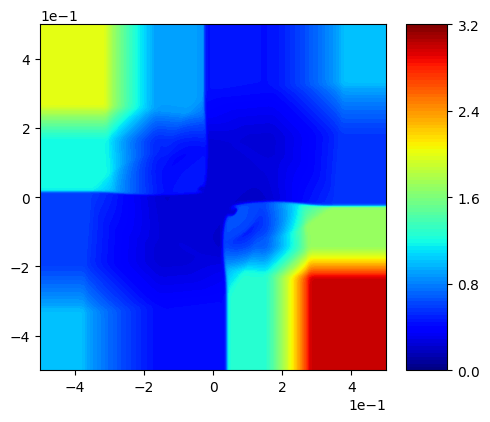}
        \caption{Case 3}
        \label{fig:riemann_2d_sol_case3}
    \end{subfigure}
    \centering
    \begin{subfigure}[b]{0.47\textwidth}
        \centering
        \includegraphics[width=\textwidth]{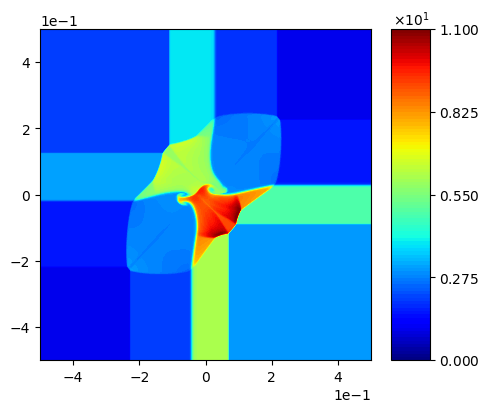}
        \caption{Case 4}
        \label{fig:riemann_2d_sol_case4}
    \end{subfigure}
    \caption{The four 2D Riemann solution from AVIP using HLLC-M RK2.}
    \label{fig:2d_riemann_sols}
\end{figure}

Results of the four 2D Riemann cases using HLLC MUSCL RK2 are shown in Fig.~\ref{fig:2d_riemann_sols}. Interfaces and shocks are well captured by the scheme for all cases. The RK1 time integration produces wiggles similar to the ones observed for the convective vortex test case especially for the first case where vortices appear.

\subsection{One-dimensional flames}
The capability of the Riemann solvers to simulate flames is critical for plasma assisted combustion simulations. An atmospheric pressure one-dimensional flame at equivalence ratio $\phi = 0.8$ is simulated in AVIP using a converged solution from CANTERA \cite{cantera}. A simple two-step chemical scheme with six species is used \cite{franzelli_twostep_bfer}:

\begin{gather}
    \ce{CH4 + 1.5 O2 -> CO + 2H2O} \\
    \ce{CO + 0.5 O2 -> CO2}
\end{gather}

The 1D domain is 2 cm long and contains 500 quadrangular cells or 1000 triangular cells which have the same shape as Fig.~\ref{fig:elements_1d}. Boundary conditions and the geometry are summarized in Fig.~\ref{fig:flame_1d_sketch}: inlet and outlet NSCBC boundary conditions are applied at the left and right parts of the domain while symmetry conditions are applied at the top and bottom. The inlet velocity of \SI{0.2815}{\metre\per\second} is the laminar flame speed velocity computed using the GRI-3.0 mechanism \cite{gri30}.

\begin{figure}[htbp]
    \centering
    \begin{tikzpicture}

    \draw[thick] (0, 0) rectangle (5, 1);
    \draw[<->] (0, -0.7) -- (5, -0.7);
    \node[anchor=center] at (2.5, -1) {2 cm};

    \node[anchor=center] at (-1.4, 0.8) {Inlet};
    \node[anchor=center] at (-1.4, 0.2) {$u = \SI{0.32}{\metre\per\second}$};
    \draw[->] (0, 0.75) -- (0.8, 0.75);
    \draw[->] (0, 0.5) -- (0.8, 0.5);
    \draw[->] (0, 0.25) -- (0.8, 0.25);

    \node[anchor=center] at (6, 0.8) {Outlet};
    \node[anchor=center] at (6, 0.2) {$p$ = 1 atm};

    \node[anchor=center] at (2.5, 1.3) {Symmetry};
    \node[anchor=center] at (2.5, -0.3) {Symmetry};

\end{tikzpicture}
    \caption{One dimensional flame configuration and boundary conditions.}
    \label{fig:flame_1d_sketch}
\end{figure}

\begin{figure}[htbp]
    \centering
    \begin{subfigure}[b]{\textwidth}
        \centering
        \includegraphics[width=0.95\textwidth]{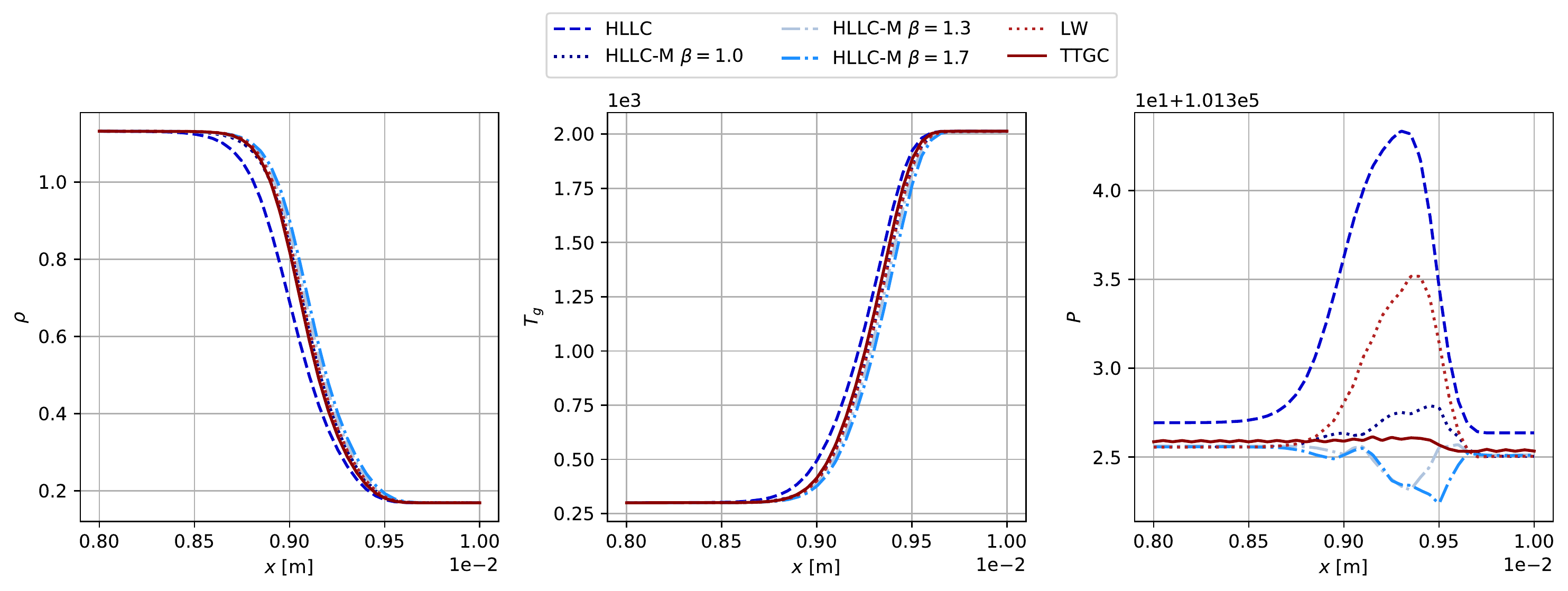}
    \caption{$t = \SI{4}{\milli\second}$}
    \label{fig:quad_1dflame_1}
    \end{subfigure}
    \begin{subfigure}[b]{\textwidth}
        \centering
        \includegraphics[width=0.95\textwidth]{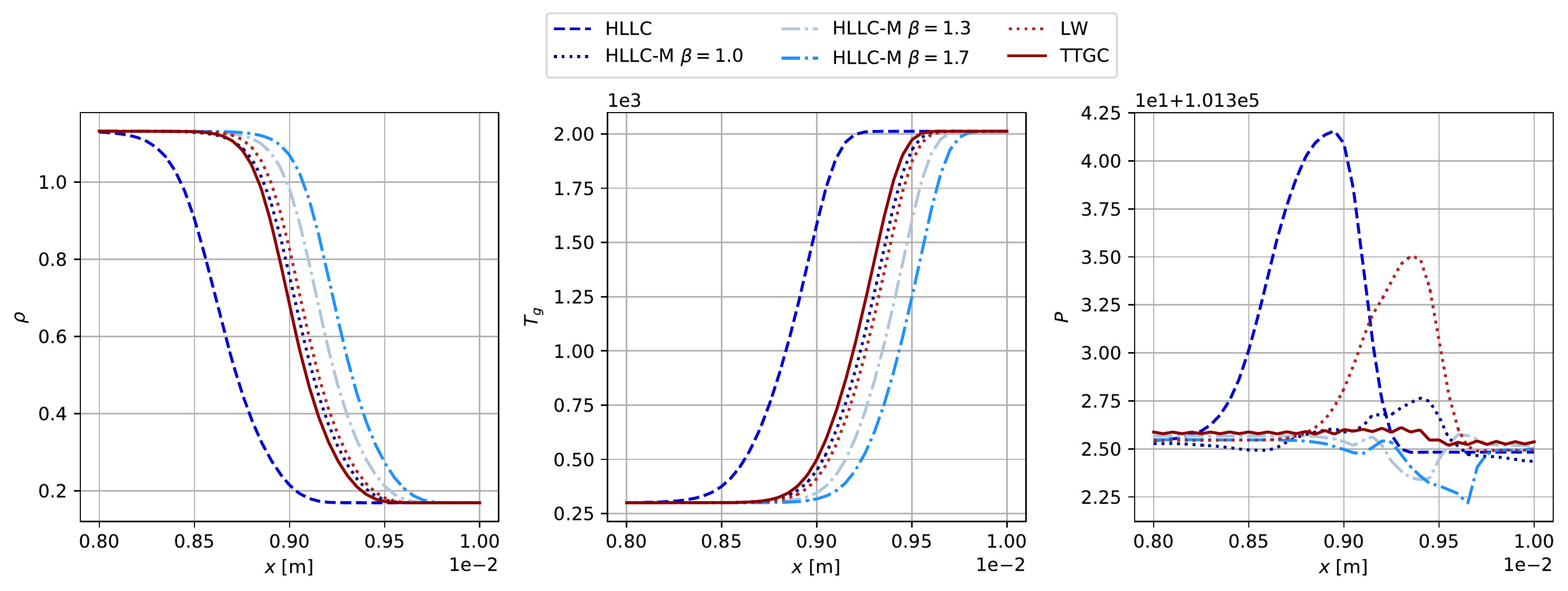}
    \caption{$t = \SI{32}{\milli\second}$}
    \label{fig:quad_1dflame_2}
    \end{subfigure}
    \begin{subfigure}[b]{\textwidth}
        \centering
        \includegraphics[width=0.95\textwidth]{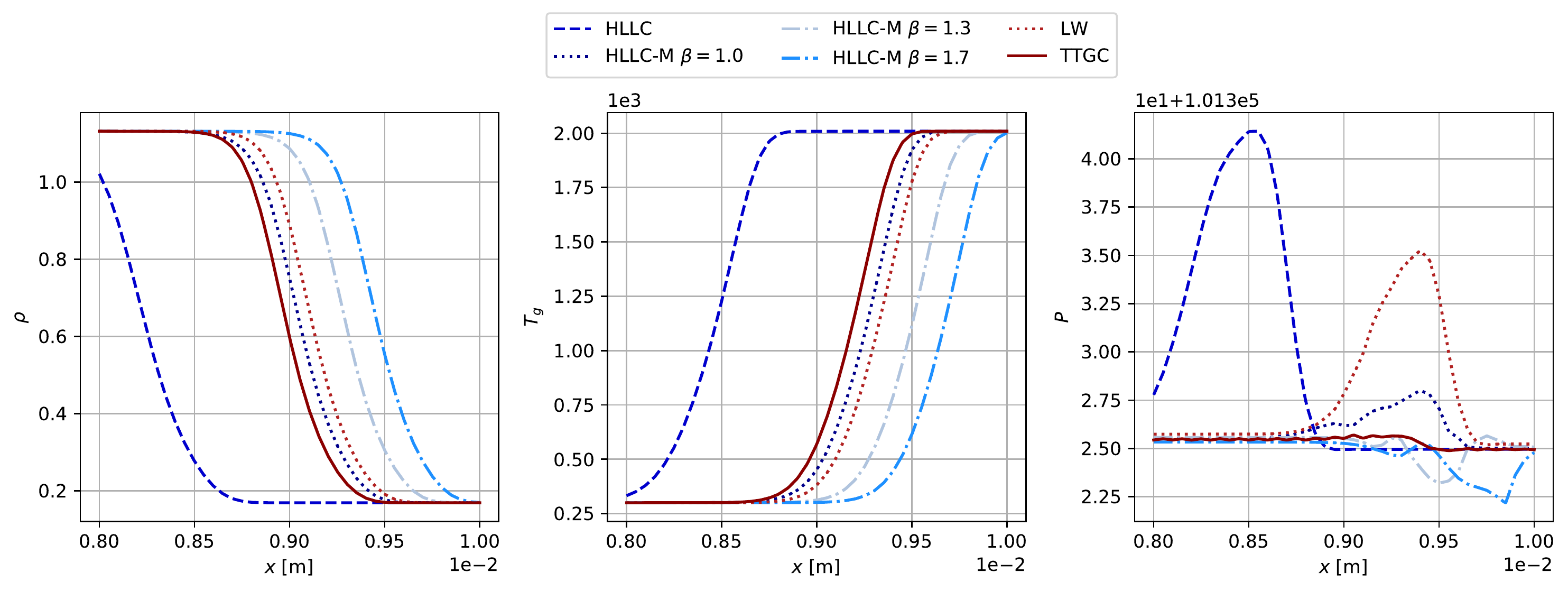}
    \caption{$t = \SI{64}{\milli\second}$}
    \label{fig:quad_1dflame_3}
    \end{subfigure}
    \caption{One-dimensional flames using Riemann solvers and classical AVBP schemes at $t=4, 32$ and 64 ms.}
    \label{fig:1d_flame}
\end{figure}

Simulations have been run for $10^{-1}$ s and results for HLLC and HLLC MUSCL are compared with classical schemes from AVBP that are embedded in AVIP: Lax-Wendroff and TTGC \cite{lamarquethesis}. The density, pressure and temperature for the different schemes are shown in Fig.~\ref{fig:1d_flame} for the quadrangular mesh at three different instants. A drift of the different schemes due to an error in the flame speed propagation is observed compared to the theoretical value of \SI{28.15}{\centi\metre\per\second}. TTGC converges to a value slightly above at \SI{28.16}{\centi\metre\per\second} (0.03\% error) after 0.08 s whereas LW underestimates the laminar flame speed at \SI{27.93}{\centi\metre\per\second} (0.78\% error). The diffusive scheme HLLC overestimates the flame speed by roughly 30\% making it unrelevant for flame front propagation simulation. For HLLC MUSCL RK2, the Sweby parameter has a great influence: at $\beta = 1.0$, the flame front of HLLC MUSCL RK2 is between TTGC and LW which is satisfactory but as the parameter is increased the laminar flame speed is underestimated quite significantly so that a value of $\beta$ below 1.3 should be prescribed for combustion applications.

Concerning the pressure, for this case the TTGC results are taken as reference since this scheme is the most accurate so that the right pressure jump is about 1 Pa. As expected, HLLC behaves poorly due to its diffusive nature and an overshoot of pressure of about 20 Pa is observed in the flame front while having a less steep flame front. The MUSCL procedure allows to retrieve a steeper flame front and a diminished pressure overshoot which is less than the one from LW for both quadrangular and triangular meshes. Increasing the value of the Sweby limiter $\beta$ causes the pressure overshoot to become an undershoot but the amplitude remains similar (around 4 Pa).

To conclude the Sweby parameter of the HLLC MUSCL RK2 scheme should be kept sufficiently low (below 1.3) to have proper flame propagation with reasonable flame speed and thickness.

\section{Conclusion}

The modeling of plasma species and its resolution using robust schemes have been detailed in this work as plasma discharges are very stiff phenomena. The improved Scharfetter-Gummel scheme and limited Lax-Wendroff scheme prove to be both efficient at simulating plasma discharges in quadrangular and triangular elements, respectively.

Robust schemes are also needed for the reactive Navier-Stokes equations and Riemann solvers have been implemented in both AVIP and AVBP where its ability to solve shocks could be interesting in other areas than just plasma assisted combustion.

Now that robust schemes have been developed and validated for both plasmas and gas mixtures, plasma assisted combustion can be carried out and will be released in future work.

\section*{Acknowledgments}
This work was supported by the ANR projects PASTEC (ANR-16-CE22-0005), GECCO (ANR-17-CE06-0019) and was performed using HPC resources from GENCI-TGCC (Grant 2021- A0112B10157).

\bibliographystyle{unsrt}
\bibliography{refs}

\begin{thebibliography}{10}

\bibitem{raimbault}
J-L. Raimbault.
\newblock {\em Introduction to the kinetic theory of weakly ionized plasma}.
\newblock Laboratoire de Physique des Plasmas, 2018.

\bibitem{jackson_electrodynamics}
John~David Jackson.
\newblock {\em Classical Electrodynamics}.
\newblock John Wiley \& Sons, 1999.

\bibitem{celes2008}
Sébastien Célestin.
\newblock {\em Study of the dynamics of streamers in air at atmospheric
  pressure}.
\newblock PhD thesis, 2008.
\newblock Thèse de doctorat dirigée par Bourdon, Anne et Rousseau, Antoine
  Physique Châtenay-Malabry, Ecole centrale de Paris 2008.

\bibitem{tholin2012}
Fabien Tholin.
\newblock {\em Numerical simulation of nanosecond repetitively pulsed
  discharges in air at atmospheric pressure : Application to plasma-assisted
  combustion}.
\newblock PhD thesis, 2012.
\newblock Thèse de doctorat dirigée par Bourdon, Anne Physique
  Châtenay-Malabry, Ecole centrale de Paris 2012.

\bibitem{tnc}
Thierry Poinsot and Denis Veynante.
\newblock {\em Theoretical and Numerical Combustion}.
\newblock Third edition, 2012.

\bibitem{Kulikovsky1995}
A.~A. Kulikovsky.
\newblock {A more accurate Scharfetter-Gummel algorithm of electron transport
  for semiconductor and gas discharge simulation}.
\newblock {\em Journal of Computational Physics}, 1995.

\bibitem{hirsch}
Charles Hirsch.
\newblock {\em Numerical Computation of Internal and External Flows: The
  Fundamentals of Computational Fluid Dynamics}.
\newblock Elsevier, second edition, 2007.

\bibitem{harris1998}
John Harris and Horst Stoecker.
\newblock {\em The Handbook of Mathematics and Computational Science}.
\newblock 01 1998.

\bibitem{Cheng2022}
L.~Cheng.
\newblock {\em Detailed Numerical Simulation of Multi-Dimensional Plasma
  Assisted Combustion}.
\newblock PhD thesis, Institut National Polytechnique de Toulouse, 2022.

\bibitem{lamarquethesis}
N.~Lamarque.
\newblock {\em Numerical schemes and boundary conditions for the LES of
  two-phase flows in helicopter chambers}.
\newblock PhD thesis, CERFACS, 2007.

\bibitem{Auffray2007a}
V.~Auffray.
\newblock {\em {Etude comparative de sch{\'{e}}mas num{\'{e}}riques pour la
  mod{\'{e}}lisation de ph{\'{e}}nom{\`{e}}nes diffusifs sur maillages
  multi{\'{e}}l{\'{e}}ments}}.
\newblock PhD thesis, CERFACS, 2007.

\bibitem{scharfetter1969}
D.L. Scharfetter and H.K. Gummel.
\newblock Large-signal analysis of a silicon read diode oscillator.
\newblock {\em IEEE Transactions on Electron Devices}, 16(1):64--77, 1969.

\bibitem{lax_1960}
Peter Lax and Burton Wendroff.
\newblock System of conservation laws.
\newblock {\em Comm. Pure Appl. Math}, 13:217--237, 1960.

\bibitem{sweby_1984}
P.~K. Sweby.
\newblock High resolution schemes using flux limiters for hyperbolic
  conservation laws.
\newblock {\em SIAM Journal on Numerical Analysis}, 21(5):995--1011, 1984.

\bibitem{vanleer_1974}
Bram {van Leer}.
\newblock Towards the ultimate conservative difference scheme. ii. monotonicity
  and conservation combined in a second-order scheme.
\newblock {\em Journal of Computational Physics}, 14(4):361--370, 1974.

\bibitem{toro}
Eleuterio Toro.
\newblock {\em Riemann Solvers and Numerical Methods for Fluid Dynamics: A
  Practical Introduction}.
\newblock 01 2009.

\bibitem{alauzet_2010}
F.~Alauzet and A.~Loseille.
\newblock High-order sonic boom modeling based on adaptive methods.
\newblock {\em Journal of Computational Physics}, 229(3):561--593, 2010.

\bibitem{Joncquieres2019}
V.~Joncquieres.
\newblock {\em Mod\'{e}lisation et simulation num\'{e}rique des moteurs \`{a}
  effet Hall}.
\newblock PhD thesis, Institut National Polytechnique de Toulouse, 2019.

\bibitem{poinsot1992}
T.J Poinsot and S.K Lelef.
\newblock Boundary conditions for direct simulations of compressible viscous
  flows.
\newblock {\em Journal of Computational Physics}, 101(1):104--129, 1992.

\bibitem{moureau2005}
V.~Moureau, G.~Lartigue, Y.~Sommerer, C.~Angelberger, O.~Colin, and T.~Poinsot.
\newblock Numerical methods for unsteady compressible multi-component reacting
  flows on fixed and moving grids.
\newblock {\em Journal of Computational Physics}, 202(2):710--736, 2005.

\bibitem{bagheri_benchmark}
B~Bagheri, J~Teunissen, Ute Ebert, Markus~M. Becker, S~Chen, O.~Ducasse,
  O.~Eichwald, D.~Loffhagen, Alejandro Luque, Db~Diana Mihailova, J.-M. Plewa,
  Jan van Dijk, and Mohammed Yousfi.
\newblock Comparison of six simulation codes for positive streamers in air.
\newblock {\em Plasma Sources Science and Technology}, 2018.

\bibitem{plasmachemistry}
A.~Fridman.
\newblock {\em Plasma Chemistry}.
\newblock Cambridge University Press, 2008.

\bibitem{bourdon2007}
A~Bourdon, V~P Pasko, N~Y Liu, S~C{\'{e}}lestin, P~S{\'{e}}gur, and E~Marode.
\newblock Efficient models for photoionization produced by non-thermal gas
  discharges in air based on radiative transfer and the helmholtz equations.
\newblock 16(3):656--678, aug 2007.

\bibitem{Morrow1997}
R~Morrow and J~J Lowke.
\newblock Streamer propagation in air.
\newblock {\em Journal of Physics D: Applied Physics}, 30(4):614--627, February
  1997.

\bibitem{deng2019}
Xi~Deng, Pierre Boivin, and Feng Xiao.
\newblock A new formulation for two-wave riemann solver accurate at contact
  interfaces.
\newblock {\em Physics of Fluids}, 31:046102, 04 2019.

\bibitem{cantera}
David~G. Goodwin, Raymond~L. Speth, Harry~K. Moffat, and Bryan~W. Weber.
\newblock Cantera: An object-oriented software toolkit for chemical kinetics,
  thermodynamics, and transport processes.
\newblock \url{https://www.cantera.org}, 2021.
\newblock Version 2.5.1.

\bibitem{franzelli_twostep_bfer}
Benedetta Franzelli, Riber Eleonore, Gicquel L.Y.M., and Thierry Poinsot.
\newblock Large eddy simulation of combustion instabilities in a lean partially
  premixed swirled flame.
\newblock {\em Combustion and Flame}, 159:621--637, 02 2012.

\bibitem{gri30}
Gregory~P. Smith, David~M. Golden, Michael Frenklach, Nigel~W. Moriarty, Boris
  Eiteneer, Mikhail Goldenberg, C.~Thomas Bowman, Ronald~K. Hanson, Soonho
  Song, William~C. Gardiner, Jr. Vitali~V. Lissianski, and Zhiwei Qin.

\end{thebibliography}

\end{document}